\pdfoutput=1
\documentclass[acmsmall]{acmart}

\setcopyright{rightsretained}
\acmDOI{10.1145/3632915}
\acmYear{2024}
\copyrightyear{2024}
\acmSubmissionID{popl24main-p382-p}
\acmJournal{PACMPL}
\acmVolume{8}
\acmNumber{POPL}
\acmArticle{73}
\acmMonth{1}
\received{2023-07-11}
\received[accepted]{2023-11-07}

\usepackage[utf8]{inputenc}
\usepackage{amsmath}
\usepackage{amsthm}
\usepackage{stmaryrd}
\usepackage{adjustbox}
\usepackage{multirow}
\usepackage{mathtools}
\usepackage[inline]{enumitem}
\usepackage{xspace}
\usepackage{comment}
\usepackage{xcolor}
\usepackage[ruled,vlined,resetcount,linesnumbered]{algorithm2e}
\usepackage{listings}
\usepackage[frozencache,cachedir=minted-cache]{minted}
\usepackage{caption}
\usepackage{subcaption}
\usepackage{tikz}
\usetikzlibrary{automata, positioning, arrows, shapes,shapes.geometric, calc}

\usepackage{float}

\usepackage{wrapfig}
\usepackage{tabularx}
\usepackage{contour}
\usepackage{bold-extra}

\lstdefinestyle{mystyle}{
  basicstyle=\ttfamily,
  columns=fullflexible,
  keepspaces=true,
  keywordstyle=\color{green},
}
\usepackage{breqn}
\usepackage[htt]{hyphenat}

\newcommand{\execution}[2]{
\scalebox{1}{
  \begin{tikzpicture}%
    \foreach \x in {1,...,#1}
    \node[right] at (1.2*\x+0.2,0.25) {${\x}$};
    \draw (1.2,0) -- (#1*1.1+0.9,0);%
    \pgfmathsetmacro{\y}{1};%
    #2%
    \draw (1.2,0) -- (1.2,-0.4*\y);%
    \draw (#1*1.1+0.9,0) -- (#1*1.1+0.9,-0.4*\y);%
    \foreach \x in {2,...,#1}
    \draw (1.2,-0.4*\y) -- (#1*1.1+0.9,-0.4*\y);%
  \end{tikzpicture}
}
}

\newcommand{\figev}[2]{
\pgfmathsetmacro{\y}{\y+1};
\pgfmathsetmacro{\y}{\y-1};
\node [left] at (1.25,-0.4*\y)  {\pgfmathprintnumber{\y}};%
\node at (#1*1.1 + 0.45,-0.4*\y) { #2 };%
\pgfmathsetmacro{\y}{\y+1};
}

\newcommand{\executionfull}[9]{
\scalebox{#3}{
  \begin{tikzpicture}
    \foreach \x in {1,...,#1}
    \node[right] at (#4*\x+#6, #8) {$t_{\x}$}; 
    \draw (#7,0) -- (#1*#4+#7,0); 
    \pgfmathsetmacro{\y}{1};
    #2 
    \draw (#7,0) -- (#7,-#5*\y); 
    \draw (#1*#4+#7,0) -- (#1*#4+#7,-#5*\y); 
    \draw (#7,-#5*\y) -- (#1*#4+#7,-#5*\y); 
    \ifthenelse{#9 = 1}{
      \foreach \x in {2,...,#1}
      \draw[dashed] (#4*\x+#7-#4,0) -- (#4*\x+#7-#4,-#5*\y); 
    }{}
  \end{tikzpicture}
}
}

\newcommand{\figevfull}[9]{
\ifthenelse{#7 = 1}{
  \ifthenelse{#8 = -1}{
    \node [left] at (#5,(-#4*\y))  {\pgfmathprintnumber{\y}};%
  }{
    \node [left] at (#5,(-#4*\y))  {#9};%
  }
}{}
\node at (#1*#3 + #6,(-#4*\y)) {$ #2 $};%
\pgfmathsetmacro{\y}{\y+1};
}

\newcommand{\OV}[3]{
\scalebox{1}{
  \begin{tikzpicture}%
    \node[right] at (1.3,0.25) {$A_1$};
    \draw (1.2,0) -- (2,0);%
    \pgfmathsetmacro{\y}{1};%
    #1%
    \draw (1.2,0) -- (1.2,-0.4*\y);%
    \draw (2,0) -- (2,-0.4*\y);%
    \draw (1.2,-0.4*\y) -- (2,-0.4*\y);%
    
    \node[right] at (2.2,0.25)  {$A_2$};
    \draw (2.1,0) -- (2.9,0);%
    \pgfmathsetmacro{\y}{1};%
    #2%
    \draw (2.1,0) -- (2.1,-0.4*\y);%
    \draw (2.9,0) -- (2.9,-0.4*\y);%
    \draw (2.1,-0.4*\y) -- (2.9,-0.4*\y);%
    
    \node[right] at (3.1,0.25)  {$A_3$};
    \draw (3,0) -- (3.8,0);%
    \pgfmathsetmacro{\y}{1};%
    #3%
    \draw (3,0) -- (3,-0.4*\y);%
    \draw (3.8,0) -- (3.8,-0.4*\y);%
    \draw (3,-0.4*\y) -- (3.8,-0.4*\y);%
  \end{tikzpicture}
}
}

\newcommand{\ovX}[1]{
\pgfmathsetmacro{\y}{\y+1};
\pgfmathsetmacro{\y}{\y-1};
\node at (1.2 + 0.4,-0.4*\y) { #1 };%
\pgfmathsetmacro{\y}{\y+1};
}

\newcommand{\ovY}[1]{
\pgfmathsetmacro{\y}{\y+1};
\pgfmathsetmacro{\y}{\y-1};
\node at (2.1 + 0.4,-0.4*\y) { #1 };%
\pgfmathsetmacro{\y}{\y+1};
}

\newcommand{\ovZ}[1]{
\pgfmathsetmacro{\y}{\y+1};
\pgfmathsetmacro{\y}{\y-1};
\node at (3 + 0.4,-0.4*\y) { #1 };%
\pgfmathsetmacro{\y}{\y+1};
}

\newenvironment{myparagraph}[1]{\vspace*{0.05in}\noindent{\bf #1.}}{}
\newcommand{\figlabel}[1]{\label{fig:#1}}

\theoremstyle{definition}
\newtheorem{example}{Example}[section]
\newtheorem{definition}{Definition}
\newtheorem{problem}{Problem}[section]
\newtheorem{theorem}{Theorem}[section]
\newtheorem{lemma}{Lemma}[section]
\newtheorem{proposition}{Proposition}[section]
\newtheorem{corollary}{Corollary}[section]
\theoremstyle{remark}

\newcommand{\figref}[1]{Figure~\ref{fig:#1}}
\newcommand{\seclabel}[1]{\label{sec:#1}}
\newcommand{\secref}[1]{Section~\ref{sec:#1}}
\newcommand{\exlabel}[1]{\label{ex:#1}}
\newcommand{\exref}[1]{Example~\ref{ex:#1}}
\newcommand{\deflabel}[1]{\label{def:#1}}
\newcommand{\defref}[1]{Definition~\ref{def:#1}}
\newcommand{\problabel}[1]{\label{prob:#1}}
\newcommand{\probref}[1]{Problem~\ref{prob:#1}}
\newcommand{\thmlabel}[1]{\label{thm:#1}}
\newcommand{\thmref}[1]{Theorem~\ref{thm:#1}}
\newcommand{\proplabel}[1]{\label{prop:#1}}
\newcommand{\propref}[1]{Proposition~\ref{prop:#1}}
\newcommand{\lemlabel}[1]{\label{lem:#1}}
\newcommand{\lemref}[1]{Lemma~\ref{lem:#1}}

\newcommand{\equlabel}[1]{\label{eq:#1}}
\newcommand{\equref}[1]{Equation~(\ref{eq:#1})}
\newcommand{\applabel}[1]{\label{app:#1}}
\newcommand{\appref}[1]{Appendix~\ref{app:#1}}
\newcommand{\algolabel}[1]{\label{algo:#1}}
\newcommand{\algoref}[1]{Algorithm~\ref{algo:#1}}
\newcommand{\tablabel}[1]{\label{tab:#1}}
\newcommand{\tabref}[1]{Table~\ref{tab:#1}}
\newcommand{\linlabel}[1]{\nllabel{nl:#1}}
\newcommand{\linref}[1]{line~\ref{nl:#1}}
\newtheorem*{rep@theorem}{\rep@title}
\newcommand{\newreptheorem}[2]{%
\newenvironment{rep#1}[1]{%
 \def\rep@title{#2 \ref{##1}}%
 \begin{rep@theorem}}%
 {\end{rep@theorem}}}
\makeatother
\newreptheorem{theorem}{Theorem}
\newreptheorem{lemma}{Lemma}

\newcommand{\set}[1]{\{#1\}}
\newcommand{\setpred}[2]{\set{#1 \,|\, #2}}
\newcommand{\nats}{\mathbb{N}}
\newcommand{\tuple}[1]{\langle#1\rangle}
\newcommand{\sqtuple}[1]{[#1]}
\newcommand{\proj}[2]{#1|_{#2}}
\newcommand{\lin}[1]{\mathsf{Lin}(#1)}
\renewcommand{\emptyset}{\varnothing}
\newcommand{\emptystr}{\varepsilon}

\newcommand{\tr}{\sigma}
\newcommand{\evid}[1]{\tuple{#1}}
\newcommand{\ev}[1]{\sqtuple{#1}}

\newcommand{\creorder}[1]{\textsf{CReorderings}(#1)}
\newcommand{\events}[1]{\textsf{Events}_{#1}}
\newcommand{\trord}[1]{<^{#1}}

\newcommand{\alphabet}{\Sigma}
\newcommand{\lbl}[1]{\mathsf{lab}(#1)}
\newcommand{\id}[1]{\mathsf{id}(#1)}
\newcommand{\threads}{\mathcal{T}}
\newcommand{\ops}{\mathcal{O}}
\newcommand{\vars}{\mathcal{X}}
\newcommand{\locks}{\mathcal{L}}
\newcommand{\lk}{\ell}
\newcommand{\rd}{\mathtt{r}}
\newcommand{\wt}{\mathtt{w}}
\newcommand{\acq}{\mathtt{acq}}
\newcommand{\rel}{\mathtt{rel}}
\newcommand{\fork}{\mathtt{fork}}

\newcommand{\indep}{\mathcal{I}}
\newcommand{\dep}{\mathcal{D}}
\newcommand{\wdthtext}[1]{\mathsf{width}(#1)}
\newcommand{\wdth}{\alpha}
\newcommand{\mazeq}[1]{\equiv_{#1}}
\newcommand{\mazpo}[2]{\leq^{#2}_{#1}}
\newcommand{\nmazpo}[2]{\not\leq^{#2}_{#1}}
\newcommand{\eqcl}[2]{[#2]_{#1}}
\newcommand{\ideal}{\mathsf{X}}
\newcommand{\pideal}[1]{\mathsf{PIdeal}(#1)}
\newcommand{\mxml}[1]{\mathsf{max}(#1)}
\newcommand{\pre}[1]{\mathsf{pre}(#1)}

\newcommand{\NP}{\textsf{NP}}

\newcommand{\poly}[1]{\textsf{poly}(#1)}
\newcommand{\SETH}{\textsf{SETH}\xspace}
\newcommand{\ovp}[1]{$#1$-\textsf{OV}\xspace}

\newcommand{\cd}[1]{\texttt{#1}}

\definecolor{eventColor}{rgb}{0.86, 0.86, 0.95} 
\definecolor{babyblueeyes}{rgb}{0.63, 0.79, 0.95}
\definecolor{lightskyblue}{rgb}{0.7, 0.81, 0.98}
\definecolor{lavender}{rgb}{0.9, 0.9, 0.98}
\definecolor{classColor}{rgb}{0.13, 0.53, 0.69}
\definecolor{methodColor}{rgb}{0.01, 0.16, 0.48}
\definecolor{callFuncColor}{rgb}{0.42,0.49,0.14}

\newcommand{\clsName}[1]{\textcolor{blue}{\textbf{\texttt{#1}}}}
\newcommand{\mthdName}[1]{\textcolor{blue}{\texttt{#1}}}
\newcommand{\fldName}[1]{\texttt{#1}}

\def\mymul#1#2{\color{#1}\underline{{\color{black}#2}}\color{black}}

\contourlength{0.3pt}

\newcommand{\patt}[2]{\alphabet^*{#1}_1\alphabet^*\dots\alphabet^*{#1}_{#2}\alphabet^*}
\newcommand{\patts}[1]{\texttt{Patt}_{#1}}
\newcommand{\patre}{\mathtt{PATT}}
\newcommand{\seq}[2]{{#1}_1, \dots, {#1}_{#2}}
\newcommand{\seqsup}[3]{{#1}^{#3}_{1}, \dots, {#1}^{#3}_{#2}}
\newcommand{\patwit}{{locally admissible tuple}\xspace}

\newcommand{\sort}[2]{\mathsf{sort}({#1}, {#2})}
\newcommand{\aft}[2]{\mathsf{After}_{#1}(#2)}
\newcommand{\aftvar}{\mathbb{A}}
\newcommand{\adset}[2]{{\partial}\mathsf{AdTpls}({#1}, {#2})}
\newcommand{\vc}{\mathsf{VC}}

\newcommand{\bertoni}{\textsc{B.ni}\xspace}
\newcommand{\ptrack}{\textsc{P.Tk}\xspace}
\newcommand{\pattrack}{\textsc{PatternTrack}\xspace}
\newcommand{\fbertoni}{\textsc{Bertoni}\xspace}

\newcommand{\fasttrack}{\textsc{FastTrack}\xspace}

\newtoggle{appendix}
\toggletrue{appendix}

\begin{document}

\title{Predictive Monitoring against Pattern Regular Languages}

\author{Zhendong Ang}
\orcid{0009-0002-0214-3462}
\affiliation{%
  \institution{National University of Singapore}
  \city{Singapore}
  \country{Singapore}
}
\email{zhendong.ang@u.nus.edu}

\author{Umang Mathur}
\orcid{0000-0002-7610-0660}
\affiliation{%
  \institution{National University of Singapore}
  \city{Singapore}
  \country{Singapore}
}
\email{umathur@comp.nus.edu.sg}


\begin{abstract}
    \text{\color{red} (Note: The proof of Theorem 3.2 has been revised. Results unchanged.)}
    
    \noindent
    While current bug detection techniques for concurrent software focus on
    unearthing low-level issues such as data races or deadlocks,
    they often fall short of discovering
    more intricate temporal behaviours that can arise even in the absence of such low-level issues.
    In this paper, we focus on the problem of dynamically analysing 
    concurrent software against high-level temporal specifications such as LTL.
    Existing techniques for runtime monitoring against such specifications are primarily designed for sequential software and
    remain inadequate in the presence of concurrency --- violations may be observed only in intricate thread interleavings, requiring many re-runs of
    the underlying software in conjunction with the analysis.
    Towards this, we study the problem of \emph{predictive runtime monitoring},
    inspired by the analogous problem of \emph{predictive data race detection} studied extensively recently.
    The predictive runtime monitoring question asks, given an execution $\tr$,
    if it can be soundly \emph{reordered} to expose violations of a specification.
    In general, this problem may become easily intractable when either the specifications
    or the notion of reorderings used is complex.
    
    In this paper, we focus on specifications that are given in regular languages.
    Our notion of reorderings is \emph{trace equivalence},
    where an execution is considered a reordering of another
    if it can be obtained from the latter 
    by successively commuting adjacent independent actions.
    We first show that, even in this simplistic setting,
    the problem of predictive monitoring admits a super-linear lower bound of $O(n^\wdth)$,
    where $n$ is the number of events in the execution,
    and $\wdth$ is a parameter describing the degree of commutativity,
    and typically corresponds to the number of threads in the execution. 
    As a result,  predictive runtime monitoring even in this setting
    is unlikely to be efficiently solvable,
    unlike in the non-predictive setting where the problem can be checked using
    a deterministic finite automaton (and thus, a constant-space streaming linear-time algorithm).
    
    Towards this, we identify a sub-class of regular languages, called
    \emph{pattern languages} (and their extension \emph{generalized pattern languages}).
    Pattern languages can naturally express specific ordering of some number of (labelled) events,
    and have been inspired by popular empirical hypotheses underlying
    many concurrency bug detection approaches such as the ``small bug depth'' hypothesis.
    More importantly, we show that for pattern (and generalized pattern) languages,
    the predictive monitoring problem can be solved 
    using a constant-space streaming linear-time algorithm.
    We implement and evaluate our algorithm \pattrack 
    on benchmarks from the literature and show that it is effective
    in monitoring large-scale applications.
\end{abstract}

\ccsdesc[500]{Software and its engineering~Software verification and validation}
\ccsdesc[300]{Theory of computation~Theory and algorithms for application domains}
\ccsdesc[300]{Theory of computation~Program analysis}

\keywords{concurrency, dynamic analysis, predictive monitoring, complexity}


\maketitle


\section{Introduction}

Writing reliable concurrent programs remains a challenge to date.
Subtle bugs, arising due to intricate choreography of threads, 
often evade rigorous testing but appear in deployment under intense workloads.
The first line of defence against such bugs is tools that enable developers 
to find concurrency errors automatically. 
Towards this, both static and dynamic analysis approaches have been proposed.
Static approaches~\cite{RacerD2018,racerx2003,Chord2006} 
are typically geared towards certifying program correctness
and can overwhelm developers with excessive false positive reports~\cite{Sadowski2018}.
Dynamic analysis techniques, on the other hand, 
remain the preferred approach for the task of automated bug detection. 
Here one executes the underlying program, observes its behaviour, 
and infers the presence or absence of bugs based on this observed behaviour. 
Even though such techniques are implicitly incomplete,
they remain popular in practice, thanks to the inherent
desirable properties such as low overhead and soundness of bug reports.
Unsurprisingly, such techniques enjoy more widespread adoption~\cite{threadsanitizer}.

Traditional dynamic analysis approaches for detecting concurrency bugs 
often cater to implicit specifications such as data race freedom, deadlock freedom
or atomicity of code blocks. 
However, such techniques do not attempt to expose undesirable
behaviours due to faulty order of interactions of threads,
that are nevertheless symptomatic of serious failures.
In practice, developers often rely on specific properties such as
class invariants or temporal behaviours such as ``file \cd{f} cannot 
be closed in between the creation of \cd{f} and access to \cd{f}'' 
to reason about the overall safety of their software.
Nevertheless, the validity of such invariants 
often relies on complex synchronizations to be put in place,
a task that even the most expert developers struggle to get right.
Tools such as data race detectors are not helpful either; they fail to expose such high-level problems.

\begin{figure}[t]
\begin{subfigure}[b]{0.4\textwidth}
\begin{minted}[fontsize=\scriptsize,linenos]{Java}
public class DBPlayer {
  AtomicInteger count = new AtomicInteger(0);
  ConcurrentHashMap.KeySetView<Integer, Boolean> inputs = ConcurrentHashMap.newKeySet();

  public void play(int event, int lastPos) {
    inputs.add(event);
    count.set(lastPos);
  }

  public void reset() {
    inputs.clear();
    count.set(0);
  }

  public static void main(String[] args) 
    throws InterruptedException{
    final DBPlayer player = new DBPlayer();

    Thread t1 = new Thread(new Runnable() {
      public void run() {
        player.reset();
    }});

    Thread t2 = new Thread(new Runnable() {
      public void run() {
        player.play(365, 5);
    }});

    t1.start();
    t2.start();
  }
}


\end{minted}
\caption{Example Java program $P$}
\figlabel{example-program}
\end{subfigure}
\begin{subfigure}[b]{0.24\textwidth}
\centering
\footnotesize
\begin{align*}
\arraycolsep=1.4pt\def\arraystretch{1.2}
\begin{array}{rl}
e_{1}:& \ev{t_{\sf main},\fork(t_1)} \\
e_{2}:& \ev{t_{\sf main},\fork(t_2)} \\
e_{3}:& \ev{t_1,{\mthdName{reset}}_{\sf Call}} \\
e_{4}:& \mymul{red}{\ev{t_1,{\mthdName{clear}}_{\sf Call}(\fldName{inputs})}} \\
e_{5}:& \ev{t_1,{\tt write}(\fldName{inputs})} \\
e_{6}:& \ev{t_1,{\mthdName{clear}}_{\sf Return}} \\
e_{7}:& \mymul{red}{\ev{t_1,\mthdName{set}(\fldName{count})}} \\
e_{8}:& \ev{t_1,{\mthdName{reset}}_{\sf Return}} \\
e_{9}:& \ev{t_2,{\mthdName{play}}_{\sf Call}} \\
e_{10}:& \mymul{red}{\ev{t_2,{\mthdName{add}}_{\sf Call}(\fldName{inputs})}} \\
e_{11}:& \ev{t_2,{\tt write}(\fldName{inputs})} \\
e_{12}:& \ev{t_2,{\mthdName{add}}_{\sf Return}} \\
e_{13}:& \mymul{red}{\ev{t_2,\mthdName{set}(\fldName{count})}} \\
e_{14}:& \ev{t_2,{\mthdName{play}}_{\sf Return}} \\
\\
\end{array}
\end{align*}
\caption{Execution $\tr_{\sf safe}$ of $P$}
\figlabel{example-safe-exec}
\end{subfigure}
\begin{subfigure}[b]{0.24\textwidth}
\centering
\footnotesize
\begin{align*}
\arraycolsep=1.4pt\def\arraystretch{1.2}
\begin{array}{rl}
e_{1}:& \ev{t_{\sf main},\fork(t_1)} \\
e_{2}:& \ev{t_{\sf main},\fork(t_2)} \\
e_{9}:& \ev{t_2,{\mthdName{play}}_{\sf Call}} \\
e_{10}:& \mymul{red}{\ev{t_2,{\mthdName{add}}_{\sf Call}(\fldName{inputs})}} \\
e_{3}:& \ev{t_1,{\mthdName{reset}}_{\sf Call}} \\
e_{4}:& \mymul{red}{\ev{t_1,{\mthdName{clear}}_{\sf Call}(\fldName{inputs})}} \\
e_{5}:& \ev{t_1,{\tt write}(\fldName{inputs})} \\
e_{6}:& \ev{t_1,{\mthdName{clear}}_{\sf Return}} \\
e_{7}:& \mymul{red}{\ev{t_1,\mthdName{set}(\fldName{count})}} \\
e_{8}:& \ev{t_1,{\mthdName{reset}}_{\sf Return}} \\
e_{11}:& \ev{t_2,{\tt write}(\fldName{inputs})} \\
e_{12}:& \ev{t_2,{\mthdName{add}}_{\sf Return}} \\
e_{13}:& \mymul{red}{\ev{t_2,\mthdName{set}(\fldName{count})}} \\
e_{14}:& \ev{t_2,{\mthdName{play}}_{\sf Return}} \\
\\
\end{array}
\end{align*}
\caption{Execution $\tr_{\sf fail}$ of $P$}
\figlabel{example-fail-exec}
\end{subfigure}
\caption{Example Java program P and its executions. 
Interleaving \mthdName{play}{\tt()} and \mthdName{reset}{\tt()} results in the values of \fldName{count} and \fldName{inputs} to be in an inconsistent state.}
\figlabel{motivating-example}
\end{figure}

Consider, for example, the simplistic Java program $P$ in~\figref{example-program}. 
This is a code snippet inspired from the GitHub project {\tt antlrworks}~\cite{antlrworks}.
The class \clsName{DBPlayer} contains two fields, \fldName{count} and \fldName{inputs},
and two methods, \mthdName{play} and \mthdName{reset}, both of which write to both fields.
In order to keep \fldName{count} and \fldName{inputs} in a consistent state,
executions of method \mthdName{play} and \mthdName{reset} should behave atomically.
Observe that this is a high-level data race, instead of a classical data race property~\cite{artho2003high}, which are prevented because of the use of thread-safe data structure and atomic integer. 
This type of violation is also studied in~\cite{vaziri2006associating}.

The problematic behaviour in the above example can, in fact,
be expressed, in a temporal specification formalism like LTL,
or as a regular language $L_{\sf fail}$, 
that expresses the order in which the events $\mthdName{add}_{\sf Call}$,
$\mthdName{clear}_{\sf Call}$,
on the \fldName{inputs} field,
and the two \mthdName{set} operations on the \fldName{count} field are made.
Such specification languages have been extensively studied in the
context of runtime verification and efficient algorithms and tools have been 
developed to monitor such specifications. 
Such techniques, however, are not well-suited for monitoring concurrent programs.
Thanks to non-determinism due to thread scheduling,
even the most well-engineered LTL-monitoring technique 
may require many re-executions of the program under test to eventually observe a violation of a desirable property. 
Consider again the program $P$ be from~\figref{example-program},
and the execution $\tr_{\sf safe}$ generated when monitoring $P$ (\figref{example-safe-exec}).
As such, $\tr_{\sf safe} \not\in L_{\sf fail}$ fails to expose the violation
encoded as $L_{\sf fail}$, 
even though the very similar execution $\tr_{\sf fail}$ (\figref{example-fail-exec}) of the same program $P$
can very well expose the problematic behaviour.
This highlights the lack of robustness in traditional monitoring
approaches that simply check whether an execution observed during dynamic analysis
witnesses a specification.

To tackle this challenge, we borrow wisdom from
\emph{predictive} approaches~\cite{wcp2017,Pavlogiannis2019,SyncP2021,Kalhauge2018,RVPredict2014,Said11,cp2012, huang2015gpredict}, that are
capable of exposing low-level bugs such as data races even starting from executions that do not explicitly observe the concurrency bug under question. 
Towards this, we consider the analogous \emph{predictive runtime monitoring} 
question --- given
a specification, encoded as a language $L$ and given 
an execution $\tr$ of some concurrent program $P$,
can $\tr$ be \emph{reordered}, in a sound manner, to execution $\tr'$
so that $\tr' \in L$?
Observe that an efficient solution to the predictive runtime monitoring
problem has the potential to enhance coverage of traditional 
runtime verification approaches that only focus on the observed execution, when used for concurrent software.

Even in the context of the simplest specification --- data races --- the 
predictive question is intractable, in general~\cite{Mathur2020}. 
Nevertheless, tractable and linear-time algorithms 
for predictive data race detection have been proposed recently~\cite{SyncP2021,wcp2017}. 
GPredict~\cite{huang2015gpredict} also uses predictive analysis to detect high-level concurrency properties.
However, it is SMT-based and not scalable in practice.
In this work, we aim to develop efficient algorithms for
predictive monitoring against richer specifications.
In the context of data races, 
the key insight underlying efficient algorithms is to restrict the search space of reorderings,
and is also central to our work presented here.

We consider reorderings described by \emph{Mazurkiewicz trace equivalence}~\cite{Mazurkiewicz1987}.
Here, one fixes an \emph{independence relation}, consisting of
pairs of actions that can be commuted when present next to each other in any context.
With such a commutativity specification,
two executions are deemed equivalent if they can be reached from each other by successive commutations of independent actions. 
Indeed, the most popular approaches for race detection --- those based on the \emph{happens-before} partial order~\cite{Djit1999,fasttrack2009} ---
 essentially employ this reasoning principle and, as a result, can be implemented using a fast linear-time algorithm.

In this paper, we study the problem of predictive trace monitoring against a regular language $L$ --- given an execution $\tr$, is $\tr$ Mazurkiewicz trace equivalent to an execution that belongs to $L$?
The problem has previously been studied in a theoretical setting~\cite{Bertoni1989}, 
where the authors proposed an algorithm that runs in time $O(n^\wdth)$
for executions with $n$ events; $\wdth$ is a parameter given by the independence relation,
and is typically equal to the number of threads in the execution.
Such an algorithm is unlikely to be practical
for large-scale software applications where executions typically contain millions of events. 
Unfortunately, as we show in our paper, 
in general, this problem cannot be solved using a faster algorithm --- we
show a matching conditional lower bound of $\Omega(n^\wdth)$
using techniques from fine-grained complexity~\cite{williams2018}.

To this end, we identify a class of specifications that we 
call \emph{pattern} regular languages for which predictive 
monitoring can be performed efficiently. 
This class of pattern languages has been inspired by systematic
concurrency bug detection approaches that often rely on empirical hypotheses
such as the \emph{small bug depth hypothesis} --- 
``\emph{Empirically, though, many bugs in programs depend on the 
precise ordering of a small number of events $\ldots$ 
there is some constant $d$ and a subset $\ldots$ 
of events such that some ordering of these $d$ events already exposes the bug no matter how all other events are ordered.}''~\cite{chistikov2016}.
Pattern languages are the natural class of specifications
under such a hypothesis --- a language $L$ is a pattern
language if it is of the form $\patts{\seq{a}{d}} = \patt{a}{d}$,
where $\alphabet$ is the alphabet labelling events, and $\seq{a}{d} \in \alphabet$
are $d$ symbols (not necessarily distinct from each other).
We also propose an extension, the class  of \emph{generalized}
pattern languages which are unions of pattern languages.
We expect that this class of specifications, previously deployed in domains such as
pattern mining~\cite{pattern1995} or root cause analysis~\cite{Murali2021},
will be useful in advancing the state of the art in concurrency testing,
given the success of past approaches based on similar empirical hypotheses
including context bounding~\cite{Musuvathi2007,Burckhardt2010,Emmi2011},
small bug depth~\cite{Burckhardt2010,chistikov2016}, 
delay bounding~\cite{Emmi2011},
coarse interleaving hypothesis~\cite{Kasikci2017} for testing Java-like 
multi-threaded software programs, event-driven and asynchronous programs,
distributed systems, and more recently for testing and model checking
of weak memory behaviours~\cite{Abdulla2019,Gao2023}.

The main result of this work is that the class of (generalized) pattern languages 
admits a constant-space linear-time algorithm
for the predictive monitoring problem under trace equivalence.
This algorithm relies on several interesting insights specific to pattern
languages that enable our efficient monitoring algorithm,
and is of independent interest.

We show that, in fact, we can use \emph{vector clocks} to
further develop a fast and scalable predictive monitoring algorithm \pattrack.
We implement \pattrack in Java and evaluate its performance
over a set of benchmark Java applications derived from prior works.
Our comparison of \pattrack against the classical algorithm due to~\cite{Bertoni1989}
reveals the effectiveness of both a specialized class of specifications
and a linear-time algorithm for predictive monitoring.

The rest of the paper is organized as follows.
In~\secref{prelims}, we discuss background on modelling events, executions,
and questions in runtime monitoring.
In~\secref{trace-languages}, we recall trace equivalence and present our hardness result
for predictive trace monitoring against regular languages (\thmref{ov-hardness}).
In \secref{pattern-languages}, we formally define the class of pattern and generalized pattern languages, 
and provide language-theoretic characterizations of these languages.
In \secref{algo-pattern}, we present our main result --- a constant-space
streaming algorithm for predictive trace monitoring against generalized pattern regular languages,
and a more practical algorithm \pattrack that uses vector clocks (\secref{algo-pattern}) for the same task.
We present our experimental evaluation in~\secref{experiments},
discuss related work in \secref{related}
and conclude in \secref{conclusions}.


\section{Preliminaries}
\seclabel{prelims}

In this section, we present relevant background and notations 
useful for the rest of the paper.



\myparagraph{Events and Executions}
Our primary subject of study is concurrent programs and their executions with
a focus on developing monitoring algorithms for them.
In this regard, executions can be modelled as sequences of events.
For this, we will consider an \emph{event alphabet}, or simply, alphabet $\alphabet$, which
will intuitively represent a set of relevant events observed during program execution.
As an example, $\alphabet$ can include information about instructions corresponding
to method invocations/returns, or even loads or stores of memory locations.
For the purpose of modelling concurrent programs, 
the symbols in our alphabet $\alphabet$ also contain information about thread identifiers,
belonging to a fixed set $\threads$.
An event is a unique occurrence of a symbol from $\alphabet$.
Formally, an event is a tuple $e = \evid{id, a}$,
where $id$ is the unique identifier of $e$, 
while $a \in \alphabet$ is the label of $e$;
we will denote the identifiers and labels of $e$ using
the notation $\lbl{e} = a$ and $\id{e} = id$.
In the context of concurrent executions, 
the labels of events will be of the form $a = \ev{t, op} \in \alphabet$
to denote information about the thread $t \in \threads$ that 
performs the corresponding event and the operation
$op$ performed in that event.
We will often omit the identifier $id$ of events when it is clear from the context.
An execution can then be modelled as a sequence of events with labels from $\alphabet$.
We use $\events{\tr}$ to denote the set of events occurring in an execution $\tr$.
We will use $\trord{\tr}$ to denote the total order induced by $\tr$, 
i.e., $e_1\trord{\tr} e_2$ iff the event $e_1$ appears before $e_2$ in the sequence $\tr$.
The word constructed by concatenating the labels of $\tr$ will be denoted
by $\lbl{\tr} \in \alphabet^*$, and will often confuse $\tr$ with $\lbl{\tr}$
when the identifiers of events are not important.
We will use $|\tr|$ to denote the number of events in $\tr$.

\myparagraph{Runtime Monitoring}
Runtime verification has emerged as a powerful technique for enhancing
reliability of software and hardware systems 
by augmenting traditional software testing methods 
with more expressive specifications that can be 
checked or \emph{monitored} during the execution.
One of the primary components in a traditional runtime monitoring workflow 
is the choice of specifications to be monitored at runtime.
In general, many different classes of specifications have been considered 
in the literature and practice, including temporal logics like
LTL~\cite{Pnueli1977} and its variants~\cite{Giacomo2013}
with quantitative~\cite{Maler2004} and timing aspects~\cite{MTL1990}.
More expressive variants such as extensions with calls and returns~\cite{CaReT2004},
extended regular expressions or vanilla regular languages have also been considered.
A large class of these specifications can be described by a
language (or a set of executions) over a chosen alphabet $\alphabet$ of interest.
Examples include LTL, FOLTL, extended-regular expressions~\cite{RV2003}, etc, 
and the more general class of regular languages.
For a language $L$ over the alphabet $\alphabet$, 
the runtime monitoring problem then asks --- given an execution
$\tr \in \alphabet^*$ coming from a program $P$,
does $\tr$ satisfy the specification $L$, i.e., $\tr \in L$?
When $L$ represents an undesired (resp. desired) behaviour, 
a positive (resp. negative) answer to the membership question 
$\tr \in L$ can then be flagged to developers or system
engineers who can fix the erroneous behaviour (or in some cases, refine the specification).
In this paper, we study the monitoring problem in the context of 
concurrent programs, with a focus on developing more robust solutions to the monitoring problem.
Next, we motivate the \emph{predictive} monitoring
problem and formally describe it subsequently.

\begin{example}
\exlabel{example-motivating}
Consider again the Java program $P$, shown in~\figref{example-program},
that implements class 
\clsName{DBPlayer}
with fields, \fldName{count} and \fldName{inputs}, and
member methods, \fldName{play} (for adding event indices to \fldName{inputs} and log position into \fldName{count})
and \mthdName{reset} (for clearing \fldName{inputs} and set \fldName{count} as 0).
The function \mthdName{main} creates a \clsName{DBPlayer} object, 
and then calls \mthdName{play} and \mthdName{reset} concurrently in different threads.
Suppose, we are interested in observing the behaviours of the program
that lead to the inconsistent state of fields \fldName{count} and \fldName{inputs}.
The events thus generated
belong to the alphabet $\alphabet_{\sf Ex} = \setpred{\ev{t, op}}{t \in \threads, op \in \ops}$,
where $\threads = \set{t_{\sf main}, t_1, t_2}$ is the set of threads corresponding to the main function and two forked threads,
while $\ops = \setpred{\fork(t)}{t \in \threads} \cup \set{{\tt write(inputs)}} 
\cup \set{{\mthdName{add}}_{\sf Call}(\fldName{inputs}), {\mthdName{add}}_{\sf Return},
{\mthdName{clear}}_{\sf Call}(\fldName{inputs}), {\mthdName{clear}}_{\sf Return}, 
\mthdName{set}(\fldName{count})}$.
Observe that an execution $\tr$ of this program might leave the fields in an inconsistent state
if it belongs to the following language $L_{\sf fail}$ that constrains the order of calls to 
\textcolor{blue}{\tt add}, \textcolor{blue}{\tt clear}, and two {\tt \textcolor{blue}{set}}s to {\tt count}:
\[
\alphabet_{\sf Ex}^* {\cdot} \ev{t_2, {\mthdName{add}}_{\sf Call}(\fldName{inputs})} {\cdot} \alphabet_{\sf Ex}^* {\cdot} \ev{t_1, {\mthdName{clear}}_{\sf Call}(\fldName{inputs})} {\cdot} \alphabet_{\sf Ex}^* {\cdot} \ev{t_1, \mthdName{set}(\fldName{count})} {\cdot} \alphabet_{\sf Ex}^* {\cdot} \ev{t_2, \mthdName{set}(\fldName{count})} {\cdot} \alphabet_{\sf Ex}^*
\]

In~\figref{example-safe-exec}, we show
an execution $\tr_{\sf safe}$ of $P$, which is a sequence of $14$ events $e_1, \ldots, e_{14}$, 
labelled with $\alphabet_{\sf Ex}$.
Observe that $\tr_{\sf safe} \not\in L_{\sf fail}$ fails to
witness the buggy behaviour of $P$, and would deem the task of
runtime monitoring against $L_{\sf fail}$ unsuccessful.
On the other hand, the bug would be exposed successfully
if the execution $\tr_{\sf fail}$ (\figref{example-fail-exec})
was observed when monitoring $P$.
More importantly, observe that, $\tr_{\sf fail}$ can, in fact, 
be obtained by \emph{reordering} the events of $\tr_{\sf safe}$, 
hinting at the possibility of enhancing the effectiveness of runtime monitoring
via \emph{prediction}!
Notice that an execution belonging to $L_{\sf fail}$ 
may not immediately leave the data in an inconsistent state.
Indeed, $\tr_{\sf fail}$ here is actually benign, 
but it hints towards another execution that orders $e_{11}$ before $e_5$, 
which is invalid.
\end{example}


\exref{example-motivating} illustrates the challenge of naively adapting the
traditional runtime monitoring workflow for the case when the software under test
exhibits concurrency --- even though the underlying software may violate a specification,
exposing the precise execution that witnesses such a violation
is like finding a needle in a haystack.
This is because, even under the same input,
not all thread interleavings may witness a violation of a temporal specification of interest,
and the precise interleaving that indeed witnesses the violation
may require careful orchestration from the thread scheduler.
As a result, the bug may remain unexposed even after several attempts to execute the program.
The emerging class of \emph{predictive analysis} techniques~\cite{wcp2017,cp2012,SyncP2021,Pavlogiannis2019}
attempts to address this problem in the context of concurrency bugs such as data races and deadlocks. 
Here, one observes a single execution of a program, and \emph{infers} additional
executions (from the observed one) that are guaranteed to be generated by 
the same underlying program, and also witness a violation of a property of interest.
Such techniques thus partially resolve the dependence on the often demonic
non-determinism due to thread scheduler, and proactively increase coverage 
when testing concurrent programs.

\myparagraph{Correct Reorderings}
In order to formalize the predictive analysis framework,
we require a notion of the set of \emph{feasible} or \emph{correct}
reorderings of a given execution~\cite{MaximalCausalModel2013}.
For a given execution $\tr$ over $\alphabet$,
the set $\creorder{\tr}$ of correct reorderings of $\tr$
represents the set of executions that will be generated
by any program $P$ that generates $\tr$.
The formal notion of correct reorderings can often
be obtained by formally modelling the programming language,
together with the semantics of concurrent objects~\cite{linearlizability1990} such as
threads, shared memory and synchronization objects such as 
locks~\cite{MaximalCausalModel2013}.
In the case of multi-threaded programs, it is customary to 
include symbols corresponding to reads and writes to all shared objects
and lock acquisition and release events
in the alphabet $\alphabet$.
Prior works have developed several different (yet related)
definitions to precisely capture this notion;
the most prominent notion is due to~\cite{cp2012},
which we describe next.

Given a well-formed\footnote{An execution is well-formed if it can be generated by a program. This means it obeys the semantics of the underlying programming language. As an example, when executions contain events corresponding to acquire and release operations of locks, well-formedness asks that critical sections on the same lock do not overlap.} execution $\tr \in \alphabet^*$, 
we have that an execution $\rho \in \creorder{\tr}$
iff the following conditions hold ---
\begin{enumerate*}
    \item The set of events in $\rho$ is a subset of those in $\tr$,
    \item $\rho$ is well-formed; this means, for example, critical sections on the same lock do not overlap in $\rho$.
    \item $\rho$ preserves the \emph{program-order} of $\tr$, i.e., 
    for every thread $t$, the sequence of events $\proj{\rho}{t}$
    obtained by projecting $\rho$ to events of thread $t$ is a prefix of the
     sequence $\proj{\tr}{t}$ obtained by projecting $\tr$ to $t$.
    \item $\rho$ preserves control flow of $\tr$, i.e., every read event $e$ in $\rho$ observes the same write event $e'$ as in $\tr$.
\end{enumerate*}
Observe that, for programming languages such as Java or C/C++,
$\rho \in \creorder{\tr}$ implies that indeed any control flow
undertaken by $\tr$ can also be undertaken by $\rho$,
and thus, any program that generates $\tr$
can also generate $\rho$, possibly under a different thread interleaving.
In general, the more permissive the set $\creorder{\tr}$,
the higher the computational complexity involved in reasoning with it.
For example, for the precise definition due to~\cite{cp2012}, 
the race \emph{prediction} question --- given an execution $\tr$,
check if there is a $\rho \in \creorder{\tr}$ that exhibits a 
data race --- is an $\NP$-hard problem~\cite{Mathur2020}.
On the other hand, for simpler (and less permissive) notions such as trace equivalence,
which we recall in \secref{trace-languages}, 
the corresponding race prediction question becomes linear-time (and also constant-space) checkable!

\myparagraph{Predictive Monitoring}
We are now ready to formalize the predictive monitoring framework,
by generalizing related notions such as predictive data race detection~\cite{cp2012, Said11,wcp2017,kulkarni2021dynamic,SyncP2021}
or predictive deadlock detection~\cite{tuncc2023sound}.
For a specification language $L$, the predictive monitoring question asks, 
given an execution $\tr \in \alphabet^*$, is there an execution 
$\rho$ such that $\rho \in \creorder{\tr} \cap L$, i.e.,
$\rho$ is a correct reordering of $\tr$ that also witnesses the specification $L$.
In the context of \exref{example-motivating},
while the execution $\tr_{\sf safe}$ (\figref{example-safe-exec})
is a negative instance of the monitoring problem against the language $L_{\sf fail}$,
it is indeed a positive instance of \emph{predictive} monitoring problem
against the same language, because the witnessing execution $\tr_{\sf fail} \in L_{\sf fail}$
is a \emph{correct reordering} of $\tr_{\sf safe}$.
In fact, notions such as predictive data race detection can be
easily formulated in this framework as follows.
Let us fix the set of threads to be $\threads$,
and also a set of memory locations/objects $\vars$.
Consider the alphabet of read and write events: $\alphabet_{\sf RW} = \setpred{\ev{t, \wt(x)}, \ev{t, \rd(x)}}{t \in \threads, x \in \vars}.
\equlabel{alphabet_r_w}$
The following language over $\alphabet_{\sf RW}$
then represents the set of all executions with a data race.
\[L_{\sf race} = 
\bigcup\limits_{\begin{subarray}{c}t \neq t' \in \threads \\ x \in \vars\end{subarray}} 
\alphabet_{\sf RW}^* \bigg( \ev{t, \wt(x)} \ev{t', \wt(x)} + \ev{t, \wt(x)} \ev{t', \rd(x)} + \ev{t, \rd(x)} \ev{t', \wt(x)}  \bigg) \alphabet_{\sf RW}^*
\equlabel{lang_race}\]
The race prediction question then asks to check --- given an execution
$\tr$, is there an execution $\rho \in \creorder{\tr} \cap  L_{\sf race}$?
Observe that, $L_{\sf race}$ is a regular language over $\alphabet_{\sf RW}$.
Likewise, the deadlock prediction question can also be formulated
analogously using a regular language $L_{\sf deadlock}$
over an alphabet that also contains information about lock
acquire and release events: 
\begin{equation}
    \alphabet_{\sf RWL} = \alphabet_{\sf RW} \cup \setpred{\ev{t, \acq(\lk)}, \ev{t, \rel(\lk)}}{t \in \threads, \lk \in \locks}, \equlabel{alphabet_r_w_l}
\end{equation}
where $\locks$ is a fixed set of lock identifiers.
We skip the precise definition of $L_{\sf deadlock}$ here.


\section{Trace Languages and Predictive Monitoring}
\seclabel{trace-languages}

In this section, we will recall trace\footnote{Readers must note that
the use of the term \emph{trace} in this paper is specifically distinguished
from its more contemporary use (to denote a specific log or sequence of
events that happen during an execution). 
The usage of \emph{trace} in this paper is derived from
the more traditional language theoretic notion of Mazurkiewicz \emph{traces}
which denote (equivalence) \emph{classes} of strings over some alphabet, 
instead of a single string modelling a single execution.}
languages~\cite{Mazurkiewicz1987}, and discuss their membership question, 
its connections to predictive monitoring and prior complexity-theoretic results.
We will finally present our hardness result, which is the first main contribution of this work.


\myparagraph{Mazurkiewicz Trace Equivalence}
Trace theory, introduced by A. Mazurkiewicz~\cite{Mazurkiewicz1987}
is a simple yet systematic framework for reasoning about the computation of concurrent programs.
The broad idea of trace theory is to characterize when two executions of a concurrent
programs must be deemed equivalent, based on the notion of commutativity 
of independent actions (or labels).
Formally an \emph{independence} (or \emph{concurrency}) relation 
over an alphabet of actions $\alphabet$ is
an irreflexive and symmetric relation
$\indep \subseteq \alphabet \times \alphabet$ denoting all pairs of actions
that can intuitively be deemed pairwise independent.
For simplicity, we will introduce the dependence relation $\dep = \alphabet \times \alphabet \setminus \indep$,
induced from a fixed independence relation $\indep$,
to denote all pairs of dependent (i.e., not independent) actions.
Together, the pair $(\alphabet, \indep)$ constitutes a \emph{concurrent alphabet}.
Then, the \emph{trace equivalence} induced by $\indep$,
denoted by $\mazeq{\indep}$,
is the smallest equivalence class over $\alphabet^*$
such that for every $(a, b) \in \indep$ and for every
two words $w_1, w_2 \in \alphabet^*$, we have
\[w_1 \cdot a \cdot b \cdot w_2 \mazeq{\indep} w_1 \cdot b \cdot a \cdot w_2.\]
For a string $w \in \alphabet^*$, the trace equivalence class of
$w$ is $\eqcl{\indep}{w} = \setpred{w'}{w \mazeq{\indep} w'}$.

\myparagraph{Trace Partial Order}
An alternative characterization of trace equivalence is often
given in terms of the partial order induced due to the concurrent alphabet.
Formally, given a concurrent alphabet $(\alphabet, \indep)$,
with the dependence relation $\dep = \alphabet \times \alphabet \setminus \indep$,
and a sequence of events $\tr$ labelled with symbols from $\alphabet$,
the partial order induced due to $\tr$, denoted by $\mazpo{\dep}{\tr}$, 
is the smallest partial order over $\events{\tr}$
such that for any two events $e_1, e_2 \in \events{\tr}$, 
if $e_1$ appears before $e_2$ (i.e., $e_1\trord{\tr} e_2$) in the sequence $\tr$ and
$(\lbl{e_1}, \lbl{e_2}) \in \dep$, then
$e_1 \mazpo{\dep}{\tr} e_2$.
One can then show that, the set of linearizations 
$\lin{\mazpo{\dep}{\tr}}$ of this partial order precisely captures
the set of words that are trace equivalent to the label of $\tr$:
\begin{proposition}
\proplabel{po-eq-trace}
Let $\tr$ be an execution over $\alphabet$.
We have,
$\eqcl{\indep}{\lbl{\tr}} = \setpred{\lbl{\tr'}}{ \tr' \in \lin{\mazpo{\dep}{\tr}}}$.
\end{proposition}
As a consequence of \propref{po-eq-trace}, we lift the notion
of trace equivalence from words over $\alphabet$ to executions over $\alphabet$
as follows.
Given two executions $\tr$ and $\tr'$ over the same set of events
(i.e., $\events{\tr} = \events{\tr'}$), we say that
$\tr \mazeq{\indep} \tr'$ if $\tr' \in \lin{\mazpo{\dep}{\tr}}$.
Likewise, we use $\eqcl{\indep}{\tr}$ to denote the set of executions $\tr'$
for which $\tr \mazeq{\indep} \tr'$.

The partial order view of trace equivalence has, in fact,
been routinely exploited in program analysis for concurrency. 
Dynamic analysis techniques such as those designed for 
data race detection~\cite{Djit1999,fasttrack2009} construct a partial order, 
namely the \emph{happens-before} partial order, computed using timestamps~\cite{Mattern89,Fidge91,TreeClocks2022}, which essentially characterizes 
the Mazurkiewicz trace equivalence of an appropriately defined concurrency alphabet.
Likewise, optimization techniques employed in bounded model checking such as 
dynamic partial order reduction~\cite{FlanaganGodefroid2005}
are rooted in Mazurkiewicz trace theory in a similar precise sense.



\myparagraph{Mazurkiewicz Equivalence v/s Correct Reorderings}
Mazurkiewicz trace equivalence provides a sound (but incomplete)
characterization of the space of correct reorderings in the following sense.
We first fix an independence relation over the alphabet $\alphabet$
that soundly characterizes the  commutativity 
induced by the underlying programming languages.
Consider, for example, the alphabet $\alphabet_{\sf RWL}$ 
described previously in~\equref{alphabet_r_w_l}
over some set of threads $\threads$, memory locations $\vars$
and locks $\locks$.
Let $\mathcal{C}_{\sf RW} = \setpred{(a_1(x), a_2(x))}{a_1 = \wt\text{ or }a_2 = \wt,\, \text{ and } x \in \vars}$
be the set of conflicting memory operations,
and let $\mathcal{C}_{\sf L} = \setpred{(a_1(\lk), a_2(\lk))}{a_1, a_2 \in \set{\acq, \rel} \text{ and } \lk \in \locks}$
be the set of conflicting lock operations.
Now consider the independence relation defined as follows:
\begin{align*}
\indep_{\sf RWL} =& \setpred{\big(\ev{t_1, o_1}, \ev{t_2, o_2}\big)}{ t_1 \neq t_2 \text{ and } (o_1, o_2) \not\in \mathcal{C}_{\sf RW} \cup \mathcal{C}_{\sf L}} 
\end{align*}
The above definition of independence relation has been carefully
crafted to ensure that the resulting trace equivalence satisfies two properties.
First, the relative order between a read event $e_\rd$ and a conflicting
write event $e_\wt$ (i.e., $(\lbl{e_\rd}, \lbl{e_\wt}) \in \mathcal{C}_{\sf RW}$)
does not get flipped, ensuring that for any two Mazurkiewicz equivalent executions $\tr$
and $\tr'$, the control flow taken by $\tr$ will also be taken by $\tr'$.
Furthermore, the relative order of conflicting lock operations
does not change, ensuring that if $\tr$ was well-formed
(i.e., critical sections on the same lock do not overlap in $\tr$),
then so is $\tr'$.
This gives us the following; as before, we use $\dep_{\sf RWL} = \alphabet_{\sf RWL}\times \alphabet_{\sf RWL}\setminus \indep_{\sf RWL}$.
\begin{proposition}
Let $\tr$ be an execution over $\alphabet_{\sf RWL}$.
We have $\lin{\mazpo{\dep_{\sf RWL}}{\tr}} \subseteq \creorder{\tr}$,
\end{proposition}
We remark that data race detection techniques like \fasttrack~\cite{fasttrack2009}
are thus implicitly predictive and \emph{sound} because they reason about
trace equivalence induced by the concurrent alphabet $(\alphabet_{\sf RWL}, \indep_{\sf RWL})$, at least until the first data race they report~\cite{SHB2018}.

\begin{example}
\exlabel{Maz-trace}
Consider again the program $P$ (\figref{example-program}) from \exref{example-motivating}.
Let $\mathsf{Dep}_{\sf Ex}$ be the set of dependent instructions,
namely those pairs $(op_1, op_2)$ such that
there is an object ${\tt o}\in\set{{\tt count, inputs}}$ such that 
$(op_1, op_2) \in \set{
\big({\tt write(o)}, {\tt write(o)}\big)}$.
Using this, we can define the independence relation as
$\indep_{\sf Ex} = \setpred{\big(\ev{t_1, op_1}, \ev{t_2, op_2}\big)}{t_1 \neq t_2 \text{ and } (op_1, op_2) \not\in \mathsf{Dep}_{\sf Ex}}$.
Observe that any two consecutive instructions which are independent
according to $\indep_{\sf Ex}$ can be commuted
without affecting the control flow of the program.
Thus, equivalence induced by $\indep_{\sf Ex}$ soundly captures correct reorderings.
Finally, observe that the two executions from \figref{motivating-example}
are also deemed equivalent according to the independence relation
defined here: $\tr_{\sf safe} \mazeq{\indep_{\sf Ex}} \tr_{\sf fail}$.
\end{example}

\subsection{Predictive Monitoring for Trace Equivalence}

The framework of Mazurkiewicz traces is well-equipped
to study the predictive monitoring problem defined 
henceforth in the context of trace equivalence.

\begin{definition}[Predictive Trace Monitoring]
Fix a concurrent alphabet $(\alphabet, \indep)$ and a language $L \subseteq \alphabet^*$.
Given an execution $\tr$ over $\alphabet$ as input, the predictive monitoring problem
asks to check if there is an execution $\tr'$ such that
$\tr' \mazeq{\indep} \tr$ and $\lbl{\tr'} \in L$.
\end{definition}
We remark that the above problem can be equivalently formulated
in terms of words (instead of executions) --- 
given an execution $\tr$ specified as the corresponding
word $w = \lbl{\tr}$, check if $\eqcl{\indep}{w} \cap L \neq \emptyset$.
As an example, for the alphabet $\alphabet_{\sf Ex}$ from \exref{example-motivating}
and the independence relation $\indep_{\sf Ex}$ defined
in \exref{Maz-trace}, the predictive monitoring question would return
YES for the input $\tr_{\sf safe}$ because $\tr_{\sf fail} \in \eqcl{\indep_{\sf Ex}}{\tr_{\sf safe}} \cap L_{\sf fail}$.

\myparagraph{Predictive Trace Monitoring against Regular Languages}
Bertoni, Mauri and Sabadini~\cite{Bertoni1989} studied the problem of
predictive trace monitoring against the class of regular (and context-free) languages
and proposed an algorithm whose time complexity is given in terms 
of a parameter of the concurrency alphabet, which is defined as follows.
The \emph{width} $\wdthtext{\alphabet, \indep}$ 
of the concurrency alphabet $(\alphabet, \indep)$
is the size of the largest clique of the undirected graph whose vertices are 
elements of $\alphabet$ and edges are elements of $\indep$.
We next recall the result of Bertoni et al. relevant to this exposition,
and provide a high-level overview of their algorithm subsequently.

\begin{theorem}[Theorem 6.2 in~\cite{Bertoni1989}]
\thmlabel{bertoni}
The predictive monitoring problem
against regular languages can be solved using an algorithm
that uses  $O(n^\alpha)$ time and space
for the input execution of size $n$, 
where $\wdth = \wdthtext{\alphabet, \indep}$ is the width of the concurrency alphabet.
\end{theorem}

\noindent
\underline{\em Ideals.}
The algorithm due to Bertoni et al. relies on the observation
that the set of \emph{prefixes} of an equivalence class of a given execution
can be defined using an \emph{ideal}.
An ideal $\ideal$ of $\tr$ is a subset $\ideal \subseteq \events{\tr}$
such that for every $e, e' \in \events{\tr}$ such that $e \mazpo{\dep}{\tr} e'$,
if $e' \in \ideal$, we also have $e \in \ideal$.
An event $e \in X$ is said to be \emph{maximal} in $X$
if for every $f \in \events{\tr}$ different from $e$ such that
$e \mazpo{\dep}{\tr} f$, we have $f \not\in X$.
We use $\mxml{\ideal}$ to denote the (unique) set of maximal events of $\ideal$.
We will use $\lin{\mazpo{\dep}{\tr}, \ideal}$ to denote
the set of linearizations of the set $\ideal$ consistent with $\mazpo{\dep}{\tr}$
i.e., each $\rho \in \lin{\mazpo{\dep}{\tr}, \ideal}$ is such that
$\events{\rho} = \ideal$ and for every $e_1 \mazpo{\dep}{\tr} e_2$,
if $e_1, e_2 \in \ideal$, then $e_1 \trord{\rho} e_2$.
Given two ideals $\ideal$ and $\mathsf{Y}$ of $\tr$,
we say that $\ideal$ is an immediate predecessor of $\mathsf{Y}$
if there is an event $e \in \mathsf{Y}$ such that $\mathsf{Y} = \ideal \uplus \set{e}$
(i.e., $e \not\in \ideal$, and $e \in \mxml{\mathsf{Y}}$).
We use $\pre{\ideal}$ to denote the set of all immediate predecessors of $\ideal$.
The principal ideal generated by an event $e \in \events{\tr}$, 
denoted $\pideal{e}$, is the smallest ideal of $\tr$ that contains $e$.
There is a one-to-one correspondence between
an ideal $\ideal$ and the set of its maximal elements.
In particular, an ideal $\ideal$
can be uniquely represented as $\ideal = \bigcup\limits_{e \in \mxml{\ideal}} \pideal{e}$.
The number of maximal elements of an ideal of $\tr$ is bounded
by $\wdth$.
As a result, the number of ideals of $\tr$ is bounded by $O(|\tr|^\wdth)$.

We now recall the algorithm due to Bertoni et al.
Let $\wdth = \wdthtext{\alphabet, \indep}$ be the width of the concurrency alphabet.
Let $\mathcal{A}_L = (Q, Q_0, \delta, F)$ be the non-deterministic finite automaton
corresponding to the regular language $L$ against which we perform
predictive trace monitoring; the size of $\mathcal{A}_L$ is assumed to be $O(1)$.
Let us also fix the input execution $\tr \in \alphabet^*$.
The algorithm computes, for every ideal $\ideal$ of $\tr$,
the set of states $S_\ideal = \setpred{\delta^*(q_0, w)}{\exists \rho \in \lin{\mazpo{\dep}{\tr}, \ideal} \text{ such that } \lbl{\rho} = w}$ that can be reached by the automaton $\mathcal{A}_L$
on some linearization of $\ideal$.
Observe that, $S_\ideal \subseteq Q$ has constant size for any ideal $\ideal$.
This information can be defined inductively using immediate predecessor ideals ---
$S_\ideal = Q_0$ if $\ideal = \emptyset$,
and otherwise,
$S_\ideal = \setpred{q'}{\exists \ideal', q, f \text{ such that } \ideal' \in \pre{\ideal}, q \in S_{\ideal'}, \set{f} = \ideal \setminus \ideal' \text{ and } (q, \lbl{f}, q') \in \delta}$.
The computation of the sets $S_\ideal$ can be performed incrementally in order
of the size of the ideals, starting from ideals of the smallest size.
The final check for determining if  $\eqcl{\indep}{\lbl{\tr}} \cap L \neq \emptyset$ simply corresponds to checking if $F \cap \ideal_\tr \neq \emptyset$,
where $\ideal_\tr = \events{\tr}$ is the largest ideal that includes all events.
The time spent for each ideal $\ideal$ is $O(1)$, and the total number
of ideals is $O(|\tr|^\wdth)$, giving us the desired bound of 
\thmref{bertoni}.

\subsection{Hardness of Predictive Trace Monitoring}
\seclabel{lower-bound}

The algorithm in~\cite{Bertoni1989} runs in time (and space) $O(n^\wdth)$
for executions of size $n$ and where $\wdth$ is the width of the concurrent alphabet,
in the case when the target language is regular.
Even though this is polynomial time (assuming $\wdth$ is constant),
it can become prohibitive when deployed for predictive monitoring
for executions coming from large-scale concurrent software systems,
which routinely generate executions with millions of events.
In sharp contrast to this is the case of non-predictive monitoring,
which boils down to checking membership in a regular language.
The non-predictive monitoring problem for regular languages thus admits
a linear-time, one-pass constant space streaming algorithm, the holy grail of runtime verification.
The question we ask here is --- in the case of predictive trace monitoring,
is there a more efficient algorithm than the $O(n^\wdth)$ algorithm proposed
by Bertoni et al.~\cite{Bertoni1989}? 
Is the exponential dependence on $\wdth$ necessary?

At first glance, predictive trace monitoring, in general, 
does not admit a constant-space (automata-theoretic) algorithm.
Consider, for example, the alphabet $\alphabet = \set{\cd{a}_1, \cd{a}_2, \cd{a}_3}$
and the independence relation $\indep = \setpred{(\cd{a}_i, \cd{a}_j)}{i \neq j \in \alphabet}$,
i.e., all pairs of distinct letters are deemed independent by $\indep$.
Now consider the language $L = (\cd{a}_1 \cd{a}_2 \cd{a}_3)^*$.
Checking if $\eqcl{\indep}{w} \cap L \neq \emptyset$, amounts to checking if
$w \in L' = \setpred{u}{\text{the number of occurrences of } \cd{a}_1, \cd{a}_2 \text{ and } \cd{a}_3 \text{ is equal in } u}$.
Since $L'$ is not a regular language (not even context-free), the
predictive trace monitoring problem does not admit 
a constant-space linear-time algorithm for $L$.
In this work, we establish an even tighter (conditional) lower bound (\thmref{ov-hardness}).
Our hardness result is, in fact, stronger in that it applies even 
for the case of \emph{star-free} regular languages~\cite{McNaughton1971, Schutzenberger1965}.
Further, our result also establishes that
the $O(n^\wdth)$ algorithm due to Bertoni et al. is in fact optimal.

\myparagraph{Complexity Theoretic Conjectures}
One of the most widely believed complexity-theoretic conjectures
is the Strong Exponential Time Hypothesis (\SETH)~\cite{SETH2009}, which states that
for every $\epsilon > 0$,
no (deterministic or randomized) algorithm determines
the satisfiability of a 3SAT formula over $n$ propositions 
in time $O(2^{(1-\epsilon)n})$.
In the proof of our result~\thmref{ov-hardness}, we will show a reduction
from the orthogonal vectors problem \ovp{k}.
An instance of \ovp{k} is $k$ sets $\set{A_1, \ldots, A_k}$,
each of which is a set of $n$ Boolean vectors over $d$ dimensions,
i.e., $A_i \subseteq \set{0, 1}^d$ and $|A_i| = n$.
The \ovp{k} problem asks to check if there are $k$ orthogonal vectors,
i.e., if there are vectors
$a_1 \in A_1, a_2 \in A_2, \ldots, a_k \in A_k$ such that
the norm of their pointwise product vector is $0$, i.e.,
$\sum_{j=1}^d \prod_{i=1}^k a_i[j] = 0$.
Under \SETH, there is no $O(n^{k-\epsilon}\cdot \poly{d})$
algorithm for $k$-OV (no matter what $\epsilon > 0$ we pick)~\cite{williams2018}.

\begin{theorem}
\thmlabel{ov-hardness}
Assume SETH holds.
For each $\epsilon > 0$, there is no algorithm that solves 
the predictive trace monitoring problem against star-free regular languages
(over a concurrent alphabet with width $\wdth$)
in time $O(n^{\wdth-\epsilon})$ for input words of length $n$.
\end{theorem}


\newcommand{\aij}[2]{\cd{a}_{#1, #2}}
\newcommand{\sep}[1]{\#_{#1}}

\begin{proof}
We show a fine-grained reduction from \ovp{k} to the problem of 
predictive trace monitoring against star-free language.
We fix $k \in \nats$.
Then we construct the concurrent alphabet $(\alphabet, \indep)$,
and the star-free language $L$ against which we check 
predictive trace monitoring, as follows.
First, the alphabet $\alphabet$ is partitioned into $k$ disjoint 
alphabets, i.e., $\alphabet = \biguplus_{i=1}^k \alphabet_i$, where
\begin{align*}
\alphabet_i = \Delta_i \cup \set{\sep{i}}, \Delta_i = \set{0_i, 1_i}
\end{align*}
The independence relation is such that all symbols across partitions are deemed
independent, while those within the same partition are deemed dependent,
That is,
\begin{align*}
\indep = \bigcup_{i \neq j} \alphabet_i \times \alphabet_j
\end{align*}
Thus the width of the concurrency alphabet is the number of partitions, i.e., $\wdth = k$.
To define the language $L$, We first introduce a shortcut to denote a finite set of strings, where every string is from $\Delta_1\times \dots \times \Delta_k$ and contains at least one $0_i$:
\begin{align*}
\texttt{AtLeast1Zero} = \Delta_1\times \dots \times \Delta_k - \set{1_1\cdot \dots \cdot 1_k}
\end{align*}
The language $L$ is given by the following regular expression:
\begin{align*}
r = \alphabet^* \cdot \sep{1} \cdot \dots \cdot \sep{k} \cdot \texttt{AtLeast1Zero}^* \cdot \sep{1} \cdot \dots \cdot \sep{k} \cdot \alphabet^*
\end{align*}
Observe that $|\alphabet| = 3k$, $\indep = 9k(k-1)/2$
and $|r| = O(2^k)$.
Also, observe that $L$ is a star-free language because 
$\texttt{AtLeast1Zero}^+$ can be written as
\[\texttt{AtLeast1Zero}^+ = (\texttt{AtLeast1Zero}\cdot \alphabet^* \cap \alphabet^*\cdot \texttt{AtLeast1Zero}) - \bigcup_{\sigma \in \alphabet^k - \texttt{AtLeast1Zero}} \alphabet^* \cdot \Delta_k \cdot \sigma \cdot \alphabet^*, \]
and then, $\texttt{AtLeast1Zero}^* = \texttt{AtLeast1Zero}^+ \cup \set{\epsilon}$, 
where $\set{\epsilon} = \alphabet^* - \bigcup_{a\in \alphabet} a\cdot \alphabet^*$ is a star-free language.

\myparagraph{Reduction}
Given an instance $\mathcal{A} = \set{A_1, A_2, \ldots, A_k}$
of \ovp{k} with $|A_i| = n$ and $A_i \subseteq \set{0, 1}^d$ 
for every $i$, we construct an execution $\tr$ such that 
$\mathcal{A}$ is a positive instance of \ovp{k} iff 
$\eqcl{\indep}{\lbl{\tr}} \cap L \neq \emptyset$.
Our construction ensures that $|\tr| = k(d+1)n+k$.
For each $i$, let us denote the vectors of $A_i$ using 
$\set{v^{(i)}_1, v^{(i)}_2, \dots, v^{(i)}_n}$.
The execution $\tr$ is then a concatenation, obtained by successively appending
smaller sub-executions, one for each set $A_i$:
\[
\tr = \tr_1 \circ \tr_2 \circ \dots \circ \tr_k.
\]
Further, for $1 \leq i \leq k$, the sub-execution $\tr_i$ is a concatenation
of smaller sub-executions corresponding to each vector in $A_i$:
\[
\tr_i = \sep{i} \circ \tr_{i, 1} \circ \sep{i} \circ \tr_{i, 2} \circ \sep{i} \cdots \sep{i} \circ \tr_{i, n} \circ \sep{i}
\]
Further, for each $1 \leq i \leq k$ and $1 \leq j \leq n$, 
the trace $\tr_{i, j}$ encodes the vector $v^{(i)}_j$:
\[
\tr_{i, l} = b_{i,j,1} \circ b_{i,j,2}\circ \cdots \circ b_{i,j,d}
\]
where
\[ 
b_{i,j,l} = 
\begin{cases}
        0_i & \text{if } v^{(i)}_j[l]=0\\
        1_i & \text{otherwise}
\end{cases}
\]
Intuitively, $\sigma$ encodes the instance $\mathcal{A}$ by sequentializing all vectors with special separator symbols $\sep{i}$s. \figref{reduction} illustrates our construction.
The string $\tr'$ (\figref{UMPTL-evidence})
is equivalent to $\tr$ (\figref{UMPTL-input})  and also belongs to $L$; 
symbols marked red highlight membership in $L$ and correspond to the
vectors whose dot product is $0$ (also marked in red in \figref{3-OV-instance}).


\begin{figure}[t]
\begin{subfigure}[b]{0.32\textwidth}
\centering
\scalebox{0.7}{
\OV{
\ovX{101}
\ovX{110}
\ovX{$\color{red} 010$}
}{
\ovY{111}
\ovY{011}
\ovY{$\color{red} 110$}
}{
\ovZ{011}
\ovZ{$\color{red} 101$}
\ovZ{111}
}
}
\caption{\ovp{3} instance with $n=3$, $d=3$}
\figlabel{3-OV-instance}
\end{subfigure}
\begin{subfigure}[b]{0.3\textwidth}
\centering
\scalebox{0.7}{
\execution{3}{
    \figev{1}{$\sep{1}$}
    \figev{1}{$1_1$}
	\figev{1}{$0_1$}
    \figev{1}{$1_1$}
    \figev{1}{$\sep{1}$}
	\figev{1}{$1_1$}
	\figev{1}{$1_1$}
    \figev{1}{$0_1$}
    \figev{1}{$\sep{1}$}
	\figev{1}{$\color{red} 0_1$}
    \figev{1}{$\color{red} 1_1$}
	\figev{1}{$\color{red} 0_1$}
    \figev{1}{$\sep{1}$}
    \figev{2}{$\sep{2}$}
    \figev{2}{$1_2$}
	\figev{2}{$1_2$}
    \figev{2}{$1_2$}
    \figev{2}{$\sep{2}$}
	\figev{2}{$0_2$}
	\figev{2}{$1_2$}
    \figev{2}{$1_2$}
    \figev{2}{$\sep{2}$}
	\figev{2}{$\color{red}1_2$}
    \figev{2}{$\color{red}1_2$}
    \figev{2}{$\color{red}0_2$}
    \figev{2}{$\sep{2}$}
    \figev{3}{$\sep{3}$}
    \figev{3}{$0_3$}
    \figev{3}{$1_3$}
    \figev{3}{$1_3$}
    \figev{3}{$\sep{3}$}
	\figev{3}{$\color{red}1_3$}
    \figev{3}{$\color{red}0_3$}
    \figev{3}{$\color{red}1_3$}
    \figev{3}{$\sep{3}$}
    \figev{3}{$1_3$}
    \figev{3}{$1_3$}
    \figev{3}{$1_3$}
    \figev{3}{$\sep{3}$}
}
}
\caption{Input execution $\tr$}
\figlabel{UMPTL-input}
\end{subfigure}
\begin{subfigure}[b]{0.32\textwidth}
\centering
\scalebox{0.7}{
\execution{3}{
    \figev{1}{$\sep{1}$}
    \figev{1}{$1_1$}
	\figev{1}{$0_1$}
    \figev{1}{$1_1$}
    \figev{1}{$\sep{1}$}
	\figev{1}{$1_1$}
	\figev{1}{$1_1$}
    \figev{1}{$0_1$}
    \figev{2}{$\sep{2}$}
    \figev{2}{$1_2$}
	\figev{2}{$1_2$}
    \figev{2}{$1_2$}
    \figev{2}{$\sep{2}$}
	\figev{2}{$0_2$}
	\figev{2}{$1_2$}
    \figev{2}{$1_2$}
    \figev{3}{$\sep{3}$}
    \figev{3}{$0_3$}
    \figev{3}{$1_3$}
    \figev{3}{$1_3$}
    \figev{1}{$\sep{1}$}
    \figev{2}{$\sep{2}$}
    \figev{3}{$\sep{3}$}
	\figev{1}{$\color{red} 0_1$}
    \figev{2}{$\color{red}1_2$}
    \figev{3}{$\color{red}1_3$}
    \figev{1}{$\color{red} 1_1$}
    \figev{2}{$\color{red}1_2$}
    \figev{3}{$\color{red}0_3$}
    \figev{1}{$\color{red} 0_1$}
    \figev{2}{$\color{red}0_2$}
    \figev{3}{$\color{red}1_3$}
    \figev{1}{$\sep{1}$}
    \figev{2}{$\sep{2}$}
    \figev{3}{$\sep{3}$}
    \figev{3}{$1_3$}
    \figev{3}{$1_3$}
    \figev{3}{$1_3$}
    \figev{3}{$\sep{3}$}
}
}
\caption{Execution $\tr'$ with $\tr'\in \eqcl{\indep}{\tr} \cap L$}
\figlabel{UMPTL-evidence}
\end{subfigure}
\caption{Reduction from \ovp{3}. 
Symbols $0_i, 1_i$ appear in the $i^{\text{th}}$ column in 
(b) and (c).
Symbols appearing in different columns are independent.
$L=\alphabet^* \cdot \sep{1} \cdot \dots \cdot \sep{k} \cdot \texttt{AtLeast1Zero}^* \cdot \sep{1} \cdot \dots \cdot \sep{k} \cdot \alphabet^*$.
}
\figlabel{reduction}
\end{figure}

\myparagraph{Correctness}
 We now argue for the correctness of our reduction.

\noindent
($\Rightarrow$)
Consider $v^{(1)}_{t_1}\in A_1, v^{(2)}_{t_2}\in A_2,\dots, v^{(k)}_{t_k} \in A_k$ 
such that $v^{(1)}_{t_1}\cdot v^{(2)}_{t_2}\cdot \dots \cdot v^{(k)}_{t_k} = 0$.
This means, for every $l\in\set{1, \dots, d}$, 
there is at least one $i\in\{1, \dots, k\}$ such that 
$v^{(i)}_{t_i}[l]=0$. 
$\tr'$ can now be described as follows:
\[\tr' = \tr'_{\sf pre} \circ \tr'_{\sf mid} \circ \tr'_{\sf post}.\]
The prefix $\tr'_{\sf pre}$ (resp. $\tr'_{\sf post}$) is 
obtained by successively concatenating
the first $t_i-1$ (resp. last $n-t_i$) sub-executions of $\tr_i$:
\begin{align*}
\begin{array}{lcr}
\tr'_{\sf pre} &=& \overleftarrow{\tr_1}[1\ldots t_1-1]\circ\dots\circ \overleftarrow{\tr_k}[1\ldots t_k-1] \\
\tr'_{\sf post} &=& \overrightarrow{\tr_1}[t_1+1\ldots n]\circ\dots\circ \overrightarrow{\tr_k}[t_k+1\ldots n]
\end{array}
\end{align*}
where $\overleftarrow{\tr_i}[p\ldots q] = \sep{i} \circ \tr_{i, p} \circ  \cdots \sep{i} \circ \tr_{i, q}$ and $\overrightarrow{\tr_i}[p\ldots q] = \tr_{i, p} \circ \sep{i} \circ  \cdots \sep{i} \circ \tr_{i, q} \circ \sep{i}$ and .
The sub-execution $\tr'_{\sf mid}$ is obtained by
successive concatenation of sub-executions corresponding to the
$j^{\text{th}}$ components of each vector $v^{(i)}_{t_i}$:
\begin{align*}
\tr'_{\sf mid} = \sep{1} \circ \sep{2} \cdots \sep{k} \circ \tr'_{{\sf mid}, 1} \circ \tr'_{{\sf mid}, 2} \circ \ldots \circ \tr'_{{\sf mid}, d} \circ \sep{1} \circ \sep{2} \cdots \sep{k}
\end{align*}
where for each $1 \leq l \leq d$,
\begin{align*}
\tr'_{{\sf mid}, l} = b_{1, t_1, l} \circ b_{2, t_2, l} \circ \ldots \circ b_{k, t_k, l}
\end{align*}
First, observe that $\tr'$ does not flip the order of any two events
that are dependent. 
Thus, $\tr' \mazeq{\indep} \tr$.
Second, each $\tr'_{{\sf mid}, l}$ matches $\texttt{AtLeast1Zero}$ by the definition of $t_1 \dots t_k$.
Thus, $\tr' \in L$.\\

\noindent
($\Leftarrow$) 
Consider an equivalent execution $\tr' \mazeq{\indep} \tr$ 
such that $\tr'\in L$. 
Clearly, $\tr'$ must contain a substring 
$\tr'_{\sf mid}$ 
which belongs to the language 
$\sep{1} \cdot \dots \cdot \sep{k} \cdot \texttt{AtLeast1Zero}^* \cdot \sep{1} \cdot \dots \cdot \sep{k}$.
Let $\tau'_{{\sf mid}} = \tr'_{\sf mid}[k \dots |\tr'_{\sf mid}| - k]$ be the subsequence
of $\tr'_{\sf mid}$ that matches $\texttt{AtLeast1Zero}^*$.

First, notice that every string belonging to $\texttt{AtLeast1Zero}$ is of length $k$.
We denote the length of $\tau'_{{\sf mid}}$ as $mk$.
Let $\tau'_{{\sf mid}, i}$ be the subsequence of $\tau'_{{\sf mid}}$ obtained by projected to $\Delta_i$.
Then, notice that $\tau'_{{\sf mid}, i}$ is in fact, 
$\tau'_{{\sf mid}}[i-1]\cdot \tau'_{{\sf mid}}[k + i-1] \cdot \dots \cdot \tau'_{{\sf mid}}[(m-1)k + i-1]$, which is of length $m$.
Finally, notice that $\sep{i} \cdot \tau'_{{\sf mid}, i}\cdot \sep{i}$, which is a subsequence of $\tr'_{\sf mid}$, is also a subsequence of $\tr$ and $\tau'_{{\sf mid}, i}$ corresponds to a vector $v^{(i)}_{t_i}\in A_i$, 
because $\sep{i} \cdot \tau'_{{\sf mid}, i}\cdot \sep{i} \in \alphabet_i^*$, where symbols in $\alphabet_i$ are dependent with each other.
So, we can conclude that 
\begin{enumerate*}
    \item $\tau'_{{\sf mid}, i}$ is of length $d$, thus $\tau'_{{\sf mid}}$ is of length $dk$, 
    \item $\tau'_{{\sf mid}}[(l-1)k + i - 1]$, $1 \le l \le d$, corresponds to the bit $v^{(i)}_{t_i}[l]$.
\end{enumerate*}
Since $\tau'_{{\sf mid}}[(l-1)k \dots (l-1)k + (k - 1)]$ belongs to $\texttt{AtLeast1Zero}$, for all $1 \le l \le d$, there is at least one $0$ in $\set{v^{(1)}_{t_1}[l], \dots, v^{(k)}_{t_k}[l]}$. 
Thus, the vectors $v^{1}_{t_1} \in A_1, \ldots, v^{k}_{t_k} \in A_k$
are orthogonal.

\myparagraph{Time Complexity} 
Recall that $|\alphabet| = 3k$, $|\indep| = 9k(k-1)/2$, 
$\wdth=k$, $|r| = O(2^k)$, and $|\tr| = kn(d+1) + k$. 
The time taken to construct $\tr$ is also $O(n\cdot k\cdot(d+1))$.
If predictive trace monitoring against $L$ can be decided in time 
$|\tr|^{\wdth-\epsilon}\cdot f{(|\alphabet|\cdot|r|)} = (kn(d+1) + k)^{k-\epsilon}\cdot f{(3k\cdot O(2^k))} = n^{k-\epsilon}\texttt{poly}(d)$, for some function $f$, then
\ovp{k} with set size $n$ can also be solved in time 
$n^{k-\epsilon}\poly{d}$, which refutes \SETH 
(since \SETH implies the absence of such an algorithm).
\end{proof}



\section{Pattern and generalized Pattern Langauges}
\seclabel{pattern-languages}

In this section, we describe the class of \emph{generalized pattern languages},
which is the central object of our study.
Later, in \secref{algo-pattern} we show that for this class of languages,
the predictive trace monitoring problem becomes highly tractable,
in that, it admits a constant-space streaming algorithm that works in linear time.

\begin{definition}[Pattern Languages]
\deflabel{patt-lang}
Let $\alphabet$ be an alphabet.
A language $L$ is said to be a pattern language of dimension $d \in \nats$ over $\alphabet$,
if it is of the form 
\[L = \patt{a}{d},\]
where $a_1, a_2, \ldots, a_d \in \alphabet$;
we use $\patts{\seq{a}{d}}$ to denote the above pattern language.
The class of all pattern languages will be denoted by $\patre$.
\end{definition}

\noindent
The above definition of pattern languages has been inspired from their
potential target application in dynamic analysis for detecting
bugs in concurrent and distributed applications.
Over the years, a large class of concurrency bug-detection techniques has been 
proposed that enhance the effectiveness
of otherwise exhaustive enumeration-style testing, by leveraging the hypothesis that 
``\emph{many bugs in programs depend on the precise ordering of 
a small number of events \ldots}''~\cite{chistikov2016}.
The additional knowledge of ``small bug depth'' can then
be leveraged to simplify testing, in that, 
only a small number of behaviours need to be analysed.
This has been the key to tackling the interleaving explosion problem encountered during
testing for concurrent software~\cite{Burckhardt2010,Musuvathi2007}.
In our setting, where the events are labelled from an alphabet (denoting, say
instructions, function entry and exit points, synchronization operations, or even load and store operations),
the natural analogue of this hypothesis asks to determine
if there is some small $d$ and labels $a_1, \ldots, a_d$
so that some $d$ events $e_1, e_2, \ldots, e_d$ (not necessarily observed in this order) 
with these labels $\lbl{e_1} = a_1$, \ldots, $\lbl{e_d} = a_d$,
can be organized in the precise order where $e_i$ is followed by $e_{i+1}$ for each $i \leq d$.
This is precisely the predictive monitoring question against the pattern language
$\patts{\seq{a}{d}} = \patt{a}{d}$ and highlights our motivation behind the class of specifications we introduce in \defref{patt-lang}.
As a final remark, observe that the language $L_{\sf fail}$
from \exref{example-motivating} is indeed a pattern language (of dimension $d=4$).

In the following, we define a more general class of languages,
thereby enhancing the class of specifications we consider.
\begin{definition}[Generalized Pattern Languages]
Let $\alphabet$ be an alphabet.
A language $L \subseteq \alphabet^*$ is said to be a 
generalized pattern language if 
it is a finite union
$L = L_1 \cup L_2 \cdots \cup L_m$,
where for each $L_i$ ($1 \leq i \leq m$), we have
\begin{enumerate*}[label=(\alph*)]
	\item $L_i = \emptyset$, or
	\item $L_i = \set{\emptystr}$, or
	\item $L_i \in \patre$ is a pattern language.
\end{enumerate*}
\end{definition}


\subsection{Properties of Pattern and Generalized Pattern Languages}
\seclabel{closure-properties}

In this section, we study language theoretic properties
of our proposed class of languages.

\myparagraph{Topological Characterization}
Pattern languages and generalized pattern languages
are clearly regular languages.
Here, we provide a finer characterization --- they are also star-free languages.
Recall that star-free languages over $\alphabet$ are those that can be constructed by using $\emptyset$,
$\set{\emptystr}$, $\set{a}$ (where $a \in \alphabet$ is some symbol),
and inductively composing them using concatenation,
union, intersection, and complementation.
We remark that \thmref{ov-hardness} shows that the complete class of star-free regular languages
are not amenable to efficient predictive trace monitoring.

\begin{proposition}
Every generalized pattern language is a star-free regular language.
\end{proposition}

Let us now examine the closure properties of our proposed class of languages.

\myparagraph{Closure Under Union}
Let us first consider closure under union for $\patre$.
Consider the two patterns $\patts{a}$ and $\patts{b}$,
where $a \neq b \in \alphabet$.
We can show using contradiction that there is no pattern $\patts{\seq{c}{k}}$
such that $\patts{\seq{c}{k}} = \patts{a} \cup \patts{b}$.
Suppose on the contrary that this holds. 
Then, $a, b \in \seq{c}{k}$ as otherwise $\big(\alphabet \setminus \set{a, b}\big) \cap \patts{\seq{c}{k}} \neq \emptyset$.
This means that $a \not\in \patts{\seq{c}{k}}$ since every string
in $\patts{\seq{c}{k}}$ must contain $b$, thereby giving us a contradiction.
On the other hand, the class of generalized pattern languages
is clearly closed under finite union.

\myparagraph{Closure Under Intersection}
Consider two pattern languages $L_1 = \patts{\seq{a}{d}}$
and $L_2 = \patts{\seq{b}{l}}$.
Then, a word $w \in L_1 \cap L_2$ must have a subsequence
$\seq{a}{d}$ as well as a sub-sequence $\seq{b}{l}$.
Let us use the notation $u \odot v$ to denote the shortest words that contain $u$ and $v$ as subsequences.
Let us use the notation $\sqsubseteq$ ($\sqsubset$) to denote subsequence (strict subsequence) relation between two words.
\[u\odot v = \setpred{w \in \Sigma^*}{u\sqsubseteq w,\, v\sqsubseteq w,\, \forall w'\sqsubset w, u\not\sqsubseteq w'\text{ or }v\not\sqsubseteq w'}\]
Then, we remark that 
$L_1 \cap L_2 = \bigcup_{u \in \seq{a}{d} \odot \seq{b}{l}}\patts{u}$.
The class of generalized pattern languages is closed under intersection, 
since intersection distributes over union.

\myparagraph{Closure under Complementation}
The class of pattern languages is not closed under complementation.
Consider the pattern language $L = \patts{a}$.
Observe that $L^c = (\alphabet \setminus \set{a})^*$
which cannot be a pattern language.
Assume on the contrary that there was a pattern language
$\patts{\seq{b}{l}} = L^c$; observe that $a\cdot b_1 \cdot \dots \cdot b_l \in \patts{\seq{b}{l}}$
giving us a contradiction right away.
In fact, the language $L^c$ here is different from $\set{\emptystr}$,
and $\emptyset$, and is also not a finite union of multiple pattern languages
(each of which will otherwise contain a string containing $a$).
Thus, even the class of generalized pattern languages is not closed under complementation.

\myparagraph{Closure Under Concatenation}
Consider two pattern languages $L_1 = \patts{\seq{a}{d}}$
and $L_2 = \patts{\seq{b}{l}}$.
Observe that the language $L = \patts{\seq{a}{d} \circ \seq{b}{l}}$ is indeed
the concatenation of $L_1$ and $L_2$, i.e.,
$L = L_1 \circ L_2$.
As a result, the class $\patre$ is closed under concatenation.
It is easy to see that 
the class of generalized pattern languages is also closed under concatenation.

\myparagraph{Closure Under Kleene Star}
Pattern languages are certainly not closed under Kleene star because
$\emptystr \not\in \patts{a}$, but $\emptystr \in (\patts{a})^*$.
However, the class of generalized pattern languages is closed under
the Kleene star operation, because of the observation that
for a pattern language $L \in \patre$,
we have $L^* = L \cup \set{\emptystr}$ which can be proved inductively.

To summarize, we have the following closure properties.
The formal proof is presented in 
\iftoggle{appendix}{\appref{closure-property}}{the extended version of our paper~\cite{ang2023predictive}}.

\begin{theorem}[Closure properties of Pattern Languages]
	\thmlabel{pat-clos}
The class of pattern languages is closed under finite concatenation, but not under union, intersection, complementation, and Kleene star.
\end{theorem}

\begin{theorem}[Closure properties of generalized Pattern Languages]
	\thmlabel{gener-clos}
The class of generalized pattern languages is closed under finite union, finite intersection, finite concatenation, and Kleene star but not under complementation.
\end{theorem}

\section{Predictive Monitoring against Generalized Pattern Languages}
\seclabel{algo-pattern}

In \secref{lower-bound}, we have established that 
the problem of predictive trace monitoring against regular languages 
does not admit an algorithm that runs faster than $O(n^\wdth)$,
assuming \SETH holds.
In this section, we show that, the case of pattern languages
and generalized pattern languages (defined in \secref{pattern-languages})
admit a more efficient --- constant-space linear-time 
streaming --- algorithm.


\subsection{Overview of the Algorithm}

\newcommand{\pat}{\patts{\seq{a}{d}}}

Recall that a generalized pattern language is a union
of pattern languages, $\set{\emptystr}$ or $\emptyset$,
and predictive monitoring against the latter two can trivially be performed in constant space.
We instead show a constant-space algorithm for predictive monitoring against pattern languages,
which would essentially imply a constant-space algorithm
for the larger class of generalized pattern languages.
Our algorithm (for pattern languages) is based on several crucial insights: 
here we briefly discuss them and give details in subsequent sections.
In order to give intuitions behind these insights, 
we will define some subproblems (\probref{candidate}, \probref{constant}).
We fix the concurrent alphabet $(\alphabet, \indep)$.
The corresponding dependence relation is $\dep = \alphabet \times \alphabet \setminus \indep$.
We also fix the pattern language $\pat$ of dimension $d$ and the input execution $\tr$ over $(\alphabet, \indep)$.
Let us now define some useful notations.

Let $\tau = \tuple{e_1, \dots, e_m}$ be a tuple of events of $\tr$ 
such that $e_1\trord{\tr} \dots \trord{\tr} e_m$.
Let $\seq{b}{m}$ be a sequence of $m$ labels from $\alphabet$.
$\tau$ is said to be a \emph{shuffle of} $\seq{b}{m}$ 
if $\tuple{\lbl{e_{1}}, \dots, \lbl{e_{m}}}$ is a permutation of $\tuple{\seq{b}{m}}$.
$\tau$ is said to be a \emph{partial candidate} tuple with respect to our pattern $\pat$,
if it is a shuffle of some subsequence of $\seq{a}{d}$.
$\tau$ is said to be a \emph{(complete) candidate} tuple with respect to our pattern $\pat$,
if it is a shuffle of $\seq{a}{d}$.

\begin{definition}[Partial and Complete Admissible Tuples]
    Let $\tau$ be a tuple of $m$ events such that $\tau$ is a partial candidate tuple.
    $\tau$ is said to be admissible with respect to $\pat$
    if there is an execution $\tr'\mazeq{\indep}\tr$ and 
    a permutation $\tau'=\tuple{e_{i_1}, \dots, e_{i_m}}$ of $\tau$, 
    such that $\tuple{\lbl{e_{i_1}}, \dots, \lbl{e_{i_m}}}$ is a 
    subsequence of $\tuple{a_1, \dots, a_d}$
    and $\tau'$ is subsequence of $\tr'$.
    As a special case, $\tau$ is a complete admissible tuple 
    if it is a \emph{complete} candidate and also, admissible.
\end{definition}

We remark that the admissibility of a tuple $\tau$, 
if true, must be witnessed 
as a special permutation of $\tau$.
Assume that $\tau$ is a partial candidate tuple, 
which is a shuffle of $a_{j_1}, \dots, a_{j_m}$, a subsequence of $\seq{a}{d}$.
Then, there is a \emph{special} permutation 
$\tau^\dagger=\tuple{e_{i^\dagger_1}, \dots, e_{i^\dagger_m}}$ of $\tau$,
such that $\tuple{\lbl{e_{i^\dagger_1}}, \dots, \lbl{e_{i^\dagger_m}}} = \tuple{a_{j_1}, \dots, a_{j_m}}$;
more importantly, if $\tau$ is admissible,
then for every $\tr'\mazeq{\indep}\tr$ that witnesses the admissibility of $\tau$,
$\tau^\dagger$ is a subsequence of $\tr'$.
Observe that this special permutation can be obtained 
by sorting $\tau$ according to $\tuple{a_1, \dots, a_d}$
while additionally ensuring that events with the same label do not get flipped.
Henceforth, we will use the notation 
$\sort{\tau}{\tuple{a_{j_1}, \dots, a_{j_m}}}$. 

The first subproblem we consider pertains to checking if a given sequence of events is admissible.

\begin{problem}
    \problabel{candidate}
    Given an execution $\tr$ and a candidate tuple $\tau$ of events from $\tr$,
    check whether $\tau$ is admissible with respect to $\pat$.
\end{problem}

A naive approach to solving this problem would enumerate all $O(|\tr|!)$ linearizations 
and check membership in $\pat$ for each of them
in total time $O(|\tr|\cdot|\tr|!)$, which is prohibitive.
A different and better approach instead relies on the partial order view of Mazurkiewicz traces.
Here, one considers the graph $G_\tr = (V_\tr, E_\tr)$ corresponding to the partial order $\mazpo{\dep}{\tr}$,
where $V_\tr=\events{\tr}$ and $E_\tr$ captures immediate edges of $\mazpo{\dep}{\tr}$.
One can get a graph $G'_\tr$ by adding additional edges $\set{(e_{i^\dagger_1}, e_{i^\dagger_2}), \dots, (e_{i^\dagger_{d-1}}, e_{i^\dagger_d})}$ 
derived from the special permutation $\tuple{e_{i^\dagger_1}, \dots, e_{i^\dagger_d}}=\sort{\tau}{\tuple{a_1, \dots, a_d}}$.
This reduces the admissibility of $\tau$ to the acyclicity of $G'_\tr$,
which can be checked in time $O(|\tr|)$
since the degree of each vertex is constant.
However, the space utilization of this graph-theoretic algorithm is also linear.
In \secref{candidate}, we will show that \probref{candidate} can be solved
using a streaming constant-space algorithm, 
that uses \emph{after sets} (\defref{after-set}).
After sets succinctly capture causality induced by the corresponding partial order 
and can be used to check the admissibility of a candidate tuple in a streaming fashion.

The second subproblem asks whether there exists an admissible tuple in the given execution.

\begin{problem}
    \problabel{constant}
    Given an execution $\tr$, check whether 
    there is a candidate tuple $\tau$ of events from $\tr$,
    such that $\tau$ is admissible with respect to $\pat$.
\end{problem}

In other words, \probref{constant} asks there is a $\tr'\mazeq{\indep}\tr$ 
such that $\lbl{\tr'}\in \pat$. 
How do we solve \probref{constant}?
Again, a naive approach would enumerate all $O(|\tr|^d)$ candidate tuples and check their admissibility 
by repeatedly invoking our algorithm for \probref{candidate}.
This is again prohibitive, in terms of both its time and space usage.
On the other hand, in \secref{alg-correct}, 
we design a linear-time streaming algorithm for \probref{constant} 
whose overall space usage is constant.
The high-level insights behind this algorithm are as follows.
First, our algorithm runs in a streaming fashion, 
and tracks not just complete candidate tuples,
but also \emph{partial} tuples which can potentially be extended to complete candidate tuples.
Second, our algorithm is \emph{eager}, in that,
it determines whether a partial tuple is admissible,
i.e., if it can be extended to a complete admissible tuple,
and proactively discards all other partial tuples.
The problem of checking if a partial tuple is admissible can,
in fact, be solved in constant space using an algorithm similar to \algoref{candidate} 
that solves \probref{candidate}.
Nevertheless, even the number of partial admissible tuples 
is still large ($O(|\tr|^d)$).
Our third and most important insight, formalized in~\lemref{exMax}, tackles this problem --- 
we only need to track constantly many partial admissible tuples,
those that are maximum in some precise sense. 

Equipped with the above insights, 
the high-level description of our algorithm for predictive monitoring against pattern languages 
is the following.
The algorithm processes one event at a time and 
tracks the maximum partial admissible tuple of each kind.
When processing a new event of the given execution, 
it checks whether any existing partial tuple can be 
extended to another (partial or complete) admissible tuple.
If at any point, a complete admissible tuple can be constructed, 
the algorithm terminates and declares success.
Otherwise, there is no complete admissible tuple in the entire execution.

In \secref{candidate}, we discuss the details of how we solve \probref{candidate} in constant space.
In \secref{alg-correct}, we present the details of our overall algorithm for 
predictive trace monitoring against pattern regular languages (\probref{constant}).


\subsection{Checking Admissibility of Candidate Tuples}
\seclabel{candidate}

Our first observation towards solving \probref{candidate}
is that in order to check the admissibility of a candidate tuple $\tau$, 
it suffices to examine only pairs of events that appear in $\tau$,
which, thankfully, is a local property of $\tau$.
We formalize this using the notion of \patwit.
Intuitively, $\tau$ is locally admissible when 
for every pair of events $(e, e')$ in $\tau$,
if the order in which $e$ and $e'$ appear in $\tau$
is different from their order in the target $\tau^\dagger = \sort{\tau}{\tuple{a_1,\dots,a_d}}$,
then they are incomparable according to $\mazpo{\dep}{\tr}$.
This observation can be generalized to partial candidate tuples as well:

\begin{definition}[Locally Admissible Tuple]
    \deflabel{localadm}
    Let $\tau = \tuple{e_1, \dots, e_m}$ be a partial candidate tuple of an execution $\tr$ with respect to $\pat$,
    so that it is a shuffle of $a_{j_1},\dots,a_{j_m}$, a subsequence of $\seq{a}{d}$.
    Let $\tuple{e_{i^\dagger_1}, \dots, e_{i^\dagger_m}}=\sort{\tau}{\tuple{a_{j_1}, \dots, a_{j_m}}}$.
    We say that $\tau$ is partially locally admissible with respect to $\pat$
    if for all $i^\dagger_r\neq i^\dagger_s$, we have
    $s < r \Rightarrow e_{i^\dagger_r}\nmazpo{\dep}{\tr}e_{i^\dagger_s}.$
    As a special case, $\tau$ is completely locally admissible if it is additionally a complete candidate.
\end{definition}

A simple intuitive implication of \defref{localadm} is the following. 
Consider again the directed acyclic graph $G_\tr$ of induced partial order $\mazpo{\dep}{\tr}$.
The local admissibility of a tuple $\tau = \tuple{e_1, \dots, e_d}$ essentially means that 
for every two vertices $e_{i^\dagger_r}, e_{i^\dagger_s}$ that need to be flipped, 
the addition of the edge $(e_{i^\dagger_r}, e_{i^\dagger_s})$ alone does not introduce any cycle in this graph.

In the following, we establish a clear connection between admissibility and local admissibility.
The proof of \lemref{witness} can be found in 
\iftoggle{appendix}{\appref{app-ad}}{~\cite{ang2023predictive}}.

\begin{lemma} \lemlabel{witness}
    Let $\tau = \tuple{e_1, \dots, e_m}$ be a partial candidate tuple of an execution $\tr$ with respect to $\pat$,
    which is a shuffle of $a_{j_1},\dots,a_{j_m}$, a subsequence of $\seq{a}{d}$.
    We have, 
    \vspace{-0.1cm}
    \[\tau\text{ is admissible }\quad\iff\quad \tau\text{ is locally admissible}\]
\end{lemma}

The above result is interesting, yet subtle. 
The intuitive implication of this lemma is that 
if each flipped edge $(e_{i^\dagger_r}, e_{i^\dagger_s})$ individually does not introduce any cycle in $G_\tr$,
then, in fact, the simultaneous addition of all such edges also does not introduce a cycle.

While the criterion of local admissibility and its equivalence with the admissibility 
are important ingredients towards our solution for \probref{candidate},
it is not immediately amenable to a space-efficient algorithm.
In particular, we cannot afford to construct a linear-sized graph and check the reachability 
in order to decide local admissibility.
Fortunately, our next ingredient (\defref{after-set}) effectively tackles this challenge.

\begin{definition}
    \deflabel{after-set}
    For an execution $\tr$ and an event $e\in\events{\tr}$,
    the after set of $e$ in a prefix $\rho$ of $\tr$ is defined as 
    $\aft{\rho}{e} = \setpred{\lbl{f}}{f\in\events{\rho} \text{ and } e\mazpo{\dep}{\tr}f}.$
\end{definition}

Observe that, for any event $e$ and prefix $\rho$, $\aft{\rho}{e}\subseteq \alphabet$,
thus this set is of constant size.
A direct consequence of \defref{after-set} is that 
after sets are sufficient to check causality between two events.

\begin{proposition}
    \proplabel{afterset}
    Let $\tr$ be an execution and $\rho$ be a prefix of $\tr$ that ends at an event $f$.
    Let $e$ be an event such that $e\trord{\rho}f$.
    We have, $\lbl{f}\in \aft{\rho}{e}\quad \iff\quad e\mazpo{\dep}{\tr}f$.
\end{proposition}

Our algorithm maintains after sets of the events participating in the candidate tuple.
Observe that the after set of an event $e$ is parameterized on a prefix $\rho$ of a prefix of $\tr$.
The next crucial observation, formally stated in \lemref{aft-incre}, 
is that this set can be updated incrementally 
(see \iftoggle{appendix}{\appref{app-ad}}{~\cite{ang2023predictive}} for the proof).

\begin{lemma}
    \lemlabel{aft-incre}
    Let $\tr$ be an execution and $\rho,\rho\circ f$ be prefixes of $\tr$ for some event $f$.
    Let $e$ be an event in $\rho$.
    We have 
    \begin{align*}
        \aft{\rho\circ f}{e}=\begin{cases}
            \aft{\rho}{e}\cup \set{\lbl{f}} & \text{ if }\exists a\in \aft{\rho}{e} \text{ s.t. } (a,\lbl{f})\in\dep,\\
            \aft{\rho}{e} &\text{ otherwise}.
        \end{cases}
    \end{align*}
\end{lemma}


\begin{algorithm}[t]
    \KwIn{$\tr\in\alphabet^*$, candidate tuple $\tau=\tuple{\seq{e}{d}}$}
    \KwOut{YES if $\tau$ is admissible; NO otherwise}

    Let $\tau^\dagger=\sort{\tau}{\tuple{\seq{a}{d}}}$\;

    \lForEach{$e\in\tau$}{$\aftvar_e\gets \emptyset$}

    \For{$f\in\tr$}{
        \ForEach(\tcp*[f]{\textcolor{blue}{Update after sets}}){$e\in\tau$}{\linlabel{update-afterset}
            \lIf{$\exists a \in \aftvar_e\text{ s.t. }(a, \lbl{f})\in\dep$}{
                $\aftvar_e\gets \aftvar_e\cup\set{\lbl{f}}$
            }
        }
        \If{$f\in\tau$}{
            $\aftvar_f \gets \set{\lbl{f}}$

            \ForEach(\tcp*[f]{\textcolor{blue}{Check local admissibility}}){$e\in\tau$}{\linlabel{check-loc-ad}
                \lIf{$f$ occurs before $e$ in $\tau^\dagger$ and $\lbl{f}\in\aftvar_e$}{
                    \Return{NO}
                }
            }
        }
    }
    \Return{YES}\;
    \caption{Constant-Space Algorithm for Checking Admissibility}
    \algolabel{candidate}
\end{algorithm}

Equipped with \lemref{witness}, \propref{afterset}, and \lemref{aft-incre},
we can now design the constant-space streaming algorithm for \probref{candidate} in \algoref{candidate}.
The algorithm first constructs the target tuple $\tau^\dagger$
which is uniquely determined from $\tau$.
It then scans the execution by processing one event at a time in the order of the given execution.
At each event, it updates the after sets of all the events occurring in $\tau$ (\linref{update-afterset}).
Then, when we see an event from $\tau$,
we also check the violations of local admissibility proactively (\linref{check-loc-ad}).


\subsection{Checking Existence of Admissible Tuples}
\seclabel{alg-correct}

Recall that the naive algorithm which enumerates all $O(n^d)$ $d$-tuples
consumes $O(n^d)$ time and space,
even though we have a constant-space algorithm to check their admissibility.
Our algorithm for solving \probref{constant} 
instead runs in a streaming fashion and uses constant space.
To do so, the algorithm tracks not only complete candidates
but also partial ones.
It incrementally extends those partial candidates as it processes new events.
A distinct and crucial feature of our algorithm is that 
it eagerly discards those partial candidates 
that cannot be extended to a complete admissible tuple.
In other words, it tracks only partially admissible tuples.
Observe that \algoref{candidate} can be easily adjusted to 
check partial admissibility.

Notice that, compared with the number of partial candidates,
the number of partially admissible tuples is not reduced significantly,
and is still $O(n^d)$.
However, tracking only partially admissible tuples 
(instead of all partial candidates)
allows us to bound the number of tuples we track.
In particular, for a set of partially admissible tuples with the same label,
we can afford to track only the maximum element.
We observe that the number of all kinds of labels is  
the number of all permutations of all subsequences of $\seq{a}{d}$, 
which is $1! + \dots + d!$, a constant number.
This is also the number of partially admissible tuples we track 
at any point when processing an execution.
The next lemma (\lemref{exMax}) shows that the maximum element exists and is unique.

Before we formally show the existence of the maximum element,
let us fix some helpful notations.
Let $\tau_1 = \tuple{\seqsup{e}{m}{1}}, \tau_2 = \tuple{\seqsup{e}{m}{2}}$ 
be two sequences of events in the execution $\tr$,
such that $\lbl{\tau_1} = \lbl{\tau_2}$.
We say that $\tau_1\unlhd\tau_2$ if for all $1\le i \le m$, 
$e^1_i\trord{\tr}e^2_i$ or $e^1_i = e^2_i$.
Also, we use $\tau_1\bigtriangledown\tau_2$ to denote the tuple $\tuple{\seq{e}{m}}$,
where for $1\le i\le m$, $e_i = e^2_i$ if $e^1_i\trord{\tr}e^2_i$, 
and $e^1_i$ otherwise.
Let $\tuple{\seq{b}{m}}$ be a permutation of a subsequence of $\tuple{\seq{a}{d}}$.
We use $\adset{\tr}{\tuple{\seq{b}{m}}}$ to denote the set of all partially admissible tuples $\tau$ in $\tr$, such that $\lbl{\tau} = \tuple{\seq{b}{m}}$.

\begin{lemma}
    \lemlabel{exMax}
    Let $\tr$ be an execution and $\rho$ be a prefix of $\tr$.
    Let $\tuple{\seq{b}{m}}$ be a permutation of a subsequence of $\tuple{\seq{a}{d}}$.
    Then the set $\adset{\rho}{\tuple{\seq{b}{m}}}$ has a unique maximum tuple $\tau_0$ with regard to $\unlhd$.
    That is, for every $\tau \in \adset{\rho}{\tuple{\seq{b}{m}}}$,
    we have $\tau\unlhd \tau_0$.
\end{lemma}

The above lemma states the  uniqueness of the maximum element.
But observe that existence implies uniqueness.
Consider $\tau_1$ and $\tau_2$ that both are maximum, 
then $\tau_1\unlhd\tau_2$ and $\tau_2\unlhd\tau_1$. 
From the definition, this implies that $\tau_1 = \tau_2$.
The proof of existence relies on two observations.
First, the set $\adset{\rho}{\tuple{\seq{b}{m}}}$ is closed under $\bigtriangledown$,
which is formalized in \lemref{clojoin}.
Second, from the definition, $\tau=\tau_1\bigtriangledown\tau_2$ implies that $\tau_1\unlhd\tau$ and $\tau_2\unlhd\tau$.
Combined the above two observations, 
we know that $\bigtriangledown_{\tau\in \adset{\rho}{\tuple{\seq{b}{m}}}}\tau$ 
is in $\adset{\rho}{\tuple{\seq{b}{m}}}$
and also, is the unique maximum element.

\begin{lemma}[Closure Property under Join]
    \lemlabel{clojoin}
    Let $\tr$ be an execution and $\rho$ be a prefix of $\tr$. 
    Let $\tuple{\seq{b}{m}}$ be a permutation of a subsequence of $\tuple{\seq{a}{d}}$.
    The set $\adset{\rho}{\tuple{\seq{b}{m}}}$ is closed under operation $\bigtriangledown$.
    To be specific, for all $\tau_1, \tau_2 \in \adset{\rho}{\tuple{\seq{b}{m}}}$,
    we have $\tau_1\bigtriangledown\tau_2\in \adset{\rho}{\tuple{\seq{b}{m}}}$.
\end{lemma}

Our final ingredient is the observation that 
for any $\rho$, a prefix of the execution $\tr$,
the maximum partially admissible tuple of each kind in $\rho$ can be 
computed incrementally using the maximum ones 
computed in the immediate prefix of $\rho$.
In the following \lemref{comp-max}, we formally establish this observation.
From now on, we use $\max(S)$ to denote 
the maximum element of the set $S$ of partially admissible tuples.

\begin{lemma}
    \lemlabel{comp-max}
    Let $\tr$ be an execution. 
    Let $\rho' = \rho\circ f$ be a prefix of $\tr$, where
    $\rho$ is also a prefix of $\tr$ and $f$ is an event.
    Let $\tuple{\seq{b}{m}}$ be a permutation of a subsequence of $\tuple{\seq{a}{d}}$.
    If $\lbl{f} = b_m$ and $\max(\adset{\rho}{\tuple{\seq{b}{m-1}}}) \circ f$ is partially admissible,
    then we have 
    \[\max(\adset{\rho\circ f}{\tuple{\seq{b}{m}}}) = \max(\adset{\rho}{\tuple{\seq{b}{m-1}}}) \circ f.\]
    Otherwise, we have 
    \[\max(\adset{\rho\circ f}{\tuple{\seq{b}{m}}}) = \max(\adset{\rho}{\tuple{\seq{b}{m}}}).\] 
    
\end{lemma}

The above lemma ensures the feasibility of our streaming algorithm.
Recall that we only keep track of the maximum element of each kind for a processed prefix.
When processing a new event, 
the algorithm can compute the new maximum partially admissible tuples easily 
from those old ones tracked so far.
The proofs of Lemmas~\ref{lem:exMax},~\ref{lem:clojoin} and~\ref{lem:comp-max} are presented in \iftoggle{appendix}{\appref{app-alg-correct}}{~\cite{ang2023predictive}}.


\begin{algorithm}[t]
  \KwIn{$\tr\in\alphabet^*$}
    \KwOut{YES if there exists an admissible $\tau$; NO otherwise}
  Let ($M$: Permutations of prefixes of $\seq{a}{d}
  \rightarrow $ Partial admissible tuples) be an empty map\linlabel{init}\;
  \For{$f\in\tr$}{
    \If{$\lbl{f}\in \tuple{\seq{a}{d}}$}{\linlabel{update}
        \ForEach{$\pi$ which is a permutation of a prefix of $\seq{a}{d}$ ending with $\lbl{f}$}{
            Let $\pi' = \pi[:-1]$\;
            \If{$M$ contains $\pi'$ and 
            $M(\pi')\circ f$ is partially admissible}{
                \tcp{\textcolor{blue}{compute the new maximum partial admissible tuple}}
                $M(\pi) \gets M(\pi')\circ f$\;
            }
        }
        \If{$M$ contains a key of length $d$}{\linlabel{yes}
            \Return{YES}\;
        }
      }
    }
    \Return{NO}\linlabel{no}\;
  \caption{Constant-Space Algorithm for Checking Existence of Admissible Tuples}
  \algolabel{constant}
\end{algorithm}

Equipped with \lemref{exMax} and \lemref{comp-max}, 
we now present a constant-space streaming algorithm for \probref{constant} in \algoref{constant},
where we omit some details about how to maintain after sets and check partial local admissibility.
A complete algorithm which combines \algoref{candidate} and \algoref{constant}
will be discussed in \secref{comb}.
We use a map $M$ to track maximum partially admissible tuples with their labels as keys,
and is initialized to be empty (\linref{init}).
The algorithm processes events in the execution one after another.
When processing an event whose label is in $\tuple{\seq{a}{d}}$,
we update the maximum partially admissible tuples we track if needed (\linref{update}), 
as stated in \lemref{comp-max}.
If a complete admissible tuple is constructed (\linref{yes}), which is of length $k$,
the algorithm returns ``YES'' to claim the existence of admissible tuples.
On the other hand, when finishing processing the whole execution and
no complete admissible tuple is constructed (\linref{no}), 
it returns ``NO'' to refute the existence.

\subsection{Algorithm for Predictive Monitoring against Pattern Languages}
\seclabel{comb}

We now present in \algoref{detail} 
the one-pass constant-space streaming algorithm 
for solving predictive monitoring against pattern languages, 
by combining \algoref{candidate} and \algoref{constant}.
In this section, we first introduce the data structure used in the algorithm,
then illustrate how the algorithm works.
In particular, we will elaborate on how to check partial admissibility 
using \algoref{candidate} and the data structure, 
which is omitted in \secref{alg-correct}.


\begin{algorithm}[t]
    \KwIn{$\tr\in\alphabet^*$}
    \KwOut{YES if there exists a $\tr'\mazeq{\indep}\tr$ such that $\lbl{\tr'}\in \pat$; NO otherwise}
    
    Let ($M$: Permutations of prefixes of $\seq{a}{d}\rightarrow $ Sequence of After Sets) be an empty map\linlabel{detail-init}\;

    \For{$f\in\tr$}{
        \ForEach(\tcp*[f]{\textcolor{blue}{Update after sets}}){$\aftvar_e\in M$}{\linlabel{detail-update-afterset}
            \lIf{$\exists a \in \aftvar_e\text{ s.t. }(a, \lbl{f})\in\dep$}{\linlabel{update-afterset-end}
                $\aftvar_e\gets \aftvar_e\cup\set{\lbl{f}}$
            }
        }
        \If{$\lbl{f}\in \tuple{\seq{a}{d}}$}{\linlabel{detail-update}
            \ForEach{$\pi$ which is a permutation of a prefix of $\seq{a}{d}$ ending with $\lbl{f}$}{
                Let $\pi' = \pi[:-1]$\;
                
                \If{$M$ contains $\pi'$}{\linlabel{check-par-adm}
                    Let $\aftvar_f = \set{\lbl{f}}$\;\linlabel{detail-af}
                    Let $\mathsf{PartCandTuple} = M(\pi')\circ \aftvar_f$\;\linlabel{detail-af2}
                    Let $\mathsf{PartCandTuple}^\dagger = \sort{\mathsf{PartCandTuple}}{\pi}$\; \linlabel{get-dagger}
                    \tcp{\textcolor{blue}{Check partial admissibility}}
                    \If{$\forall \aftvar_e\in M(\pi')$,
                        $\aftvar_f$ occurs after $\aftvar_e$ in $\mathsf{PartCandTuple}^\dagger$ or $\lbl{f}\not\in\aftvar_e$
                    }{ \linlabel{detail-cond}
                        $M(\pi) \gets M(\pi')\circ \set{\lbl{f}}$
                    }
                }\linlabel{check-par-adm-end}
            }
            \If{$M$ contains a key of length $d$}{\linlabel{return-yes}
                \Return{YES}\;
            }
        }\linlabel{update-end}
    }
    \Return{NO}\;\linlabel{return-no}
    \caption{Constant-Space Algorithm for Predictive Monitoring against Pattern Languages}
    \algolabel{detail}
\end{algorithm}
The data structure is similar to what is used in \algoref{constant}.
First, recall that our algorithm tracks the maximum partially admissible tuple for each kind of label.
So, for any processed execution $\rho$, we use a map to store key-value pairs, 
$(\tuple{\seq{b}{m}},\max(\adset{\rho}{\tuple{\seq{b}{m}}}))$, 
where $\tuple{\seq{b}{m}}$ is a permutation of a subsequence of $\tuple{\seq{a}{d}}$.
The difference lies in that we keep track of the after set of each event in maximum partially admissible tuples,
instead of events themselves.
It is sufficient to do so because 
the only information we need to check admissibility is the label and the after set of an event.
Notice that the label information is implicitly stored in the key of the map.

In \algoref{detail}, we first initialize the map as empty (\linref{detail-init}).
When processing an event $f$, 
we first update all after sets we track by accommodating $f$ 
(\linref{detail-update-afterset} - \linref{update-afterset-end}).
If $\lbl{f}$ participates in $\tuple{\seq{a}{d}}$, 
we try to update partially admissible tuples
based on \lemref{aft-incre}
(\linref{detail-update} - \linref{update-end}).
Event $f$ might participate in a maximum partially admissible tuples, 
whose labels end with $\lbl{f}$.
From \linref{check-par-adm} to \linref{check-par-adm-end},
we check whether a partially admissible tuple extended with $f$ is 
still admissible.
The algorithm first constructs the after set for $f$,
which contains only $\lbl{f}$.
Then it computes the target $\mathsf{PartCandTuple}^\dagger$ 
uniquely determined from a partial candidate tuple $\mathsf{PartCandTuple}$, 
which is a concatenation of 
a previously computed maximum partially admissible tuples and $f$ (\lemref{aft-incre}).
It finally uses \algoref{candidate} to determine partial admissibility of $\mathsf{PartCandTuple}$.
This can be done because we have maintained the label and the after set of each event.

Finally, similar to \algoref{constant},
when it witnesses a $k$-length complete admissible tuple,
the algorithm claims the existence of a $\tr'\mazeq{\indep} \tr$, 
such that $\lbl{\tr'}\in\pat$, and says ``YES'' to the predictive monitoring problem (\linref{return-yes}).
Otherwise, after the whole execution is processed, it refutes the existence of such a reordering (\linref{return-no}).

Fix $(\alphabet, \indep)$ and the pattern language $L$. The following two theorems summarize the correctness and the complexity of our \algoref{detail},
where we assume that $d, |\alphabet|\in O(1)$.
The proof of \thmref{algo-correct} is presented in
\iftoggle{appendix}{\appref{comb}}{~\cite{ang2023predictive}}.

\begin{theorem}[Correctness]
    \thmlabel{algo-correct}
    On input execution $\tr$, \algoref{detail} declares ``YES''
    iff there is a $\tr'\mazeq{\indep} \tr$ such that $\tr'\in L$.
\end{theorem}

\begin{theorem}[Complexity]
    \thmlabel{complex}
    On input execution $\tr$,
    \algoref{detail} runs in time $O(|\tr|)$ 
    and uses constant space.
\end{theorem}

For a generalized pattern language $L$, we have the following corollary.

\begin{corollary}
    The problem of predictive trace monitoring against generalized pattern languages 
    can be solved in linear time and constant space.
\end{corollary}


\subsection{Vector Clock Algorithm}
\seclabel{algo-vc}

In \algoref{detail}, we store the after set for each event we track.
Although the size of an after set is constant and bounded by $\alphabet$,
it can become too large when $|\alphabet|$ is large.
To overcome this problem, in this section, 
we propose the use of vector timestamps
that can be used to efficiently check happens-before relation 
in the context of a concurrent execution~\cite{Mattern89,Fidge91,TreeClocks2022}.
Our resulting timestamp-based algorithm will be referred to as \pattrack.

We first provide a brief introduction to vector timestamps and vector clocks.
Formally, a vector timestamp is a mapping $\vc: \threads \rightarrow \nats$ 
from the threads to natural numbers. 
Vector clocks are variables taking values from the space of vector timestamps.
Three common operations on vector clocks are 
\emph{join}, \emph{comparison}, and \emph{update}.
We use $\vc_1\sqcup \vc_2$ to denote
the join of two vector timestamps $\vc_1$ and $\vc_2$,
which is the vector timestamp $\lambda t,\max(\vc_1(t), \vc_2(t))$.
The partial order $\sqsubseteq$ is defined so that $\vc_1 \subseteq \vc_2$ iff 
$\forall t,\vc_1(t) \le \vc_2(t)$.
The minimum timestamp $\lambda t, 0$ is denoted by $\bot$.
For $c\in \nats$, the updated timestamp $\vc[t\rightarrow c]$ is $\lambda u, \texttt{ if } u=t \texttt{ then } c \texttt{ else } \vc(u)$. 
Our algorithm will assign a timestamp $\vc_e$
for each event $e$ so that for each thread $t \in \threads$, we have
$\vc_e(t) = |\setpred{f}{f\mazpo{\dep}{\tr}e, \lbl{f} = \ev{t, op}}|$.
The following captures the relation between vector timestamps and the order $\mazpo{\dep}{\tr}$:

\begin{lemma}
    \label{vecclk}
    Let $\tr$ be an execution. Let $e, f$ be two events in $\tr$.
    $\vc_e \sqsubseteq \vc_f \iff e\mazpo{\dep}{\tr} f$.
\end{lemma}

The aforementioned lemma guarantees that 
our algorithm can utilize vector clocks to 
perform the same task as long as 
we can compute the vector timestamp for each event in a streaming fashion and constant space.
\algoref{vc} demonstrates how to compute vector timestamps by maintaining a vector clock
for each label $\ev{t, op} \in \alphabet$ and each $t\in\threads$. 


\begin{algorithm}[t]
    \KwIn{$\tr\in\alphabet^*$}
    
    Let $\vc_{\ev{t, op}} = \bot$ for every $\ev{t, op}\in\alphabet$\linlabel{vc-init}\;
    Let $\vc_{t} = \bot$ for every $t\in\threads$\;
    \For{$f\in\tr$}{
        Let $\ev{t, op} = \lbl{f}$\;
        $\vc_t\gets \vc_t[t\rightarrow (\vc_t(t) + 1)]$\;\linlabel{vc-incre}
        \For{$\ev{t', op'}\in \alphabet$ such that $(\ev{t, op}, \ev{t', op'})\not\in \indep$}{
            $\vc_t\gets \vc_t\sqcup \vc_{\ev{t', op'}}$\linlabel{vc-join}
        }
        $\vc_{\ev{t, op}} \gets \vc_t$\;
        $\vc_f \gets \vc_{\ev{t, op}}$
    }
    \caption{Constant-Space Streaming Algorithm for Computing Vector Timestamps}
    \algolabel{vc}
\end{algorithm}

In \algoref{vc},
we maintain a vector clock, for each label $a \in \alphabet$, to track the vector timestamp of the last event labeled with $a$ in the prefix $\rho$ seen so far.
Additionally, we also maintain a vector clock to track the timestamp of the latest event of each thread.
Initially, in \linref{vc-init}, 
all vector clocks for every label and thread are initialized to $\bot$.
When processing an event $f$ executed by thread $t$,
the entry corresponding to $t$ is first incremented in $\vc_t$ (\linref{vc-incre}).
Then $\vc_t$ is updated by joining it with all vector clocks corresponding to 
labels that are dependent on $\lbl{f}$,
since the last events with these labels happen before $f$ (\linref{vc-join}).
Finally, we update the clock for $\lbl{f}$ with $\vc_t$.
The vector timestamp of $f$ is precisely $\vc_{\lbl{f}}$.

Therefore, obtaining a vector-clock version of \algoref{detail} is simple.
We can replace initialization (\algoref{detail} \linref{detail-init}) and 
maintain (\algoref{detail} \linref{detail-update-afterset}) parts with 
the initialization of all vector clocks and the computation of $\vc_{\lbl{f}}$
as in \algoref{vc}.
For checking partial admissibility,
we first update $\vc_f$ to be $\vc_{\lbl{f}}$ and
compute $\mathsf{PartCandTuple}$ as $M(\pi')\circ \vc_f$ (\algoref{detail} \linref{detail-af} and \linref{detail-af2}).
Next we modify the condition in \algoref{detail} \linref{detail-cond} as follows:
``$\forall \vc_e\in M(\pi')$,
$\vc_f$ occurs after $\vc_e$ in $\mathsf{PartCandTuple}^\dagger$ or $\vc_e\not\sqsubseteq \vc_f$''.
Notice that we use the $\sqsubseteq$ relation between vector timestamps to 
determine happens-before relation.

Fix $(\alphabet, \indep)$ and the pattern language $L$. In the following, we assume that every arithmetic operation takes $O(1)$ time and $d, |\threads|, |\alphabet| \in O(1)$.

\begin{theorem}[Correctness]
     On input execution $\tr$, \pattrack declares ``YES''
    iff there is a $\tr'\mazeq{\indep} \tr$ such that $\tr'\in L$.
\end{theorem}


\begin{theorem}[Complexity]
    On input execution $\tr$,
    \pattrack runs in time $O(|\tr|)$
    and uses constant space.
\end{theorem}

\section{Implementation and Evaluation}
\seclabel{experiments}

In this section, we discuss the implementation and evaluation of our vector clock 
algorithm, which we call \pattrack,
and the algorithm due to~\cite{Bertoni1989}, which we call \fbertoni.
The goal of our evaluation is two-fold.
First, we want to understand whether 
the framework of pattern languages can express high-level temporal properties
beyond traditional concurrency bugs, such as data races and atomicity violations.
Further, we want to demonstrate whether the algorithm we propose can effectively expose those violations.
Our second objective is to understand 
the performance improvement of our linear-time algorithm \pattrack
over the classical algorithm \fbertoni 
and show the scalability of our algorithm.
For this, we collected benchmarks from the Java Grande forum benchmark suite~\cite{Smith01},
the DaCapo benchmark suite~\cite{Blackburn06},
as well as open-source GitHub projects used in prior work~\cite{Legunsen2016}. 


\subsection{Experimental Setup}
\myparagraph{Algorithm Implementation}
We implemented the vector clock version of \algoref{detail}
and the classical algorithm from~\cite{Bertoni1989}
in a public prototype tool \textsf{RAPID}~\cite{rapid}, written in Java.
Both algorithms have been implemented in a streaming fashion,
in that, they process each event as soon as it is observed.
To ensure a fair comparison,
we input the same executions to both algorithms.
For this, we first generated one or more execution logs in each benchmark (a total of $33$ executions) 
using the instrumentation and logging facility provided by {\sc RoadRunner}~\cite{RoadRunner2010},
and then, used the generated trace logs
instead of performing monitoring at runtime.
The events of the logs are related to a full program 
which may span many classes during the execution of a program.
For native operations like read and write, 
the instrumentation resolves references precisely,
while for the calls to external APIs, we used manual annotations to print the addresses of the objects.

\myparagraph{Time Reporting}
Comparing the two algorithms in a streaming way and 
with the same termination criteria ensures 
the fairness and correctness of the comparison.
Both algorithms return ``YES'' as soon as they witness that 
a processed prefix can be reordered as a member of the pattern language.
Thus, the length of execution processed by both algorithms is 
always the same, 
either the same prefix or the whole execution. 
This approach to the comparison ensures that 
any differences in performance are due to algorithmic differences 
rather than variations in the amount of data processed.
In addition, we have imposed a 3-hour timeout (TO).
Moreover, the maximum heap size of the Java Virtual Machine is set to 256 GB,
which puts a limit on the amount of memory 
that can be used during execution (OOM).

\myparagraph{Machine Configuration}
Our experiments were conducted on a 1996.250 MHz 64-bit Linux machine 
with Java 19 and 256 GB heap space.



\subsection{Bug Finding}

We demonstrate the effectiveness of the class of pattern languages
in encoding concurrency properties, and of our algorithm
to predict violations against these specifications
using five Java GitHub projects derived from~\cite{Legunsen2016}:
\texttt{logstash-logback-encoder}~\cite{logstash-logback-encoder},
\texttt{exp4j}~\cite{exp4j},
\texttt{jfreechart}~\cite{jfreechart},
\texttt{zmq-appender}~\cite{zeromq},
\textsf{antlrworks}~\cite{antlrworks}.

We examined five high-level properties that we will briefly describe.
The first property ensures consistency between two fields of a class.
The last four come from Java API specifications~\cite{Legunsen2016}.
A violation of these specifications would result in runtime errors.
In the following, we describe these properties briefly and 
present how to encode them as pattern languages.

\begin{description}
    \item[\texttt{Class\_Invariant}.] 
        A common requirement in many object-oriented software designs
        is to ensure that two (or, in general, more) fields of a given class 
        are in a consistent state after every method call.
        Consider two fields \texttt{a},\texttt{b} and also
        two methods \texttt{f1},\texttt{f2}, both of which write to both \texttt{a} and \texttt{b}.
        The design specification of such a class asks that, 
        in every execution, calls to \texttt{f1} and \texttt{f2} behave atomically.
        In other words, an interleaving of \texttt{f1} and \texttt{f2} might leave the values
        in an inconsistent state.
        We can encode one of the violations as the following pattern language:
        $\patts{\scriptsize \texttt{f1.write(a)}, \;\texttt{f2.write(a)}, \;\texttt{f2.write(b)},\; \texttt{f1.write(b)}}.$
    \item[\texttt{Buffer\_ManipulateAfterClose}.] 
        This property asks whether there is an access (write) to a buffer that follows a call to the \texttt{close()} method on the buffer.
        We can encode the violation of this property as the following pattern language:
        $\patts{\scriptsize \texttt{buf.close()}, \;\texttt{buf.write()}}.$
    \item[\texttt{UnsafeIterator}.]
        The property \texttt{Collection\_UnsafeIterator} checks if an execution modifies a collection while iterating it.
        We encode the violation as 
        $\patts{\scriptsize \texttt{iter.next()},\;\texttt{c.add()},\;\texttt{iter.next()}},$
        where \texttt{c} is a collection and \texttt{iter} is one of its iterators. 
        The property \texttt{Map\_UnsafeIterator} is almost the same 
        except that \texttt{iter} is one of the iterators of \texttt{c.entrySet()}.
    \item[\texttt{Collection\_UnsynchronizedAddAll}.]
        This property checks whether a call to \texttt{addAll()} interleaves with another modification, such as $\texttt{add()}$, to the same collection.
        Here, we encode the violation of this property as 
        $\patts{\scriptsize \texttt{c.addAll()Enter},\;\texttt{c.add()},\;\texttt{c.addAll()Exit}},$
        where \texttt{c.addAll()Enter} and \texttt{c.addAll()Exit} are
        events corresponding to the invocation
        and the return of this method respectively.
\end{description}


\begin{table}[t]
    \caption{
    \tablabel{res-bug}
    Experimental Results on Violation Prediction: 
    Columns 1 - 3 present the benchmark name, number of processed events, and number of threads.
    Column 4 shows the name of the violation that our algorithm predicts.
    Columns 5 and 6 report the processing time of \pattrack and \fbertoni.
    }
    \footnotesize
    \scalebox{0.88}{
    \begin{tabular}{ |c|c|c||c|c|c|  }
        \hline
        Program & $\mathcal{N}$ & $\threads$ & Property Violation & \pattrack & \fbertoni  \\
        \hline
        logstash-logback-encoder & 587 & 3 & \texttt{Buffer\_ManipulateAfterClose} & 23ms & 64ms\\
        \hline
        jfreechart & 2753 & 3 & \texttt{Collection\_UnsafeIterator} & 63ms & 683ms\\
        \hline
        zmq-appender & 10742 & 8 & \texttt{Map\_UnsafeIterator} & 168ms & OOM \\
        \hline
        exp4j & 1420 & 3 & \texttt{Collection\_UnsynchronizedAddAll} & 56ms & 171ms\\
        \hline
        antlrworks & 1027 & 3 & \texttt{Class\_Invariant} & 76ms & 379ms\\
        \hline
    \end{tabular}
}
\end{table}

\begin{figure}[t]
    \centering
\begin{subfigure}[b]{0.51\textwidth}
    \centering
\begin{minted}[fontsize=\scriptsize,linenos]{Java}
class JFreeChart{
  public LegendTitle getLegend(int index) {
    int seen = 0;
    Iterator iterator = this.subtitles.iterator();
    while (iterator.hasNext()) {
      Title subtitle = (Title) iterator.next();
      ...
      /* get title */ 
      ...
    }
    return null;
  }

  public void addSubtitle(Title subtitle) {
    ...
    this.subtitles.add(subtitle);
    ...
  }
}
\end{minted}
\caption{\texttt{jfreechart}}
\figlabel{jfreechart-program}
\end{subfigure}
\begin{subfigure}[b]{0.39\textwidth}
\centering
\begin{minted}[fontsize=\scriptsize,linenos]{Java}
class ReusableByteBuffer extends OutputStream {
  @Override
  public void close() {
    this.closed = true;
  }

  @Override
  public void write(byte[] data ...) {
    ...
    if (this.closed) {
      /* throw IOException */
    }
    while (length > 0) {
      ...
      /* write to buffer */
      ...
    }
  }
}
\end{minted}
\caption{\texttt{logstash}}
\figlabel{logstash-program}
\end{subfigure}
\caption{Code snippets.
Violation of \texttt{Collection\_UnsafeIterator} in \texttt{jfreechart}. 
Violation of \texttt{Buffer\_ManipulateAfterClose} in \texttt{logstash}.}
\end{figure}

We executed multi-threaded test cases of the aforementioned 5 GitHub Java projects
and logged the executions.
We ensure that these executions do not contain any pattern violations themselves
either by ensuring that no runtime exception is thrown or by manual inspection.
Nevertheless, as shown in \tabref{res-bug},
our algorithm is able to predict violations in these executions,
indicating the ability of \pattrack to find real-world bugs.
\tabref{res-bug} also demonstrates the performance improvement of \pattrack over \fbertoni.
In the execution of \texttt{zmq-appender}, we observed that \fbertoni throws an out-of-memory exception.
Recall that the space complexity of \fbertoni is $O(\mathcal{N}^{|\threads|})$ (since the width is $\wdth = |\threads|$), 
which is large when the number of processed events and the number of threads is even moderately large,
and this is evident in its memory usage.

Let us highlight some of the snippets from the source code of these benchmarks
and patterns exhibited by them that lead to violations of the mentioned properties.
First, \figref{jfreechart-program} shows a real bug violating the property \texttt{Collection\_UnsafeIterator} in \texttt{JFreeChart}.
When two different threads call the methods \texttt{getLegend} and \texttt{addSubtitle} concurrently,
a \texttt{ConcurrentModification} error might be thrown.
This is because the call to \texttt{add()} in \texttt{addSubtitle}
might interleave between two \texttt{next()} calls.
Next, consider the code snippet in~\figref{logstash-program}
derived from \texttt{logstash}.
Here, \texttt{Buffer\_ManipulateAfterClose} property might be violated.
Even though the \texttt{write()} method checks \texttt{this.closed} before writing to the buffer,
it is still possible that setting \texttt{this.closed} in \texttt{close()} occurs 
between the checking (line 10) and the writing (lines 13 - 17). 
The author of this project has claimed that this class is not thread-safe.
The last example of {\tt antlrworks} has been presented as the motivating example in \exref{example-motivating}.
In conclusion, pattern languages are expressive enough to encode
simple and intuitive real-world high-level correctness specifications,
and our algorithm can predict violations of these effectively.


\subsection{Performance Evaluation}


\begin{table}[t]
\caption{
\tablabel{pat3dim}
Experimental results on patterns of dimension 3: 
Columns 1 - 3 contain benchmark information, including name, number of events, and number of threads.
Columns 4 - 13 report the time spent until the first instance of an admissible tuple is found, or the time to process the entire execution (if there is no match)
of \pattrack and \fbertoni on five different 3-dimension patterns.
It is important to note that the patterns in the same column are different for different benchmarks.
For the sake of clarity and ease of presentation, 
we have named the five patterns as $\textsf{Pattern}$ 1-5.}
\renewcommand{\arraystretch}{1.0}
\footnotesize
\scalebox{0.9}{
\begin{tabular}{ |c|c|c||c|c||c|c||c|c||c|c||c|c|  }
\hline 
1 & 2 & 3 & 4 & 5 & 6 & 7 & 8 & 9 & 10 & 11 & 12 & 13\\
 \hline
 Program & $\mathcal{N}$ & $\threads$ & \multicolumn{2}{c||}{Pattern 1} & \multicolumn{2}{c||}{Pattern 2} & \multicolumn{2}{c||}{Pattern 3} & \multicolumn{2}{c||}{Pattern 4} & \multicolumn{2}{c|}{Pattern 5} \\
 \hline
  & & & \ptrack & \bertoni & \ptrack & \bertoni & \ptrack & \bertoni &\ptrack & \bertoni &\ptrack & \bertoni  \\
 \hline
 \textsf{exp4j-1} & 1.0K & 2 & 0.03s & 1.80s & 0.03s & 0.66s & 0.04s & 1.10s & 0.03s & 0.33s & 0.03s & 0.16s\\
\textsf{crawler-1} & 1.8K & 2 & 0.04s & 0.18s & 0.07s & 0.37s & 0.05s & 0.46s & 0.04s & 0.38s & 0.05s & 0.32s\\
\textsf{crawler-2} & 2.0K & 2 & 0.06s & 0.51s & 0.06s & 0.23s & 0.06s & 0.19s & 0.03s & 0.23s & 0.06s & 0.38s\\
\textsf{junit4-1} & 3.1K & 3 & 0.07s & 0.10s & 0.08s & 2m27s & 0.07s & 40.19s & 0.06s & 11m43s & 0.10s & OOM\\
\textsf{junit4-2} & 5.5K & 101 & 0.05s & TO & 0.04s & TO & 0.06s & TO & 0.04s & TO & 0.04s & TO\\
\textsf{junit4-3} & 5.5K & 101 & 0.05s & TO & 0.07s & TO & 0.04s & TO & 0.05s & TO & 0.05s & TO\\
\textsf{junit4-4} & 25K & 501 & 0.09s & TO & 0.08s & TO & 0.09s & TO & 0.11s & TO & 0.11s & TO\\
\textsf{metafacture} & 195K & 3 & 0.83s & 4.76s & 0.98s & 19.27s & 0.92s & 24.18s & 1.04s & 4.30s & 1.17s & 22.51s\\
\textsf{logstash-1} & 371K & 11 & 0.19s & OOM & 0.14s & OOM & 0.13s & OOM & 0.14s & OOM & 0.14s & OOM\\
\textsf{logstash-2} & 446K & 3 & 0.02s & 3.31s & 0.02s & 2.40s & 0.02s & 3.22s & 1.69s & OOM & 1.68s & OOM\\
\textsf{cassandra-1} & 1.3M & 4 & 0.10s & 0.52s & 0.08s & 1.42s & 0.08s & 1.61s & 1.78s & OOM & 1.75s & OOM\\
\textsf{cassandra-2} & 1.5M & 4 & 3.04s & OOM & 0.09s & OOM & 0.07s & OOM & 3.50s & OOM & 0.07s & OOM\\
\textsf{cassandra-3} & 1.6M & 3 & 3.66s & OOM & 0.34s & 26.57s & 3.44s & OOM & 3.26s & OOM & 3.22s & OOM\\
\textsf{cassandra-4} & 1.6M & 3 & 0.30s & 36.48s & 0.32s & 1.67s & 0.35s & 1.71s & 3.38s & OOM & 0.31s & 34.68s\\
\textsf{cassandra-5} & 1.6M & 3 & 6.14s & OOM & 0.51s & 2.11s & 0.49s & 1.46s & 3.77s & OOM & 4.33s & OOM\\
\textsf{exp4j-2} & 1.9M & 11 & 0.05s & 3.88s & 0.04s & 29.93s & 0.04s & 1.50s & 0.04s & 2.91s & 0.05s & 8m48s\\
\textsf{zookeeper-1} & 2.2M & 18 & 0.30s & OOM & 0.25s & OOM & 0.28s & OOM & 0.31s & 8m41s & 0.39s & OOM\\
\textsf{tomcat} & 2.6M & 11 & 5.39s & OOM & 5.48s & OOM & 5.41s & OOM & 5.44s & OOM & 5.31s & OOM\\
\textsf{zookeeper-2} & 5.6M & 10 & 0.76s & OOM & 0.28s & OOM & 0.30s & OOM & 14.48s & OOM & 0.32s & OOM\\
\textsf{zookeeper-3} & 7.5M & 18 & 0.29s & TO & 0.43s & TO & 0.21s & TO & 0.20s & TO & 0.33s & TO\\
\textsf{zookeeper-4} & 8.0M & 10 & 0.09s & 0.37s & 0.22s & OOM & 0.18s & OOM & 0.20s & OOM & 0.13s & 6.62s\\
\textsf{batik} & 95M & 4 & 1m34s & OOM & 1m42s & OOM & 1m34s & OOM & 1m47s & OOM & 1m8s & OOM\\
\textsf{xalan} & 248M & 3 & 3.97s & OOM & 4.97s & OOM & 4.97s & OOM & 3.39s & OOM & 6.09s & OOM\\
\textsf{sunflow} & 262M & 5 & 25.41s & OOM & 25.83s & OOM & 26.22s & OOM & 26.04s & OOM & 25.91s & OOM\\
\textsf{lusearch} & 325M & 4 & 0.76s & OOM & 0.40s & OOM & 0.98s & OOM & 2m2s & OOM & 0.48s & OOM\\
\textsf{moldyn} & 401M & 4 & 14.21s & OOM & 14.90s & OOM & 14.80s & OOM & 14.64s & OOM & 14.55s & OOM\\
\textsf{series} & 509M & 4 & 0.09s & OOM & 0.62s & OOM & 0.17s & OOM & 0.07s & OOM & 0.12s & OOM\\
\textsf{montecarlo} & 530M & 4 & 0.47s & 0.83s & 1.38s & OOM & 1.32s & OOM & 1.25s & OOM & 1.31s & OOM\\
\textsf{raytracer} & 545M & 4 & 2m45s & OOM & 2m36s & OOM & 2m34s & OOM & 2m33s & OOM & 2m34s & OOM\\
\textsf{sparsematmult} & 602M & 4 & 7m45s & OOM & 7m57s & OOM & 7m48s & OOM & 7m55s & OOM & 7m55s & OOM\\
\textsf{lufact} & 616M & 4 & 1m29s & OOM & 1m26s & OOM & 57.77s & OOM & 2m26s & OOM & 3m33s & OOM\\
\textsf{sor} & 642M & 4 & 1m39s & OOM & 1m42s & OOM & 1m40s & OOM & 18.92s & 43.08s & 1m0s & OOM\\
\textsf{avrora} & 739M & 8 & 19.09s & OOM & 31.29s & OOM & 48.93s & OOM & 52.88s & OOM & 36.23s & OOM\\
\hline
\textbf{Mean} & 168M & - & 30.29s & 2h0m & 30.75s & 2h0m & 29.86s & 2h5m & 35.19s & 2h17m & 32.59s & 2h22m\\
 \hline
\end{tabular}
}
\vspace{-0.2in}
\end{table}

\begin{table}[t]
\caption{
\tablabel{pat5dim}
Experimental results on patterns of dimension 5.
Columns 1 - 3 contain benchmark information, including name, number of events, and number of threads.
Columns 4 - 13 report the time spent until the first instance of an admissible tuple is found, or the time to process the entire execution (if there is no match)
of \pattrack and \fbertoni on five different patterns.
}
\footnotesize
\scalebox{0.88}{
\begin{tabular}{ |c|c|c||c|c||c|c||c|c||c|c||c|c|  }
    \hline 
    1 & 2 & 3 & 4 & 5 & 6 & 7 & 8 & 9 & 10 & 11 & 12 & 13\\
 \hline
 Program & $\mathcal{N}$ & $\threads$ & \multicolumn{2}{c||}{Pattern 1} & \multicolumn{2}{c||}{Pattern 2} & \multicolumn{2}{c||}{Pattern 3} & \multicolumn{2}{c||}{Pattern 4} & \multicolumn{2}{c|}{Pattern 5} \\
 \hline
  & & & \ptrack & \bertoni & \ptrack & \bertoni & \ptrack & \bertoni &\ptrack & \bertoni &\ptrack & \bertoni  \\
 \hline
 \textsf{exp4j-1} & 1.0K & 2 & 0.04s & 1.62s & 0.04s & 1.15s & 0.05s & 1.17s & 0.04s & 1.56s & 0.04s & 0.36s\\
\textsf{crawler-1} & 1.8K & 2 & 0.07s & 0.33s & 0.06s & 0.34s & 0.05s & 0.46s & 0.05s & 0.33s & 0.06s & 0.44s\\
\textsf{crawler-2} & 2.0K & 2 & 0.08s & 0.49s & 0.10s & 0.43s & 0.05s & 0.48s & 0.07s & 0.48s & 0.08s & 0.45s\\
\textsf{junit4-1} & 3.1K & 3 & 0.09s & 5m5s & 0.13s & OOM & 0.08s & OOM & 0.09s & OOM & 0.10s & OOM\\
\textsf{junit4-2} & 5.5K & 101 & 0.06s & TO & 0.04s & TO & 0.05s & TO & 0.05s & TO & 0.06s & TO\\
\textsf{junit4-3} & 5.5K & 101 & 0.06s & TO & 0.06s & TO & 0.05s & TO & 0.06s & TO & 0.06s & TO\\
\textsf{junit4-4} & 25K & 501 & 0.10s & TO & 0.09s & TO & 0.16s & TO & 0.12s & TO & 0.09s & TO\\
\textsf{metafacture} & 195K & 3 & 1.19s & 22.75s & 1.01s & 19.18s & 0.91s & 19.17s & 1.12s & 22.53s & 0.92s & 22.98s\\
\textsf{logstash-1} & 371K & 11 & 0.26s & OOM & 0.30s & OOM & 0.28s & OOM & 0.28s & OOM & 0.40s & OOM\\
\textsf{logstash-2} & 446K & 3 & 1.83s & OOM & 0.03s & 2.45s & 2.11s & OOM & 0.02s & 3.18s & 2.08s & OOM\\
\textsf{cassandra-1} & 1.3M & 4 & 1.99s & OOM & 1.99s & OOM & 1.92s & OOM & 2.05s & OOM & 0.08s & 1.24s\\
\textsf{cassandra-2} & 1.5M & 4 & 49.05s & OOM & 1m14s & OOM & 15.41s & OOM & 6.14s & OOM & 9.50s & OOM\\
\textsf{cassandra-3} & 1.6M & 3 & 3.42s & OOM & 4.11s & OOM & 7.96s & OOM & 4.56s & OOM & 4.10s & OOM\\
\textsf{cassandra-4} & 1.6M & 3 & 3.62s & OOM & 5.19s & OOM & 4.06s & OOM & 3.25s & OOM & 3.90s & OOM\\
\textsf{cassandra-5} & 1.6M & 3 & 5.19s & OOM & 20.42s & OOM & 5.42s & OOM & 4.20s & OOM & 3.87s & OOM\\
\textsf{exp4j-2} & 1.9M & 11 & 0.04s & 6m49s & 0.05s & 10m34s & 0.06s & 56.35s & 0.07s & 9m14s & 0.09s & OOM\\
\textsf{zookeeper-1} & 2.2M & 18 & 6.02s & OOM & 0.45s & OOM & 0.36s & OOM & 0.45s & OOM & 0.33s & OOM\\
\textsf{tomcat} & 2.6M & 11 & 5.33s & OOM & 5.35s & OOM & 5.42s & OOM & 5.43s & OOM & 5.48s & OOM\\
\textsf{zookeeper-2} & 5.6M & 10 & 0.39s & 19.50s & 0.33s & OOM & 18.48s & OOM & 21.51s & OOM & 28.82s & OOM\\
\textsf{zookeeper-3} & 7.5M & 18 & 0.34s & TO & 0.21s & TO & 0.23s & TO & 1.77s & TO & 0.58s & TO\\
\textsf{zookeeper-4} & 8.0M & 10 & 0.23s & OOM & 0.29s & OOM & 0.34s & OOM & 0.27s & OOM & 24.08s & OOM\\
\textsf{batik} & 95M & 4 & 2m27s & OOM & 2m25s & OOM & 2m27s & OOM & 2m26s & OOM & 2m25s & OOM\\
\textsf{xalan} & 248M & 3 & 3.59s & OOM & 6.98s & OOM & 4.71s & OOM & 7.17s & OOM & 8.14s & OOM\\
\textsf{sunflow} & 262M & 5 & 25.43s & OOM & 27.92s & OOM & 25.63s & OOM & 25.36s & OOM & 26.28s & OOM\\
\textsf{lusearch} & 325M & 4 & 0.75s & OOM & 2m5s & OOM & 1.57s & OOM & 39.54s & OOM & 3.65s & OOM\\
\textsf{moldyn} & 401M & 4 & 15.48s & OOM & 15.61s & OOM & 13.39s & OOM & 16.44s & OOM & 15.25s & OOM\\
\textsf{series} & 509M & 4 & 0.14s & OOM & 0.09s & OOM & 0.14s & OOM & 0.24s & OOM & 0.12s & OOM\\
\textsf{montecarlo} & 530M & 4 & 2.12s & OOM & 1.62s & OOM & 1.75s & OOM & 1.15s & OOM & 1.54s & OOM\\
\textsf{raytracer} & 545M & 4 & 2m36s & OOM & 2m41s & OOM & 3m49s & OOM & 51m11s & OOM & 47m11s & OOM\\
\textsf{sparsematmult} & 602M & 4 & 7m43s & OOM & 7m50s & OOM & 7m52s & OOM & 7m47s & OOM & 7m51s & OOM\\
\textsf{lufact} & 616M & 4 & 3m45s & OOM & 58.28s & OOM & 1m31s & OOM & 3m19s & OOM & 1m6s & OOM\\
\textsf{sor} & 642M & 4 & 1m43s & OOM & 2m1s & OOM & 52.96s & OOM & 1m53s & OOM & 52.18s & OOM\\
\textsf{avrora} & 739M & 8 & 1m57s & OOM & 3m20s & OOM & 42.06s & OOM & 1m28s & OOM & 53.27s & OOM\\
\hline
 \textbf{Mean} & 168M & - & 40.66s & 2h22m & 43.97s & 2h27m & 34.74s & 2h32m & 2m8s & 2h27m & 1m53s & 2h32m\\
 \hline
\end{tabular}
}
\end{table}

The goal of this section is to evaluate the performance of \pattrack
over large execution traces and a variety of patterns.
We first discuss how we obtained the patterns we monitor against.
Next, we evaluate both \pattrack as well \fbertoni~\cite{Bertoni1989}
against these specifications and compare their performance.
Finally, we investigate the impact of several parameters that may affect the practical performance
of our algorithm.
Specifically, we examine how the performance of our algorithm evolves 
with changes in the length of executions and the number of threads.

\myparagraph{Pattern Generation}
In order to ensure that the comparison is sufficient and fair,
we generated pattern languages of dimension $d=5$ and $d=3$ at random
by sampling events in each execution log.
The choice of $d=5$ and $d=3$ is in line with
prior empirical observations that most concurrent bugs in 
real-world scenarios can be observed by the ordering of a small number of events~\cite{Burckhardt2010}.
In our experiment, our pattern is a sequence of program locations $\tuple{\seq{l}{d}}$, 
instead of labels.
Each program location can be viewed as a set of program labels,
so we evaluate the membership problem against a generalized pattern language
$L = \bigcup_{a^1_i\in l_1, \dots, a^d_i\in l_d} \patts{a^1_i,\dots,a^d_i}$.

To generate a pattern, we randomly chose 5 or 3 events and logged their program locations.
We employed two policies for selecting events. 
The first policy we utilized is locality. 
We divided each execution into 100 parts of equal length and 
randomly selected one of those parts as the source of one pattern.
Events chosen from widely separated parts of the execution
are less likely to participate in a bug-inducing pattern.
The second policy is diversity.
We chose events from as many different threads as possible,
as this leads to more concurrency.


\begin{figure}[t]
    \centering
        \begin{subfigure}{0.49\textwidth}
            \centerline{\includegraphics[width=0.75\textwidth]{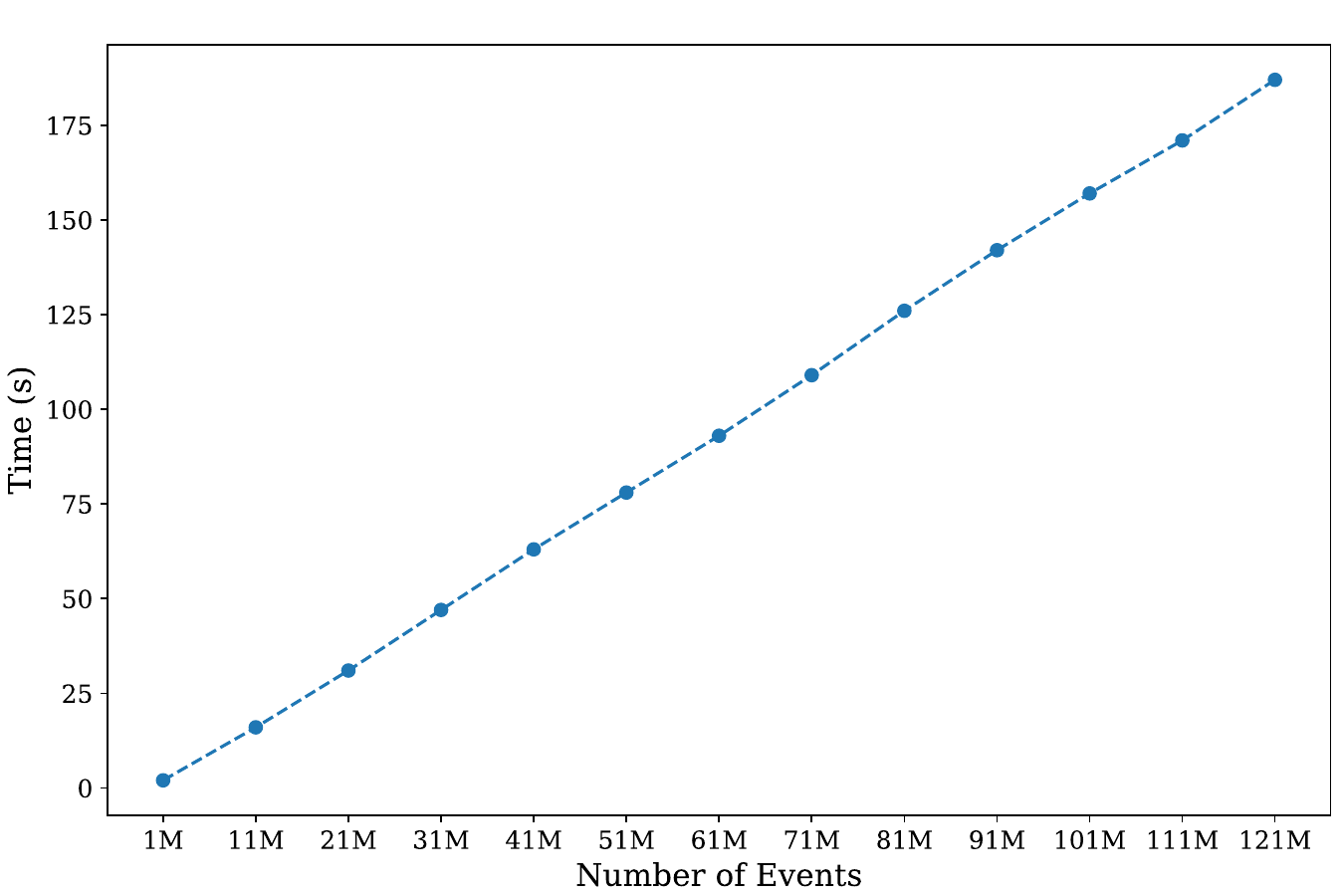}}
            \figlabel{linear_1}
            \caption{raytracer}
        \end{subfigure}
        \hfill
        \begin{subfigure}{0.49\textwidth}
            \centerline{\includegraphics[width=0.75\textwidth]{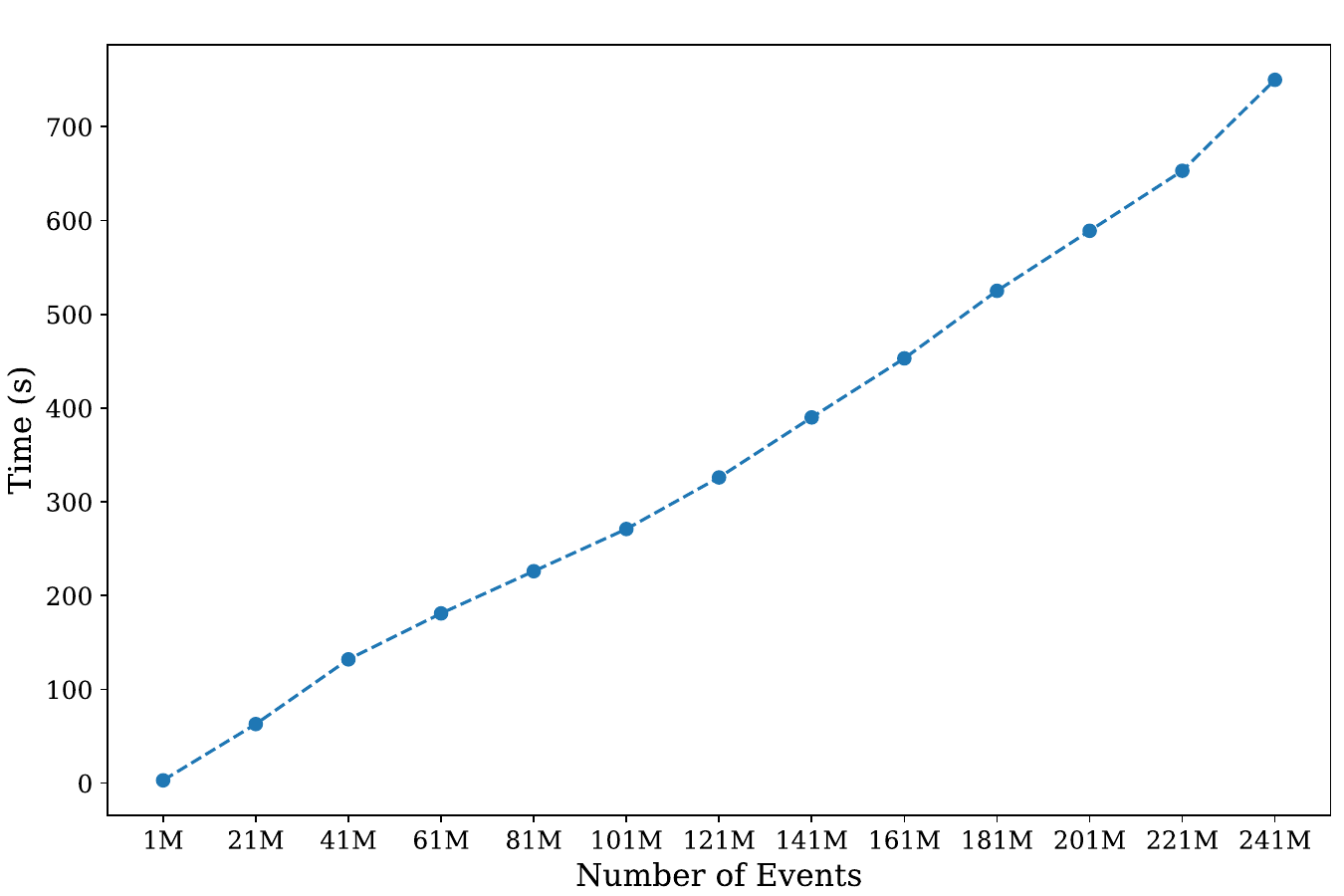}}
            \figlabel{linear_2}
            \caption{sparsematmult}
        \end{subfigure}
           \caption{Scalability with number of events in the executions}
           \figlabel{linear}
\vspace{-0.15in}
\end{figure}

\myparagraph{Speedup over \cite{Bertoni1989}}
Theoretical analysis indicates that our algorithm runs in $O(|\tr|)$ time, 
while the algorithm in~\cite{Bertoni1989} runs in $O(|\tr|^\wdth)$ time. 
As shown in \tabref{pat3dim} and \tabref{pat5dim}, 
our algorithm \pattrack exhibits significant performance advantages 
over \fbertoni in practice. 
Additionally, \fbertoni is not scalable 
when the length of the execution increases, 
as evidenced by the numerous instances of ``OOM'' (Out of Memory) cases 
in the tables. 
This is because \fbertoni needs to store all ideals of a partial order, 
which consumes $O(|\tr|^\wdth)$ space.
As a result, this limits its scalability only to short executions.
In contrast, our constant-space algorithm \pattrack, catered to pattern regular languages, is not burdened by this limitation and 
can handle executions with millions of events.

\myparagraph{Performance w.r.t. execution length}
We next investigate whether the theoretical linear-time complexity of our algorithm also translates empirically.
We selected two executions, from programs \textsf{raytracer} and \textsf{sparsematmult},
based on the fact that their length and the time taken to witness the first pattern match are both large.
This selection criterion ensures that we can gather sufficient data to
evaluate the performance across a range of execution lengths.
To accomplish this, 
we recorded the time taken by our algorithm to process
every 10 million events for \textsf{raytracer} and 
every 20 million events for \textsf{sparsematmult}. 
\figref{linear} presents the result, and is in alignment with the theoretical linear running time of \pattrack.


\begin{wrapfigure}[10]{r}{0.35\textwidth}
    \vspace{-0.2in}
    \includegraphics[width=0.35\textwidth]{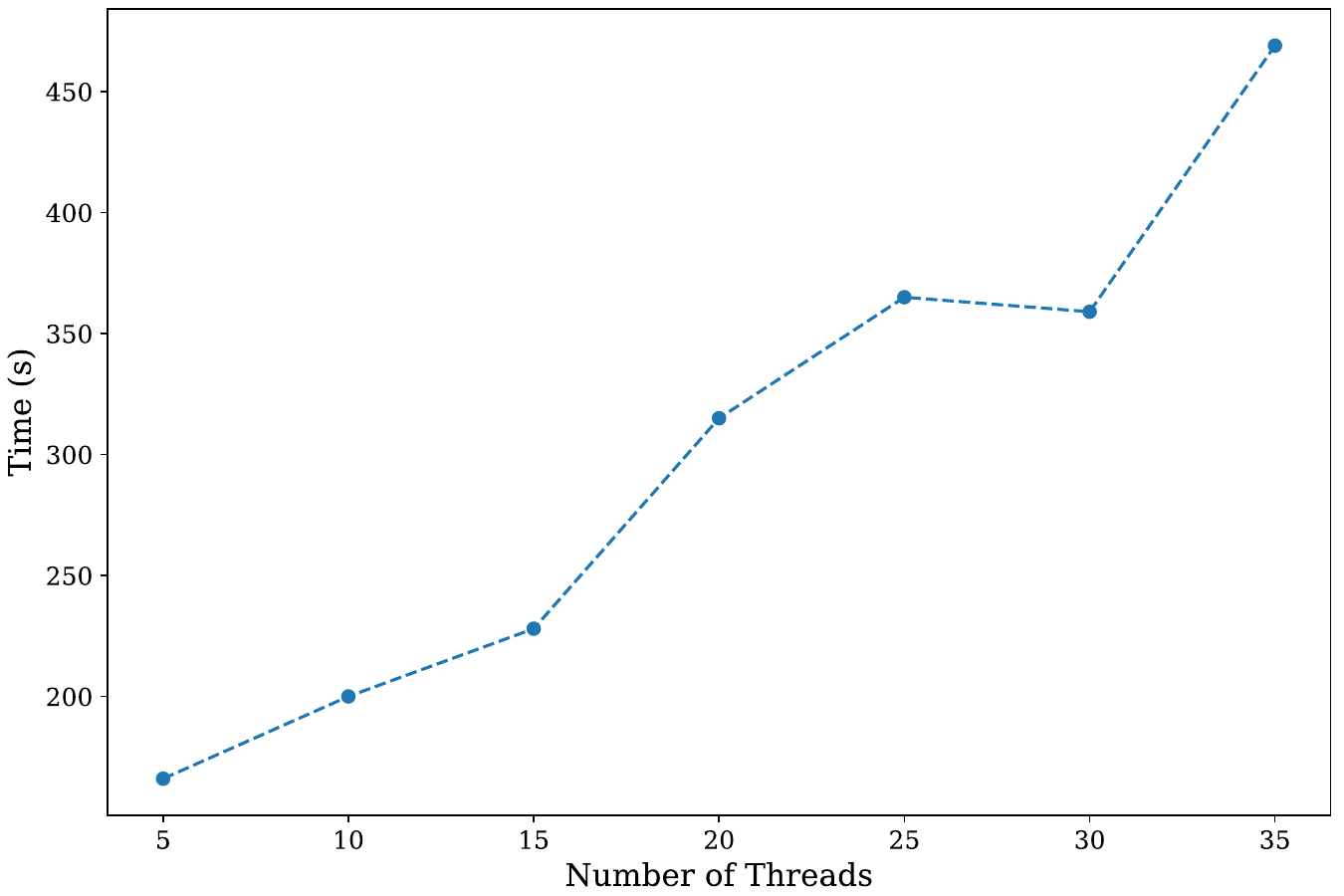}
    \caption{Performance over \#threads}
    \figlabel{thread}
\end{wrapfigure}

\myparagraph{Performance w.r.t number of threads}
The time taken by each vector clock operation 
is proportional to the number of threads, 
and we expect that \pattrack's running time is
also proportional to this number. 
To test this assumption,
we generated seven executions from benchmark \textsf{raytracer} 
with $5$, $10$, $15$, $20$, $25$, and $35$ threads respectively.
The \textsf{raytracer} application allows us to configure the number of threads when generating executions.
We then recorded the time taken by \pattrack to process $1,000,000$ events for each execution.
The results, in \figref{thread}, show that the time consumption increases 
as the number of threads increases, 
in line with \pattrack's theoretical complexity.


\section{Related Work}
\seclabel{related}

Runtime verification has emerged as a popular class of techniques 
instrumental in popularizing the adoption of lightweight formal methods in 
large-scale industrial settings. 
Several classes of specifications such as LTL~\cite{rocsu2005rewriting}, extended regular expressions~\cite{sen2003generating}, and CFG~\cite{meredith2010efficient}
and techniques to monitor the runtime behaviour of the underlying software
in conjunction with these specifications have been proposed.
Aspect-oriented monitoring frameworks have been developed to bridge the gap between formal algorithms and practical tools. 
These include Java-MaC~\cite{kim2004java} and JavaMOP~\cite{jin2012javamop}.
However, such techniques are catered towards the problem of the ``detection'' of violations of these specifications at runtime, in contrast to the ``prediction'' problem, which we address in our work when the underlying software is concurrent or multi-threaded.

Predictive techniques have recently emerged as a fundamental paradigm to enhance the efficacy of runtime verification techniques --- instead of focusing on bugs in the observed sequence of events, such techniques attempt to find bugs in alternate correct reorderings.
Most works in the runtime predictive analysis are centered around the detection of implicit concurrency-centric specifications such as the presence of data races, deadlocks, or violations of atomicity. 
Some of the early works that studied predictive analysis for data races are 
instrumental in formulating core ideas such as sound and maximal causal models and correct reorderings~\cite{Sen2005,sen2006online,chen2007parametric,chen2008jpredictor,MaximalCausalModel2013,Said11,huang2015gpredict,RVPredict2014}.
Many of these techniques, however, primarily employ heavy-weight approaches such as exhaustive enumeration or the use of SAT/SMT solvers, limiting their scalability in practice.
The focus of recent works in predictive analysis tools
~\cite{cai2021sound,wcp2017,SyncP2021,roemer2019online,roemer2020smarttrack,SHB2018,Pavlogiannis2019}
has been on efficiency and practical scalability. 
Such techniques develop carefully crafted algorithms relying on specific heuristics or partial orders and avoid heavy-weight SMT solving.
Beyond data races, recent improvements in these techniques have been used for 
predicting other concurrency bugs such as deadlocks~\cite{cai2021sound,sorrentino2010penelope,tuncc2023sound,Kalhauge2018,SHB2018}. 
Some works also propose predictive methods for higher level, 
but specific properties such as use-after-free violations~\cite{huang2018ufo}, null-pointer dereferences~\cite{farzan2012predicting}, 
and violations of atomicity specifications~\cite{farzan2008monitoring,farzan2009complexity,flanagan2008velodrome,biswas2014doublechecker,sinha2012predicting,mathur2020atomicity}. 
Our work is a step towards adapting predictive techniques to explicitly defined specifications.
GPredict~\cite{huang2015gpredict} proposes the use of SMT solving for generalized predictive monitoring over explicitly defined specifications (under the maximal causality model) but has limited scalability.

We remark that for the case of implicitly defined bugs such as data races, the computational intractability~\cite{Mathur2020,Gibbons1997} in prediction arises primarily because of the exhaustive space of correct reorderings considered in these techniques~\cite{cp2012,MaximalCausalModel2013}.
Reasoning based on Mazurkiewicz traces, though less exhaustive, is in general more scalable.
Indeed, the most widely adopted data race detection techniques crucially rely on the happens-before partial order~\cite{pozniansky2003efficient,fasttrack2009,SHB2018} based on a Mazurkiewicz dependence relation, and typically run in linear time
(assuming the alphabet is of constant size) or in low-degree polynomial time (when the alphabet is not constant~\cite{kulkarni2021dynamic}).
Our work has been inspired by this contrast in computational complexity of reasoning with Mazurkiewicz traces v/s correct reorderings. 
Unfortunately, though, our hardness result depicts that this insight does not trivially generalize to arbitrary properties beyond data races. 
Our work therefore proposes pattern languages to offer a good balance of expressiveness and algorithmic efficiency.
Our central result essentially says that the \emph{trace closure} of pattern languages is regular.
A complete characterization of regular languages whose closure is regular is that of star-connected regular languages due to~\cite{Ochmanski85}.
However, this result does not immediately yield an algorithm to translate an arbitrary star-connected languages to an automaton that recognizes its trace closure.
Languages generated by Alphabetic Pattern Constraints (APC)~\cite{BouajjaniMT01} are a subclass of star-connected language, are closed under trace closure, and are strictly more expressive than generalized pattern languages. 
However, an efficient vector clock algorithm for predictive monitoring against APCs is not known.
Our work thus fills the gap between theoretical results and practical techniques
by designing small-sized monitors for predictive monitoring against pattern languages.
The use of after sets in our constant space algorithm has been inspired from work in data race detection~\cite{ZipTrack2018}.

Soundness (absence of false positives), although highly desirable, often comes at a cost of computational complexity. 
Towards this, many runtime predictive analysis techniques often forego soundness in favour of the simplicity of analysis, 
including the Eraser algorithm based on locksets~\cite{savage1997eraser} or works on deadlock prediction~\cite{cai2020low,bensalem2005dynamic}. 
Some approaches perform post-hoc analysis to reduce false positives~\cite{roemer2018high,roemer2020smarttrack}. 
Other approaches rely on re-executions to confirm the predicted bugs~\cite{joshi2009randomized,sorrentino2010penelope} to reduce false positives.

Concurrency bug detection has been an active area of research for several decades. 
Besides runtime verification and predictive monitoring, 
there are many other techniques for the analysis of concurrent programs.
Model checking~\cite{clarke1986automatic} has emerged as a popular paradigm for finding bugs, 
thanks to advances such as dynamic partial order reduction~\cite{FlanaganGodefroid2005} and stateless model checking~\cite{abdulla2014optimal,kokologiannakis2019model,Trust2022,oberhauser2021vsync}. 
Randomized testing techniques~\cite{joshi2009randomized,Burckhardt2010,ozkan2019trace,yuan2018partial} as well as fuzz testing techniques~\cite{Razzer2019} have also been shown effective in practice. 
Static analysis techniques~\cite{voung2007relay,Chord2006,RacerD2018,racerx2003} have been developed, but their adoption is often limited by high false positive rates. 
Type systems for preventing data races~\cite{boyapati2002ownership,flanagan2008types,flanagan2003type} have been instrumental in the design of programming languages such as Rust.



\section{Conclusions}
\seclabel{conclusions}

In this work, we study the predictive monitoring problem --- given
an execution of a concurrent program, can it be reordered to witness the violation of a specification? 
We show that for specifications expressed using regular languages, and when the reorderings are restricted
to the \emph{trace equivalence} class of the observed execution,
this problem suffers a high complexity polynomial lower bound.
Towards this, we propose a sub-class of regular languages, called (generalized)
pattern languages, and show that this class of languages can be effectively monitored,
in a predictive sense,
using a constant-space linear-time algorithm.
Our experimental evaluation, using an implementation of the algorithm
\pattrack we develop, shows the effectiveness of our proposed class
of specification languages and the algorithms for (predictive) monitoring
against them.

There are many avenues for future work.
We expect pattern languages to be useful in controlled
concurrency testing techniques~\cite{Musuvathi2007, Nekara2021} and
fuzz testing for concurrent software~\cite{Razzer2019}.
Another interesting direction is to generalize the algorithms developed here to
a larger class of temporal specifications, such as APCs~\cite{BouajjaniMT01}, which remain
easier to write, are developer-friendly, and house the potential of widespread adoption.
Building optimal stateless model checking~\cite{Trust2022} algorithms for richer specifications
such as pattern languages is another interesting direction.

\begin{acks}
We thank the anonymous reviewers for several comments that helped improve the paper.
We thank Vladimir Gladshtein and Martin Mirchev for initial discussions, and  Georg Zetzsche for pointing us to the connection with the work of Edward Ochma{\'n}ski on star-connected languages.
This work was partially supported by a Singapore Ministry of Education (MoE) Academic Research Fund (AcRF) Tier 1 grant.
\end{acks}

\section*{Data Availability Statement}
The artefact for this work is available~\cite{artifact}, which contains source code and benchmarks to reproduce our evaluation in \secref{experiments}.


\bibliography{refs}


\begin{thebibliography}{103}


\ifx \showCODEN    \undefined \def \showCODEN     #1{\unskip}     \fi
\ifx \showDOI      \undefined \def \showDOI       #1{#1}\fi
\ifx \showISBNx    \undefined \def \showISBNx     #1{\unskip}     \fi
\ifx \showISBNxiii \undefined \def \showISBNxiii  #1{\unskip}     \fi
\ifx \showISSN     \undefined \def \showISSN      #1{\unskip}     \fi
\ifx \showLCCN     \undefined \def \showLCCN      #1{\unskip}     \fi
\ifx \shownote     \undefined \def \shownote      #1{#1}          \fi
\ifx \showarticletitle \undefined \def \showarticletitle #1{#1}   \fi
\ifx \showURL      \undefined \def \showURL       {\relax}        \fi
\providecommand\bibfield[2]{#2}
\providecommand\bibinfo[2]{#2}
\providecommand\natexlab[1]{#1}
\providecommand\showeprint[2][]{arXiv:#2}

\bibitem[ant(2023)]%
        {antlrworks}
 \bibinfo{year}{2023}\natexlab{}.
\newblock \bibinfo{title}{AntLRWorks}.
\newblock \bibinfo{howpublished}{\url{https://github.com/antlr/antlrworks}}.
\newblock
\newblock
\shownote{[Online; accessed 24-October-2023]}.


\bibitem[exp(2023)]%
        {exp4j}
 \bibinfo{year}{2023}\natexlab{}.
\newblock \bibinfo{title}{exp4j}.
\newblock \bibinfo{howpublished}{\url{https://github.com/fasseg/exp4j}}.
\newblock
\newblock
\shownote{[Online; accessed 24-October-2023]}.


\bibitem[jfr(2023)]%
        {jfreechart}
 \bibinfo{year}{2023}\natexlab{}.
\newblock \bibinfo{title}{JFreeChart}.
\newblock \bibinfo{howpublished}{\url{https://github.com/jfree/jfreechart}}.
\newblock
\newblock
\shownote{[Online; accessed 24-October-2023]}.


\bibitem[log(2023)]%
        {logstash-logback-encoder}
 \bibinfo{year}{2023}\natexlab{}.
\newblock \bibinfo{title}{Logstash Logback Encoder}.
\newblock \bibinfo{howpublished}{\url{https://github.com/logfellow/logstash-logback-encoder}}.
\newblock
\newblock
\shownote{[Online; accessed 24-October-2023]}.


\bibitem[zer(2023)]%
        {zeromq}
 \bibinfo{year}{2023}\natexlab{}.
\newblock \bibinfo{title}{ZeroMQ log4j appender}.
\newblock \bibinfo{howpublished}{\url{https://github.com/lusis/zmq-appender}}.
\newblock
\newblock
\shownote{[Online; accessed 24-October-2023]}.


\bibitem[Abdulla et~al\mbox{.}(2014)]%
        {abdulla2014optimal}
\bibfield{author}{\bibinfo{person}{Parosh~Aziz Abdulla}, \bibinfo{person}{Stavros Aronis}, \bibinfo{person}{Bengt Jonsson}, {and} \bibinfo{person}{Konstantinos Sagonas}.} \bibinfo{year}{2014}\natexlab{}.
\newblock \showarticletitle{Optimal dynamic partial order reduction}. In \bibinfo{booktitle}{\emph{The 41st Annual {ACM} {SIGPLAN-SIGACT} Symposium on Principles of Programming Languages, {POPL} '14, San Diego, CA, USA, January 20-21, 2014}}, \bibfield{editor}{\bibinfo{person}{Suresh Jagannathan} {and} \bibinfo{person}{Peter Sewell}} (Eds.). \bibinfo{publisher}{{ACM}}, \bibinfo{pages}{373--384}.
\newblock
\urldef\tempurl%
\url{https://doi.org/10.1145/2535838.2535845}
\showDOI{\tempurl}


\bibitem[Abdulla et~al\mbox{.}(2019)]%
        {Abdulla2019}
\bibfield{author}{\bibinfo{person}{Parosh~Aziz Abdulla}, \bibinfo{person}{Jatin Arora}, \bibinfo{person}{Mohamed~Faouzi Atig}, {and} \bibinfo{person}{Shankaranarayanan Krishna}.} \bibinfo{year}{2019}\natexlab{}.
\newblock \showarticletitle{Verification of Programs under the Release-Acquire Semantics}. In \bibinfo{booktitle}{\emph{Proceedings of the 40th ACM SIGPLAN Conference on Programming Language Design and Implementation}} (Phoenix, AZ, USA) \emph{(\bibinfo{series}{PLDI 2019})}. \bibinfo{publisher}{Association for Computing Machinery}, \bibinfo{address}{New York, NY, USA}, \bibinfo{pages}{1117–1132}.
\newblock
\showISBNx{9781450367127}
\urldef\tempurl%
\url{https://doi.org/10.1145/3314221.3314649}
\showDOI{\tempurl}


\bibitem[Agarwal et~al\mbox{.}(2021)]%
        {Nekara2021}
\bibfield{author}{\bibinfo{person}{Udit Agarwal}, \bibinfo{person}{Pantazis Deligiannis}, \bibinfo{person}{Cheng Huang}, \bibinfo{person}{Kumseok Jung}, \bibinfo{person}{Akash Lal}, \bibinfo{person}{Immad Naseer}, \bibinfo{person}{Matthew Parkinson}, \bibinfo{person}{Arun Thangamani}, \bibinfo{person}{Jyothi Vedurada}, {and} \bibinfo{person}{Yunpeng Xiao}.} \bibinfo{year}{2021}\natexlab{}.
\newblock \showarticletitle{Nekara: Generalised Concurrency Testing}. In \bibinfo{booktitle}{\emph{2021 36th IEEE/ACM International Conference on Automated Software Engineering (ASE)}}. \bibinfo{pages}{679--691}.
\newblock
\urldef\tempurl%
\url{https://doi.org/10.1109/ASE51524.2021.9678838}
\showDOI{\tempurl}


\bibitem[Agrawal and Srikant(1995)]%
        {pattern1995}
\bibfield{author}{\bibinfo{person}{Rakesh Agrawal} {and} \bibinfo{person}{Ramakrishnan Srikant}.} \bibinfo{year}{1995}\natexlab{}.
\newblock \showarticletitle{Mining Sequential Patterns}. In \bibinfo{booktitle}{\emph{Proceedings of the Eleventh International Conference on Data Engineering, March 6-10, 1995, Taipei, Taiwan}}, \bibfield{editor}{\bibinfo{person}{Philip~S. Yu} {and} \bibinfo{person}{Arbee L.~P. Chen}} (Eds.). \bibinfo{publisher}{{IEEE} Computer Society}, \bibinfo{pages}{3--14}.
\newblock
\urldef\tempurl%
\url{https://doi.org/10.1109/ICDE.1995.380415}
\showDOI{\tempurl}


\bibitem[Alur et~al\mbox{.}(2004)]%
        {CaReT2004}
\bibfield{author}{\bibinfo{person}{Rajeev Alur}, \bibinfo{person}{Kousha Etessami}, {and} \bibinfo{person}{P. Madhusudan}.} \bibinfo{year}{2004}\natexlab{}.
\newblock \showarticletitle{A Temporal Logic of Nested Calls and Returns}. In \bibinfo{booktitle}{\emph{Tools and Algorithms for the Construction and Analysis of Systems, 10th International Conference, {TACAS} 2004, Held as Part of the Joint European Conferences on Theory and Practice of Software, {ETAPS} 2004, Barcelona, Spain, March 29 - April 2, 2004, Proceedings}} \emph{(\bibinfo{series}{Lecture Notes in Computer Science}, Vol.~\bibinfo{volume}{2988})}, \bibfield{editor}{\bibinfo{person}{Kurt Jensen} {and} \bibinfo{person}{Andreas Podelski}} (Eds.). \bibinfo{publisher}{Springer}, \bibinfo{pages}{467--481}.
\newblock
\urldef\tempurl%
\url{https://doi.org/10.1007/978-3-540-24730-2\_35}
\showDOI{\tempurl}


\bibitem[Ang and Mathur(2023)]%
        {artifact}
\bibfield{author}{\bibinfo{person}{Zhendong Ang} {and} \bibinfo{person}{Umang Mathur}.} \bibinfo{year}{2023}\natexlab{}.
\newblock \bibinfo{booktitle}{\emph{Predictive Monitoring against Pattern Regular Languages}}.
\newblock
\urldef\tempurl%
\url{https://doi.org/10.5281/zenodo.8424626}
\showDOI{\tempurl}
\newblock
\shownote{Artifact}.


\bibitem[Artho et~al\mbox{.}(2003)]%
        {artho2003high}
\bibfield{author}{\bibinfo{person}{Cyrille Artho}, \bibinfo{person}{Klaus Havelund}, {and} \bibinfo{person}{Armin Biere}.} \bibinfo{year}{2003}\natexlab{}.
\newblock \showarticletitle{High-level data races}.
\newblock \bibinfo{journal}{\emph{Softw. Test. Verification Reliab.}} \bibinfo{volume}{13}, \bibinfo{number}{4} (\bibinfo{year}{2003}), \bibinfo{pages}{207--227}.
\newblock
\urldef\tempurl%
\url{https://doi.org/10.1002/STVR.281}
\showDOI{\tempurl}


\bibitem[Bensalem and Havelund(2005)]%
        {bensalem2005dynamic}
\bibfield{author}{\bibinfo{person}{Saddek Bensalem} {and} \bibinfo{person}{Klaus Havelund}.} \bibinfo{year}{2005}\natexlab{}.
\newblock \showarticletitle{Dynamic Deadlock Analysis of Multi-threaded Programs}. In \bibinfo{booktitle}{\emph{Hardware and Software Verification and Testing, First International Haifa Verification Conference, Haifa, Israel, November 13-16, 2005, Revised Selected Papers}} \emph{(\bibinfo{series}{Lecture Notes in Computer Science}, Vol.~\bibinfo{volume}{3875})}, \bibfield{editor}{\bibinfo{person}{Shmuel Ur}, \bibinfo{person}{Eyal Bin}, {and} \bibinfo{person}{Yaron Wolfsthal}} (Eds.). \bibinfo{publisher}{Springer}, \bibinfo{pages}{208--223}.
\newblock
\urldef\tempurl%
\url{https://doi.org/10.1007/11678779\_15}
\showDOI{\tempurl}


\bibitem[Bertoni et~al\mbox{.}(1989)]%
        {Bertoni1989}
\bibfield{author}{\bibinfo{person}{A. Bertoni}, \bibinfo{person}{G. Mauri}, {and} \bibinfo{person}{N. Sabadini}.} \bibinfo{year}{1989}\natexlab{}.
\newblock \showarticletitle{Membership problems for regular and context-free trace languages}.
\newblock \bibinfo{journal}{\emph{Information and Computation}} \bibinfo{volume}{82}, \bibinfo{number}{2} (\bibinfo{year}{1989}), \bibinfo{pages}{135--150}.
\newblock
\showISSN{0890-5401}
\urldef\tempurl%
\url{https://doi.org/10.1016/0890-5401(89)90051-5}
\showDOI{\tempurl}


\bibitem[Biswas et~al\mbox{.}(2014)]%
        {biswas2014doublechecker}
\bibfield{author}{\bibinfo{person}{Swarnendu Biswas}, \bibinfo{person}{Jipeng Huang}, \bibinfo{person}{Aritra Sengupta}, {and} \bibinfo{person}{Michael~D. Bond}.} \bibinfo{year}{2014}\natexlab{}.
\newblock \showarticletitle{DoubleChecker: efficient sound and precise atomicity checking}. In \bibinfo{booktitle}{\emph{{ACM} {SIGPLAN} Conference on Programming Language Design and Implementation, {PLDI} '14, Edinburgh, United Kingdom - June 09 - 11, 2014}}, \bibfield{editor}{\bibinfo{person}{Michael F.~P. O'Boyle} {and} \bibinfo{person}{Keshav Pingali}} (Eds.). \bibinfo{publisher}{{ACM}}, \bibinfo{pages}{28--39}.
\newblock
\urldef\tempurl%
\url{https://doi.org/10.1145/2594291.2594323}
\showDOI{\tempurl}


\bibitem[Blackburn et~al\mbox{.}(2006)]%
        {Blackburn06}
\bibfield{author}{\bibinfo{person}{Stephen~M. Blackburn}, \bibinfo{person}{Robin Garner}, \bibinfo{person}{Chris Hoffmann}, \bibinfo{person}{Asjad~M. Khan}, \bibinfo{person}{Kathryn~S. McKinley}, \bibinfo{person}{Rotem Bentzur}, \bibinfo{person}{Amer Diwan}, \bibinfo{person}{Daniel Feinberg}, \bibinfo{person}{Daniel Frampton}, \bibinfo{person}{Samuel~Z. Guyer}, \bibinfo{person}{Martin Hirzel}, \bibinfo{person}{Antony~L. Hosking}, \bibinfo{person}{Maria Jump}, \bibinfo{person}{Han~Bok Lee}, \bibinfo{person}{J.~Eliot~B. Moss}, \bibinfo{person}{Aashish Phansalkar}, \bibinfo{person}{Darko Stefanovic}, \bibinfo{person}{Thomas VanDrunen}, \bibinfo{person}{Daniel von Dincklage}, {and} \bibinfo{person}{Ben Wiedermann}.} \bibinfo{year}{2006}\natexlab{}.
\newblock \showarticletitle{The DaCapo benchmarks: java benchmarking development and analysis}. In \bibinfo{booktitle}{\emph{Proceedings of the 21th Annual {ACM} {SIGPLAN} Conference on Object-Oriented Programming, Systems, Languages, and Applications, {OOPSLA} 2006, October 22-26, 2006, Portland, Oregon, {USA}}}, \bibfield{editor}{\bibinfo{person}{Peri~L. Tarr} {and} \bibinfo{person}{William~R. Cook}} (Eds.). \bibinfo{publisher}{{ACM}}, \bibinfo{pages}{169--190}.
\newblock
\urldef\tempurl%
\url{https://doi.org/10.1145/1167473.1167488}
\showDOI{\tempurl}


\bibitem[Blackshear et~al\mbox{.}(2018)]%
        {RacerD2018}
\bibfield{author}{\bibinfo{person}{Sam Blackshear}, \bibinfo{person}{Nikos Gorogiannis}, \bibinfo{person}{Peter~W. O'Hearn}, {and} \bibinfo{person}{Ilya Sergey}.} \bibinfo{year}{2018}\natexlab{}.
\newblock \showarticletitle{RacerD: Compositional Static Race Detection}.
\newblock \bibinfo{journal}{\emph{Proc. ACM Program. Lang.}} \bibinfo{volume}{2}, \bibinfo{number}{OOPSLA}, Article \bibinfo{articleno}{144} (\bibinfo{date}{oct} \bibinfo{year}{2018}), \bibinfo{numpages}{28}~pages.
\newblock
\urldef\tempurl%
\url{https://doi.org/10.1145/3276514}
\showDOI{\tempurl}


\bibitem[Bouajjani et~al\mbox{.}(2001)]%
        {BouajjaniMT01}
\bibfield{author}{\bibinfo{person}{Ahmed Bouajjani}, \bibinfo{person}{Anca Muscholl}, {and} \bibinfo{person}{Tayssir Touili}.} \bibinfo{year}{2001}\natexlab{}.
\newblock \showarticletitle{Permutation Rewriting and Algorithmic Verification}. In \bibinfo{booktitle}{\emph{16th Annual {IEEE} Symposium on Logic in Computer Science, Boston, Massachusetts, USA, June 16-19, 2001, Proceedings}}. \bibinfo{publisher}{{IEEE} Computer Society}, \bibinfo{pages}{399--408}.
\newblock
\urldef\tempurl%
\url{https://doi.org/10.1109/LICS.2001.932515}
\showDOI{\tempurl}


\bibitem[Boyapati et~al\mbox{.}(2002)]%
        {boyapati2002ownership}
\bibfield{author}{\bibinfo{person}{Chandrasekhar Boyapati}, \bibinfo{person}{Robert Lee}, {and} \bibinfo{person}{Martin~C. Rinard}.} \bibinfo{year}{2002}\natexlab{}.
\newblock \showarticletitle{Ownership types for safe programming: preventing data races and deadlocks}. In \bibinfo{booktitle}{\emph{Proceedings of the 2002 {ACM} {SIGPLAN} Conference on Object-Oriented Programming Systems, Languages and Applications, {OOPSLA} 2002, Seattle, Washington, USA, November 4-8, 2002}}, \bibfield{editor}{\bibinfo{person}{Mamdouh Ibrahim} {and} \bibinfo{person}{Satoshi Matsuoka}} (Eds.). \bibinfo{publisher}{{ACM}}, \bibinfo{pages}{211--230}.
\newblock
\urldef\tempurl%
\url{https://doi.org/10.1145/582419.582440}
\showDOI{\tempurl}


\bibitem[Burckhardt et~al\mbox{.}(2010)]%
        {Burckhardt2010}
\bibfield{author}{\bibinfo{person}{Sebastian Burckhardt}, \bibinfo{person}{Pravesh Kothari}, \bibinfo{person}{Madanlal Musuvathi}, {and} \bibinfo{person}{Santosh Nagarakatte}.} \bibinfo{year}{2010}\natexlab{}.
\newblock \showarticletitle{A Randomized Scheduler with Probabilistic Guarantees of Finding Bugs}. In \bibinfo{booktitle}{\emph{Proceedings of the Fifteenth International Conference on Architectural Support for Programming Languages and Operating Systems}} (Pittsburgh, Pennsylvania, USA) \emph{(\bibinfo{series}{ASPLOS XV})}. \bibinfo{publisher}{Association for Computing Machinery}, \bibinfo{address}{New York, NY, USA}, \bibinfo{pages}{167–178}.
\newblock
\showISBNx{9781605588391}
\urldef\tempurl%
\url{https://doi.org/10.1145/1736020.1736040}
\showDOI{\tempurl}


\bibitem[Cai et~al\mbox{.}(2020)]%
        {cai2020low}
\bibfield{author}{\bibinfo{person}{Yan Cai}, \bibinfo{person}{Ruijie Meng}, {and} \bibinfo{person}{Jens Palsberg}.} \bibinfo{year}{2020}\natexlab{}.
\newblock \showarticletitle{Low-overhead deadlock prediction}. In \bibinfo{booktitle}{\emph{{ICSE} '20: 42nd International Conference on Software Engineering, Seoul, South Korea, 27 June - 19 July, 2020}}, \bibfield{editor}{\bibinfo{person}{Gregg Rothermel} {and} \bibinfo{person}{Doo{-}Hwan Bae}} (Eds.). \bibinfo{publisher}{{ACM}}, \bibinfo{pages}{1298--1309}.
\newblock
\urldef\tempurl%
\url{https://doi.org/10.1145/3377811.3380367}
\showDOI{\tempurl}


\bibitem[Cai et~al\mbox{.}(2021)]%
        {cai2021sound}
\bibfield{author}{\bibinfo{person}{Yan Cai}, \bibinfo{person}{Hao Yun}, \bibinfo{person}{Jinqiu Wang}, \bibinfo{person}{Lei Qiao}, {and} \bibinfo{person}{Jens Palsberg}.} \bibinfo{year}{2021}\natexlab{}.
\newblock \showarticletitle{Sound and efficient concurrency bug prediction}. In \bibinfo{booktitle}{\emph{{ESEC/FSE} '21: 29th {ACM} Joint European Software Engineering Conference and Symposium on the Foundations of Software Engineering, Athens, Greece, August 23-28, 2021}}, \bibfield{editor}{\bibinfo{person}{Diomidis Spinellis}, \bibinfo{person}{Georgios Gousios}, \bibinfo{person}{Marsha Chechik}, {and} \bibinfo{person}{Massimiliano~Di Penta}} (Eds.). \bibinfo{publisher}{{ACM}}, \bibinfo{pages}{255--267}.
\newblock
\urldef\tempurl%
\url{https://doi.org/10.1145/3468264.3468549}
\showDOI{\tempurl}


\bibitem[Calabro et~al\mbox{.}(2009)]%
        {SETH2009}
\bibfield{author}{\bibinfo{person}{Chris Calabro}, \bibinfo{person}{Russell Impagliazzo}, {and} \bibinfo{person}{Ramamohan Paturi}.} \bibinfo{year}{2009}\natexlab{}.
\newblock \showarticletitle{The Complexity of Satisfiability of Small Depth Circuits}. In \bibinfo{booktitle}{\emph{Parameterized and Exact Computation: 4th International Workshop, IWPEC 2009, Copenhagen, Denmark, September 10-11, 2009, Revised Selected Papers}}. \bibinfo{publisher}{Springer-Verlag}, \bibinfo{address}{Berlin, Heidelberg}, \bibinfo{pages}{75–85}.
\newblock
\showISBNx{9783642112683}
\urldef\tempurl%
\url{https://doi.org/10.1007/978-3-642-11269-0_6}
\showURL{%
\tempurl}


\bibitem[Chen and Rosu(2007)]%
        {chen2007parametric}
\bibfield{author}{\bibinfo{person}{Feng Chen} {and} \bibinfo{person}{Grigore Rosu}.} \bibinfo{year}{2007}\natexlab{}.
\newblock \showarticletitle{Parametric and Sliced Causality}. In \bibinfo{booktitle}{\emph{Computer Aided Verification, 19th International Conference, {CAV} 2007, Berlin, Germany, July 3-7, 2007, Proceedings}} \emph{(\bibinfo{series}{Lecture Notes in Computer Science}, Vol.~\bibinfo{volume}{4590})}, \bibfield{editor}{\bibinfo{person}{Werner Damm} {and} \bibinfo{person}{Holger Hermanns}} (Eds.). \bibinfo{publisher}{Springer}, \bibinfo{pages}{240--253}.
\newblock
\urldef\tempurl%
\url{https://doi.org/10.1007/978-3-540-73368-3\_27}
\showDOI{\tempurl}


\bibitem[Chen et~al\mbox{.}(2008)]%
        {chen2008jpredictor}
\bibfield{author}{\bibinfo{person}{Feng Chen}, \bibinfo{person}{Traian{-}Florin Serbanuta}, {and} \bibinfo{person}{Grigore Rosu}.} \bibinfo{year}{2008}\natexlab{}.
\newblock \showarticletitle{jPredictor: a predictive runtime analysis tool for java}. In \bibinfo{booktitle}{\emph{30th International Conference on Software Engineering {(ICSE} 2008), Leipzig, Germany, May 10-18, 2008}}, \bibfield{editor}{\bibinfo{person}{Wilhelm Sch{\"{a}}fer}, \bibinfo{person}{Matthew~B. Dwyer}, {and} \bibinfo{person}{Volker Gruhn}} (Eds.). \bibinfo{publisher}{{ACM}}, \bibinfo{pages}{221--230}.
\newblock
\urldef\tempurl%
\url{https://doi.org/10.1145/1368088.1368119}
\showDOI{\tempurl}


\bibitem[Chistikov et~al\mbox{.}(2016)]%
        {chistikov2016}
\bibfield{author}{\bibinfo{person}{Dmitry Chistikov}, \bibinfo{person}{Rupak Majumdar}, {and} \bibinfo{person}{Filip Niksic}.} \bibinfo{year}{2016}\natexlab{}.
\newblock \showarticletitle{Hitting Families of Schedules for Asynchronous Programs}. In \bibinfo{booktitle}{\emph{Computer Aided Verification - 28th International Conference, {CAV} 2016, Toronto, ON, Canada, July 17-23, 2016, Proceedings, Part {II}}} \emph{(\bibinfo{series}{Lecture Notes in Computer Science}, Vol.~\bibinfo{volume}{9780})}, \bibfield{editor}{\bibinfo{person}{Swarat Chaudhuri} {and} \bibinfo{person}{Azadeh Farzan}} (Eds.). \bibinfo{publisher}{Springer}, \bibinfo{pages}{157--176}.
\newblock
\urldef\tempurl%
\url{https://doi.org/10.1007/978-3-319-41540-6\_9}
\showDOI{\tempurl}


\bibitem[Clarke et~al\mbox{.}(1986)]%
        {clarke1986automatic}
\bibfield{author}{\bibinfo{person}{Edmund~M. Clarke}, \bibinfo{person}{E.~Allen Emerson}, {and} \bibinfo{person}{A.~Prasad Sistla}.} \bibinfo{year}{1986}\natexlab{}.
\newblock \showarticletitle{Automatic Verification of Finite-State Concurrent Systems Using Temporal Logic Specifications}.
\newblock \bibinfo{journal}{\emph{{ACM} Trans. Program. Lang. Syst.}} \bibinfo{volume}{8}, \bibinfo{number}{2} (\bibinfo{year}{1986}), \bibinfo{pages}{244--263}.
\newblock
\urldef\tempurl%
\url{https://doi.org/10.1145/5397.5399}
\showDOI{\tempurl}


\bibitem[Emmi et~al\mbox{.}(2011)]%
        {Emmi2011}
\bibfield{author}{\bibinfo{person}{Michael Emmi}, \bibinfo{person}{Shaz Qadeer}, {and} \bibinfo{person}{Zvonimir Rakamari\'{c}}.} \bibinfo{year}{2011}\natexlab{}.
\newblock \showarticletitle{Delay-Bounded Scheduling}. In \bibinfo{booktitle}{\emph{Proceedings of the 38th Annual ACM SIGPLAN-SIGACT Symposium on Principles of Programming Languages}} (Austin, Texas, USA) \emph{(\bibinfo{series}{POPL '11})}. \bibinfo{publisher}{Association for Computing Machinery}, \bibinfo{address}{New York, NY, USA}, \bibinfo{pages}{411–422}.
\newblock
\showISBNx{9781450304900}
\urldef\tempurl%
\url{https://doi.org/10.1145/1926385.1926432}
\showDOI{\tempurl}


\bibitem[Engler and Ashcraft(2003)]%
        {racerx2003}
\bibfield{author}{\bibinfo{person}{Dawson Engler} {and} \bibinfo{person}{Ken Ashcraft}.} \bibinfo{year}{2003}\natexlab{}.
\newblock \showarticletitle{RacerX: Effective, Static Detection of Race Conditions and Deadlocks}. In \bibinfo{booktitle}{\emph{Proceedings of the Nineteenth ACM Symposium on Operating Systems Principles}} (Bolton Landing, NY, USA) \emph{(\bibinfo{series}{SOSP '03})}. \bibinfo{publisher}{Association for Computing Machinery}, \bibinfo{address}{New York, NY, USA}, \bibinfo{pages}{237–252}.
\newblock
\showISBNx{1581137575}
\urldef\tempurl%
\url{https://doi.org/10.1145/945445.945468}
\showDOI{\tempurl}


\bibitem[Farzan and Madhusudan(2008)]%
        {farzan2008monitoring}
\bibfield{author}{\bibinfo{person}{Azadeh Farzan} {and} \bibinfo{person}{P. Madhusudan}.} \bibinfo{year}{2008}\natexlab{}.
\newblock \showarticletitle{Monitoring Atomicity in Concurrent Programs}. In \bibinfo{booktitle}{\emph{Computer Aided Verification, 20th International Conference, {CAV} 2008, Princeton, NJ, USA, July 7-14, 2008, Proceedings}} \emph{(\bibinfo{series}{Lecture Notes in Computer Science}, Vol.~\bibinfo{volume}{5123})}, \bibfield{editor}{\bibinfo{person}{Aarti Gupta} {and} \bibinfo{person}{Sharad Malik}} (Eds.). \bibinfo{publisher}{Springer}, \bibinfo{pages}{52--65}.
\newblock
\urldef\tempurl%
\url{https://doi.org/10.1007/978-3-540-70545-1\_8}
\showDOI{\tempurl}


\bibitem[Farzan and Madhusudan(2009)]%
        {farzan2009complexity}
\bibfield{author}{\bibinfo{person}{Azadeh Farzan} {and} \bibinfo{person}{P. Madhusudan}.} \bibinfo{year}{2009}\natexlab{}.
\newblock \showarticletitle{The Complexity of Predicting Atomicity Violations}. In \bibinfo{booktitle}{\emph{Tools and Algorithms for the Construction and Analysis of Systems, 15th International Conference, {TACAS} 2009, Held as Part of the Joint European Conferences on Theory and Practice of Software, {ETAPS} 2009, York, UK, March 22-29, 2009. Proceedings}} \emph{(\bibinfo{series}{Lecture Notes in Computer Science}, Vol.~\bibinfo{volume}{5505})}, \bibfield{editor}{\bibinfo{person}{Stefan Kowalewski} {and} \bibinfo{person}{Anna Philippou}} (Eds.). \bibinfo{publisher}{Springer}, \bibinfo{pages}{155--169}.
\newblock
\urldef\tempurl%
\url{https://doi.org/10.1007/978-3-642-00768-2\_14}
\showDOI{\tempurl}


\bibitem[Farzan et~al\mbox{.}(2012)]%
        {farzan2012predicting}
\bibfield{author}{\bibinfo{person}{Azadeh Farzan}, \bibinfo{person}{P. Madhusudan}, \bibinfo{person}{Niloofar Razavi}, {and} \bibinfo{person}{Francesco Sorrentino}.} \bibinfo{year}{2012}\natexlab{}.
\newblock \showarticletitle{Predicting null-pointer dereferences in concurrent programs}. In \bibinfo{booktitle}{\emph{20th {ACM} {SIGSOFT} Symposium on the Foundations of Software Engineering (FSE-20), SIGSOFT/FSE'12, Cary, NC, {USA} - November 11 - 16, 2012}}, \bibfield{editor}{\bibinfo{person}{Will Tracz}, \bibinfo{person}{Martin~P. Robillard}, {and} \bibinfo{person}{Tevfik Bultan}} (Eds.). \bibinfo{publisher}{{ACM}}, \bibinfo{pages}{47}.
\newblock
\urldef\tempurl%
\url{https://doi.org/10.1145/2393596.2393651}
\showDOI{\tempurl}


\bibitem[Fidge(1991)]%
        {Fidge91}
\bibfield{author}{\bibinfo{person}{Colin Fidge}.} \bibinfo{year}{1991}\natexlab{}.
\newblock \showarticletitle{Logical Time in Distributed Computing Systems}.
\newblock \bibinfo{journal}{\emph{Computer}} \bibinfo{volume}{24}, \bibinfo{number}{8} (\bibinfo{date}{Aug.} \bibinfo{year}{1991}), \bibinfo{pages}{28--33}.
\newblock
\showISSN{0018-9162}
\urldef\tempurl%
\url{https://doi.org/10.1109/2.84874}
\showDOI{\tempurl}


\bibitem[Flanagan and Freund(2009)]%
        {fasttrack2009}
\bibfield{author}{\bibinfo{person}{Cormac Flanagan} {and} \bibinfo{person}{Stephen~N. Freund}.} \bibinfo{year}{2009}\natexlab{}.
\newblock \showarticletitle{FastTrack: Efficient and Precise Dynamic Race Detection}. In \bibinfo{booktitle}{\emph{Proceedings of the 30th ACM SIGPLAN Conference on Programming Language Design and Implementation}} (Dublin, Ireland) \emph{(\bibinfo{series}{PLDI '09})}. \bibinfo{publisher}{Association for Computing Machinery}, \bibinfo{address}{New York, NY, USA}, \bibinfo{pages}{121–133}.
\newblock
\showISBNx{9781605583921}
\urldef\tempurl%
\url{https://doi.org/10.1145/1542476.1542490}
\showDOI{\tempurl}


\bibitem[Flanagan and Freund(2010)]%
        {RoadRunner2010}
\bibfield{author}{\bibinfo{person}{Cormac Flanagan} {and} \bibinfo{person}{Stephen~N. Freund}.} \bibinfo{year}{2010}\natexlab{}.
\newblock \showarticletitle{The RoadRunner Dynamic Analysis Framework for Concurrent Programs}. In \bibinfo{booktitle}{\emph{Proceedings of the 9th ACM SIGPLAN-SIGSOFT Workshop on Program Analysis for Software Tools and Engineering}} (Toronto, Ontario, Canada) \emph{(\bibinfo{series}{PASTE '10})}. \bibinfo{publisher}{Association for Computing Machinery}, \bibinfo{address}{New York, NY, USA}, \bibinfo{pages}{1–8}.
\newblock
\showISBNx{9781450300827}
\urldef\tempurl%
\url{https://doi.org/10.1145/1806672.1806674}
\showDOI{\tempurl}


\bibitem[Flanagan et~al\mbox{.}(2008b)]%
        {flanagan2008types}
\bibfield{author}{\bibinfo{person}{Cormac Flanagan}, \bibinfo{person}{Stephen~N. Freund}, \bibinfo{person}{Marina Lifshin}, {and} \bibinfo{person}{Shaz Qadeer}.} \bibinfo{year}{2008}\natexlab{b}.
\newblock \showarticletitle{Types for atomicity: Static checking and inference for Java}.
\newblock \bibinfo{journal}{\emph{{ACM} Trans. Program. Lang. Syst.}} \bibinfo{volume}{30}, \bibinfo{number}{4} (\bibinfo{year}{2008}), \bibinfo{pages}{20:1--20:53}.
\newblock
\urldef\tempurl%
\url{https://doi.org/10.1145/1377492.1377495}
\showDOI{\tempurl}


\bibitem[Flanagan et~al\mbox{.}(2008a)]%
        {flanagan2008velodrome}
\bibfield{author}{\bibinfo{person}{Cormac Flanagan}, \bibinfo{person}{Stephen~N. Freund}, {and} \bibinfo{person}{Jaeheon Yi}.} \bibinfo{year}{2008}\natexlab{a}.
\newblock \showarticletitle{Velodrome: a sound and complete dynamic atomicity checker for multithreaded programs}. In \bibinfo{booktitle}{\emph{Proceedings of the {ACM} {SIGPLAN} 2008 Conference on Programming Language Design and Implementation, Tucson, AZ, USA, June 7-13, 2008}}, \bibfield{editor}{\bibinfo{person}{Rajiv Gupta} {and} \bibinfo{person}{Saman~P. Amarasinghe}} (Eds.). \bibinfo{publisher}{{ACM}}, \bibinfo{pages}{293--303}.
\newblock
\urldef\tempurl%
\url{https://doi.org/10.1145/1375581.1375618}
\showDOI{\tempurl}


\bibitem[Flanagan and Godefroid(2005)]%
        {FlanaganGodefroid2005}
\bibfield{author}{\bibinfo{person}{Cormac Flanagan} {and} \bibinfo{person}{Patrice Godefroid}.} \bibinfo{year}{2005}\natexlab{}.
\newblock \showarticletitle{Dynamic Partial-Order Reduction for Model Checking Software}. In \bibinfo{booktitle}{\emph{Proceedings of the 32nd ACM SIGPLAN-SIGACT Symposium on Principles of Programming Languages}} (Long Beach, California, USA) \emph{(\bibinfo{series}{POPL '05})}. \bibinfo{publisher}{Association for Computing Machinery}, \bibinfo{address}{New York, NY, USA}, \bibinfo{pages}{110–121}.
\newblock
\showISBNx{158113830X}
\urldef\tempurl%
\url{https://doi.org/10.1145/1040305.1040315}
\showDOI{\tempurl}


\bibitem[Flanagan and Qadeer(2003)]%
        {flanagan2003type}
\bibfield{author}{\bibinfo{person}{Cormac Flanagan} {and} \bibinfo{person}{Shaz Qadeer}.} \bibinfo{year}{2003}\natexlab{}.
\newblock \showarticletitle{A type and effect system for atomicity}. In \bibinfo{booktitle}{\emph{Proceedings of the {ACM} {SIGPLAN} 2003 Conference on Programming Language Design and Implementation 2003, San Diego, California, USA, June 9-11, 2003}}, \bibfield{editor}{\bibinfo{person}{Ron Cytron} {and} \bibinfo{person}{Rajiv Gupta}} (Eds.). \bibinfo{publisher}{{ACM}}, \bibinfo{pages}{338--349}.
\newblock
\urldef\tempurl%
\url{https://doi.org/10.1145/781131.781169}
\showDOI{\tempurl}


\bibitem[Gao et~al\mbox{.}(2023)]%
        {Gao2023}
\bibfield{author}{\bibinfo{person}{Mingyu Gao}, \bibinfo{person}{Soham Chakraborty}, {and} \bibinfo{person}{Burcu~Kulahcioglu Ozkan}.} \bibinfo{year}{2023}\natexlab{}.
\newblock \showarticletitle{Probabilistic Concurrency Testing for Weak Memory Programs}. In \bibinfo{booktitle}{\emph{Proceedings of the 28th ACM International Conference on Architectural Support for Programming Languages and Operating Systems, Volume 2}} (Vancouver, BC, Canada) \emph{(\bibinfo{series}{ASPLOS 2023})}. \bibinfo{publisher}{Association for Computing Machinery}, \bibinfo{address}{New York, NY, USA}, \bibinfo{pages}{603–616}.
\newblock
\showISBNx{9781450399166}
\urldef\tempurl%
\url{https://doi.org/10.1145/3575693.3575729}
\showDOI{\tempurl}


\bibitem[Giacomo and Vardi(2013)]%
        {Giacomo2013}
\bibfield{author}{\bibinfo{person}{Giuseppe~De Giacomo} {and} \bibinfo{person}{Moshe~Y. Vardi}.} \bibinfo{year}{2013}\natexlab{}.
\newblock \showarticletitle{Linear Temporal Logic and Linear Dynamic Logic on Finite Traces}. In \bibinfo{booktitle}{\emph{{IJCAI} 2013, Proceedings of the 23rd International Joint Conference on Artificial Intelligence, Beijing, China, August 3-9, 2013}}, \bibfield{editor}{\bibinfo{person}{Francesca Rossi}} (Ed.). \bibinfo{publisher}{{IJCAI/AAAI}}, \bibinfo{pages}{854--860}.
\newblock
\urldef\tempurl%
\url{https://doi.org/10.5555/2540128.2540252}
\showDOI{\tempurl}


\bibitem[Gibbons and Korach(1997)]%
        {Gibbons1997}
\bibfield{author}{\bibinfo{person}{Phillip~B. Gibbons} {and} \bibinfo{person}{Ephraim Korach}.} \bibinfo{year}{1997}\natexlab{}.
\newblock \showarticletitle{Testing Shared Memories}.
\newblock \bibinfo{journal}{\emph{SIAM J. Comput.}} \bibinfo{volume}{26}, \bibinfo{number}{4} (\bibinfo{year}{1997}), \bibinfo{pages}{1208--1244}.
\newblock
\urldef\tempurl%
\url{https://doi.org/10.1137/S0097539794279614}
\showDOI{\tempurl}


\bibitem[Herlihy and Wing(1990)]%
        {linearlizability1990}
\bibfield{author}{\bibinfo{person}{Maurice Herlihy} {and} \bibinfo{person}{Jeannette~M. Wing}.} \bibinfo{year}{1990}\natexlab{}.
\newblock \showarticletitle{Linearizability: {A} Correctness Condition for Concurrent Objects}.
\newblock \bibinfo{journal}{\emph{{ACM} Trans. Program. Lang. Syst.}} \bibinfo{volume}{12}, \bibinfo{number}{3} (\bibinfo{year}{1990}), \bibinfo{pages}{463--492}.
\newblock
\urldef\tempurl%
\url{https://doi.org/10.1145/78969.78972}
\showDOI{\tempurl}


\bibitem[Huang(2018)]%
        {huang2018ufo}
\bibfield{author}{\bibinfo{person}{Jeff Huang}.} \bibinfo{year}{2018}\natexlab{}.
\newblock \showarticletitle{{UFO:} predictive concurrency use-after-free detection}. In \bibinfo{booktitle}{\emph{Proceedings of the 40th International Conference on Software Engineering, {ICSE} 2018, Gothenburg, Sweden, May 27 - June 03, 2018}}, \bibfield{editor}{\bibinfo{person}{Michel Chaudron}, \bibinfo{person}{Ivica Crnkovic}, \bibinfo{person}{Marsha Chechik}, {and} \bibinfo{person}{Mark Harman}} (Eds.). \bibinfo{publisher}{{ACM}}, \bibinfo{pages}{609--619}.
\newblock
\urldef\tempurl%
\url{https://doi.org/10.1145/3180155.3180225}
\showDOI{\tempurl}


\bibitem[Huang et~al\mbox{.}(2015)]%
        {huang2015gpredict}
\bibfield{author}{\bibinfo{person}{Jeff Huang}, \bibinfo{person}{Qingzhou Luo}, {and} \bibinfo{person}{Grigore Rosu}.} \bibinfo{year}{2015}\natexlab{}.
\newblock \showarticletitle{GPredict: Generic Predictive Concurrency Analysis}. In \bibinfo{booktitle}{\emph{37th {IEEE/ACM} International Conference on Software Engineering, {ICSE} 2015, Florence, Italy, May 16-24, 2015, Volume 1}}, \bibfield{editor}{\bibinfo{person}{Antonia Bertolino}, \bibinfo{person}{Gerardo Canfora}, {and} \bibinfo{person}{Sebastian~G. Elbaum}} (Eds.). \bibinfo{publisher}{{IEEE} Computer Society}, \bibinfo{pages}{847--857}.
\newblock
\urldef\tempurl%
\url{https://doi.org/10.1109/ICSE.2015.96}
\showDOI{\tempurl}


\bibitem[Huang et~al\mbox{.}(2014)]%
        {RVPredict2014}
\bibfield{author}{\bibinfo{person}{Jeff Huang}, \bibinfo{person}{Patrick~O'Neil Meredith}, {and} \bibinfo{person}{Grigore Rosu}.} \bibinfo{year}{2014}\natexlab{}.
\newblock \showarticletitle{Maximal Sound Predictive Race Detection with Control Flow Abstraction}. In \bibinfo{booktitle}{\emph{Proceedings of the 35th ACM SIGPLAN Conference on Programming Language Design and Implementation}} (Edinburgh, United Kingdom) \emph{(\bibinfo{series}{PLDI '14})}. \bibinfo{publisher}{Association for Computing Machinery}, \bibinfo{address}{New York, NY, USA}, \bibinfo{pages}{337–348}.
\newblock
\showISBNx{9781450327848}
\urldef\tempurl%
\url{https://doi.org/10.1145/2594291.2594315}
\showDOI{\tempurl}


\bibitem[Itzkovitz et~al\mbox{.}(1999)]%
        {Djit1999}
\bibfield{author}{\bibinfo{person}{Ayal Itzkovitz}, \bibinfo{person}{Assaf Schuster}, {and} \bibinfo{person}{Oren Zeev-Ben-Mordehai}.} \bibinfo{year}{1999}\natexlab{}.
\newblock \showarticletitle{Toward Integration of Data Race Detection in DSM Systems}.
\newblock \bibinfo{journal}{\emph{J. Parallel and Distrib. Comput.}} \bibinfo{volume}{59}, \bibinfo{number}{2} (\bibinfo{year}{1999}), \bibinfo{pages}{180--203}.
\newblock
\showISSN{0743-7315}
\urldef\tempurl%
\url{https://doi.org/10.1006/jpdc.1999.1574}
\showDOI{\tempurl}


\bibitem[Jeong et~al\mbox{.}(2019)]%
        {Razzer2019}
\bibfield{author}{\bibinfo{person}{Dae~R. Jeong}, \bibinfo{person}{Kyungtae Kim}, \bibinfo{person}{Basavesh Shivakumar}, \bibinfo{person}{Byoungyoung Lee}, {and} \bibinfo{person}{Insik Shin}.} \bibinfo{year}{2019}\natexlab{}.
\newblock \showarticletitle{Razzer: Finding Kernel Race Bugs through Fuzzing}. In \bibinfo{booktitle}{\emph{2019 IEEE Symposium on Security and Privacy (SP)}}. \bibinfo{pages}{754--768}.
\newblock
\urldef\tempurl%
\url{https://doi.org/10.1109/SP.2019.00017}
\showDOI{\tempurl}


\bibitem[Jin et~al\mbox{.}(2012)]%
        {jin2012javamop}
\bibfield{author}{\bibinfo{person}{Dongyun Jin}, \bibinfo{person}{Patrick~O'Neil Meredith}, \bibinfo{person}{Choonghwan Lee}, {and} \bibinfo{person}{Grigore Rosu}.} \bibinfo{year}{2012}\natexlab{}.
\newblock \showarticletitle{JavaMOP: Efficient parametric runtime monitoring framework}. In \bibinfo{booktitle}{\emph{34th International Conference on Software Engineering, {ICSE} 2012, June 2-9, 2012, Zurich, Switzerland}}, \bibfield{editor}{\bibinfo{person}{Martin Glinz}, \bibinfo{person}{Gail~C. Murphy}, {and} \bibinfo{person}{Mauro Pezz{\`{e}}}} (Eds.). \bibinfo{publisher}{{IEEE} Computer Society}, \bibinfo{pages}{1427--1430}.
\newblock
\urldef\tempurl%
\url{https://doi.org/10.1109/ICSE.2012.6227231}
\showDOI{\tempurl}


\bibitem[Joshi et~al\mbox{.}(2009)]%
        {joshi2009randomized}
\bibfield{author}{\bibinfo{person}{Pallavi Joshi}, \bibinfo{person}{Chang{-}Seo Park}, \bibinfo{person}{Koushik Sen}, {and} \bibinfo{person}{Mayur Naik}.} \bibinfo{year}{2009}\natexlab{}.
\newblock \showarticletitle{A randomized dynamic program analysis technique for detecting real deadlocks}. In \bibinfo{booktitle}{\emph{Proceedings of the 2009 {ACM} {SIGPLAN} Conference on Programming Language Design and Implementation, {PLDI} 2009, Dublin, Ireland, June 15-21, 2009}}, \bibfield{editor}{\bibinfo{person}{Michael Hind} {and} \bibinfo{person}{Amer Diwan}} (Eds.). \bibinfo{publisher}{{ACM}}, \bibinfo{pages}{110--120}.
\newblock
\urldef\tempurl%
\url{https://doi.org/10.1145/1542476.1542489}
\showDOI{\tempurl}


\bibitem[Kalhauge and Palsberg(2018)]%
        {Kalhauge2018}
\bibfield{author}{\bibinfo{person}{Christian~Gram Kalhauge} {and} \bibinfo{person}{Jens Palsberg}.} \bibinfo{year}{2018}\natexlab{}.
\newblock \showarticletitle{Sound Deadlock Prediction}.
\newblock \bibinfo{journal}{\emph{Proc. ACM Program. Lang.}} \bibinfo{volume}{2}, \bibinfo{number}{OOPSLA}, Article \bibinfo{articleno}{146} (\bibinfo{date}{Oct.} \bibinfo{year}{2018}), \bibinfo{numpages}{29}~pages.
\newblock
\urldef\tempurl%
\url{https://doi.org/10.1145/3276516}
\showDOI{\tempurl}


\bibitem[Kasikci et~al\mbox{.}(2017)]%
        {Kasikci2017}
\bibfield{author}{\bibinfo{person}{Baris Kasikci}, \bibinfo{person}{Weidong Cui}, \bibinfo{person}{Xinyang Ge}, {and} \bibinfo{person}{Ben Niu}.} \bibinfo{year}{2017}\natexlab{}.
\newblock \showarticletitle{Lazy Diagnosis of In-Production Concurrency Bugs}. In \bibinfo{booktitle}{\emph{Proceedings of the 26th Symposium on Operating Systems Principles}} (Shanghai, China) \emph{(\bibinfo{series}{SOSP '17})}. \bibinfo{publisher}{Association for Computing Machinery}, \bibinfo{address}{New York, NY, USA}, \bibinfo{pages}{582–598}.
\newblock
\showISBNx{9781450350853}
\urldef\tempurl%
\url{https://doi.org/10.1145/3132747.3132767}
\showDOI{\tempurl}


\bibitem[Kim et~al\mbox{.}(2004)]%
        {kim2004java}
\bibfield{author}{\bibinfo{person}{Moonzoo Kim}, \bibinfo{person}{Mahesh Viswanathan}, \bibinfo{person}{Sampath Kannan}, \bibinfo{person}{Insup Lee}, {and} \bibinfo{person}{Oleg Sokolsky}.} \bibinfo{year}{2004}\natexlab{}.
\newblock \showarticletitle{Java-MaC: {A} Run-Time Assurance Approach for Java Programs}.
\newblock \bibinfo{journal}{\emph{Formal Methods Syst. Des.}} \bibinfo{volume}{24}, \bibinfo{number}{2} (\bibinfo{year}{2004}), \bibinfo{pages}{129--155}.
\newblock
\urldef\tempurl%
\url{https://doi.org/10.1023/B:FORM.0000017719.43755.7C}
\showDOI{\tempurl}


\bibitem[Kini et~al\mbox{.}(2017)]%
        {wcp2017}
\bibfield{author}{\bibinfo{person}{Dileep Kini}, \bibinfo{person}{Umang Mathur}, {and} \bibinfo{person}{Mahesh Viswanathan}.} \bibinfo{year}{2017}\natexlab{}.
\newblock \showarticletitle{Dynamic Race Prediction in Linear-Time}. In \bibinfo{booktitle}{\emph{Proceedings of the 38th ACM SIGPLAN Conference on Programming Language Design and Implementation}} (Barcelona, Spain) \emph{(\bibinfo{series}{PLDI 2017})}. \bibinfo{publisher}{Association for Computing Machinery}, \bibinfo{address}{New York, NY, USA}, \bibinfo{pages}{157–170}.
\newblock
\showISBNx{9781450349888}
\urldef\tempurl%
\url{https://doi.org/10.1145/3062341.3062374}
\showDOI{\tempurl}


\bibitem[Kini et~al\mbox{.}(2018)]%
        {ZipTrack2018}
\bibfield{author}{\bibinfo{person}{Dileep Kini}, \bibinfo{person}{Umang Mathur}, {and} \bibinfo{person}{Mahesh Viswanathan}.} \bibinfo{year}{2018}\natexlab{}.
\newblock \showarticletitle{Data Race Detection on Compressed Traces}. In \bibinfo{booktitle}{\emph{Proceedings of the 2018 26th ACM Joint Meeting on European Software Engineering Conference and Symposium on the Foundations of Software Engineering}} (Lake Buena Vista, FL, USA) \emph{(\bibinfo{series}{ESEC/FSE 2018})}. \bibinfo{publisher}{Association for Computing Machinery}, \bibinfo{address}{New York, NY, USA}, \bibinfo{pages}{26–37}.
\newblock
\showISBNx{9781450355735}
\urldef\tempurl%
\url{https://doi.org/10.1145/3236024.3236025}
\showDOI{\tempurl}


\bibitem[Kokologiannakis et~al\mbox{.}(2022)]%
        {Trust2022}
\bibfield{author}{\bibinfo{person}{Michalis Kokologiannakis}, \bibinfo{person}{Iason Marmanis}, \bibinfo{person}{Vladimir Gladstein}, {and} \bibinfo{person}{Viktor Vafeiadis}.} \bibinfo{year}{2022}\natexlab{}.
\newblock \showarticletitle{Truly Stateless, Optimal Dynamic Partial Order Reduction}.
\newblock \bibinfo{journal}{\emph{Proc. ACM Program. Lang.}} \bibinfo{volume}{6}, \bibinfo{number}{POPL}, Article \bibinfo{articleno}{49} (\bibinfo{date}{jan} \bibinfo{year}{2022}), \bibinfo{numpages}{28}~pages.
\newblock
\urldef\tempurl%
\url{https://doi.org/10.1145/3498711}
\showDOI{\tempurl}


\bibitem[Kokologiannakis et~al\mbox{.}(2019)]%
        {kokologiannakis2019model}
\bibfield{author}{\bibinfo{person}{Michalis Kokologiannakis}, \bibinfo{person}{Azalea Raad}, {and} \bibinfo{person}{Viktor Vafeiadis}.} \bibinfo{year}{2019}\natexlab{}.
\newblock \showarticletitle{Model checking for weakly consistent libraries}. In \bibinfo{booktitle}{\emph{Proceedings of the 40th {ACM} {SIGPLAN} Conference on Programming Language Design and Implementation, {PLDI} 2019, Phoenix, AZ, USA, June 22-26, 2019}}, \bibfield{editor}{\bibinfo{person}{Kathryn~S. McKinley} {and} \bibinfo{person}{Kathleen Fisher}} (Eds.). \bibinfo{publisher}{{ACM}}, \bibinfo{pages}{96--110}.
\newblock
\urldef\tempurl%
\url{https://doi.org/10.1145/3314221.3314609}
\showDOI{\tempurl}


\bibitem[Koymans(1990)]%
        {MTL1990}
\bibfield{author}{\bibinfo{person}{Ron Koymans}.} \bibinfo{year}{1990}\natexlab{}.
\newblock \showarticletitle{Specifying Real-Time Properties with Metric Temporal Logic}.
\newblock \bibinfo{journal}{\emph{Real Time Syst.}} \bibinfo{volume}{2}, \bibinfo{number}{4} (\bibinfo{year}{1990}), \bibinfo{pages}{255--299}.
\newblock
\urldef\tempurl%
\url{https://doi.org/10.1007/BF01995674}
\showDOI{\tempurl}


\bibitem[Kulkarni et~al\mbox{.}(2021)]%
        {kulkarni2021dynamic}
\bibfield{author}{\bibinfo{person}{Rucha Kulkarni}, \bibinfo{person}{Umang Mathur}, {and} \bibinfo{person}{Andreas Pavlogiannis}.} \bibinfo{year}{2021}\natexlab{}.
\newblock \showarticletitle{Dynamic Data-Race Detection Through the Fine-Grained Lens}. In \bibinfo{booktitle}{\emph{32nd International Conference on Concurrency Theory, {CONCUR} 2021, August 24-27, 2021, Virtual Conference}} \emph{(\bibinfo{series}{LIPIcs}, Vol.~\bibinfo{volume}{203})}, \bibfield{editor}{\bibinfo{person}{Serge Haddad} {and} \bibinfo{person}{Daniele Varacca}} (Eds.). \bibinfo{publisher}{Schloss Dagstuhl - Leibniz-Zentrum f{\"{u}}r Informatik}, \bibinfo{pages}{16:1--16:23}.
\newblock
\urldef\tempurl%
\url{https://doi.org/10.4230/LIPICS.CONCUR.2021.16}
\showDOI{\tempurl}


\bibitem[Legunsen et~al\mbox{.}(2016)]%
        {Legunsen2016}
\bibfield{author}{\bibinfo{person}{Owolabi Legunsen}, \bibinfo{person}{Wajih~Ul Hassan}, \bibinfo{person}{Xinyue Xu}, \bibinfo{person}{Grigore Rosu}, {and} \bibinfo{person}{Darko Marinov}.} \bibinfo{year}{2016}\natexlab{}.
\newblock \showarticletitle{How good are the specs? a study of the bug-finding effectiveness of existing Java {API} specifications}. In \bibinfo{booktitle}{\emph{Proceedings of the 31st {IEEE/ACM} International Conference on Automated Software Engineering, {ASE} 2016, Singapore, September 3-7, 2016}}, \bibfield{editor}{\bibinfo{person}{David Lo}, \bibinfo{person}{Sven Apel}, {and} \bibinfo{person}{Sarfraz Khurshid}} (Eds.). \bibinfo{publisher}{{ACM}}, \bibinfo{pages}{602--613}.
\newblock
\urldef\tempurl%
\url{https://doi.org/10.1145/2970276.2970356}
\showDOI{\tempurl}


\bibitem[Maler and Nickovic(2004)]%
        {Maler2004}
\bibfield{author}{\bibinfo{person}{Oded Maler} {and} \bibinfo{person}{Dejan Nickovic}.} \bibinfo{year}{2004}\natexlab{}.
\newblock \showarticletitle{Monitoring Temporal Properties of Continuous Signals}. In \bibinfo{booktitle}{\emph{Formal Techniques, Modelling and Analysis of Timed and Fault-Tolerant Systems, Joint International Conferences on Formal Modelling and Analysis of Timed Systems, {FORMATS} 2004 and Formal Techniques in Real-Time and Fault-Tolerant Systems, {FTRTFT} 2004, Grenoble, France, September 22-24, 2004, Proceedings}} \emph{(\bibinfo{series}{Lecture Notes in Computer Science}, Vol.~\bibinfo{volume}{3253})}, \bibfield{editor}{\bibinfo{person}{Yassine Lakhnech} {and} \bibinfo{person}{Sergio Yovine}} (Eds.). \bibinfo{publisher}{Springer}, \bibinfo{pages}{152--166}.
\newblock
\urldef\tempurl%
\url{https://doi.org/10.1007/978-3-540-30206-3\_12}
\showDOI{\tempurl}


\bibitem[Mathur(2023)]%
        {rapid}
\bibfield{author}{\bibinfo{person}{Umang Mathur}.} \bibinfo{year}{2023}\natexlab{}.
\newblock \bibinfo{booktitle}{\emph{{RAPID}}}.
\newblock
\urldef\tempurl%
\url{https://github.com/umangm/rapid}
\showURL{%
\tempurl}
\newblock
\shownote{Accessed: 2023-10-25}.


\bibitem[Mathur et~al\mbox{.}(2018)]%
        {SHB2018}
\bibfield{author}{\bibinfo{person}{Umang Mathur}, \bibinfo{person}{Dileep Kini}, {and} \bibinfo{person}{Mahesh Viswanathan}.} \bibinfo{year}{2018}\natexlab{}.
\newblock \showarticletitle{What Happens-after the First Race? Enhancing the Predictive Power of Happens-before Based Dynamic Race Detection}.
\newblock \bibinfo{journal}{\emph{Proc. ACM Program. Lang.}} \bibinfo{volume}{2}, \bibinfo{number}{OOPSLA}, Article \bibinfo{articleno}{145} (\bibinfo{date}{oct} \bibinfo{year}{2018}), \bibinfo{numpages}{29}~pages.
\newblock
\urldef\tempurl%
\url{https://doi.org/10.1145/3276515}
\showDOI{\tempurl}


\bibitem[Mathur et~al\mbox{.}(2022)]%
        {TreeClocks2022}
\bibfield{author}{\bibinfo{person}{Umang Mathur}, \bibinfo{person}{Andreas Pavlogiannis}, \bibinfo{person}{H\"{u}nkar~Can Tun\c{c}}, {and} \bibinfo{person}{Mahesh Viswanathan}.} \bibinfo{year}{2022}\natexlab{}.
\newblock \showarticletitle{A Tree Clock Data Structure for Causal Orderings in Concurrent Executions}. In \bibinfo{booktitle}{\emph{Proceedings of the 27th ACM International Conference on Architectural Support for Programming Languages and Operating Systems}} (Lausanne, Switzerland) \emph{(\bibinfo{series}{ASPLOS '22})}. \bibinfo{publisher}{Association for Computing Machinery}, \bibinfo{address}{New York, NY, USA}, \bibinfo{pages}{710–725}.
\newblock
\showISBNx{9781450392051}
\urldef\tempurl%
\url{https://doi.org/10.1145/3503222.3507734}
\showDOI{\tempurl}


\bibitem[Mathur et~al\mbox{.}(2020)]%
        {Mathur2020}
\bibfield{author}{\bibinfo{person}{Umang Mathur}, \bibinfo{person}{Andreas Pavlogiannis}, {and} \bibinfo{person}{Mahesh Viswanathan}.} \bibinfo{year}{2020}\natexlab{}.
\newblock \showarticletitle{The Complexity of Dynamic Data Race Prediction}. In \bibinfo{booktitle}{\emph{Proceedings of the 35th Annual ACM/IEEE Symposium on Logic in Computer Science}} (Saarbr\"{u}cken, Germany) \emph{(\bibinfo{series}{LICS '20})}. \bibinfo{publisher}{Association for Computing Machinery}, \bibinfo{address}{New York, NY, USA}, \bibinfo{pages}{713–727}.
\newblock
\showISBNx{9781450371049}
\urldef\tempurl%
\url{https://doi.org/10.1145/3373718.3394783}
\showDOI{\tempurl}


\bibitem[Mathur et~al\mbox{.}(2021)]%
        {SyncP2021}
\bibfield{author}{\bibinfo{person}{Umang Mathur}, \bibinfo{person}{Andreas Pavlogiannis}, {and} \bibinfo{person}{Mahesh Viswanathan}.} \bibinfo{year}{2021}\natexlab{}.
\newblock \showarticletitle{Optimal Prediction of Synchronization-Preserving Races}.
\newblock \bibinfo{journal}{\emph{Proc. ACM Program. Lang.}} \bibinfo{volume}{5}, \bibinfo{number}{POPL}, Article \bibinfo{articleno}{36} (\bibinfo{date}{jan} \bibinfo{year}{2021}), \bibinfo{numpages}{29}~pages.
\newblock
\urldef\tempurl%
\url{https://doi.org/10.1145/3434317}
\showDOI{\tempurl}


\bibitem[Mathur and Viswanathan(2020)]%
        {mathur2020atomicity}
\bibfield{author}{\bibinfo{person}{Umang Mathur} {and} \bibinfo{person}{Mahesh Viswanathan}.} \bibinfo{year}{2020}\natexlab{}.
\newblock \showarticletitle{Atomicity Checking in Linear Time using Vector Clocks}. In \bibinfo{booktitle}{\emph{{ASPLOS} '20: Architectural Support for Programming Languages and Operating Systems, Lausanne, Switzerland, March 16-20, 2020}}, \bibfield{editor}{\bibinfo{person}{James~R. Larus}, \bibinfo{person}{Luis Ceze}, {and} \bibinfo{person}{Karin Strauss}} (Eds.). \bibinfo{publisher}{{ACM}}, \bibinfo{pages}{183--199}.
\newblock
\urldef\tempurl%
\url{https://doi.org/10.1145/3373376.3378475}
\showDOI{\tempurl}


\bibitem[Mattern(1989)]%
        {Mattern89}
\bibfield{author}{\bibinfo{person}{Friedemann Mattern}.} \bibinfo{year}{1989}\natexlab{}.
\newblock \showarticletitle{Virtual Time and Global States of Distributed Systems}.
\newblock In \bibinfo{booktitle}{\emph{Parallel and Distributed Algorithms: proceedings of the International Workshop on Parallel \& Distributed Algorithms}}, \bibfield{editor}{\bibinfo{person}{M.~Cosnard et. al.}} (Ed.). \bibinfo{publisher}{Elsevier Science Publishers B. V.}, \bibinfo{pages}{215--226}.
\newblock


\bibitem[Mazurkiewicz(1987)]%
        {Mazurkiewicz1987}
\bibfield{author}{\bibinfo{person}{Antoni Mazurkiewicz}.} \bibinfo{year}{1987}\natexlab{}.
\newblock \showarticletitle{Trace theory}. In \bibinfo{booktitle}{\emph{Petri Nets: Applications and Relationships to Other Models of Concurrency}}, \bibfield{editor}{\bibinfo{person}{W.~Brauer}, \bibinfo{person}{W.~Reisig}, {and} \bibinfo{person}{G.~Rozenberg}} (Eds.). \bibinfo{publisher}{Springer Berlin Heidelberg}, \bibinfo{address}{Berlin, Heidelberg}, \bibinfo{pages}{278--324}.
\newblock
\showISBNx{978-3-540-47926-0}


\bibitem[McNaughton and Papert(1971)]%
        {McNaughton1971}
\bibfield{author}{\bibinfo{person}{Robert McNaughton} {and} \bibinfo{person}{Seymour Papert}.} \bibinfo{year}{1971}\natexlab{}.
\newblock \showarticletitle{Counter-Free Automata}.
\newblock


\bibitem[Meredith et~al\mbox{.}(2010)]%
        {meredith2010efficient}
\bibfield{author}{\bibinfo{person}{Patrick~O'Neil Meredith}, \bibinfo{person}{Dongyun Jin}, \bibinfo{person}{Feng Chen}, {and} \bibinfo{person}{Grigore Rosu}.} \bibinfo{year}{2010}\natexlab{}.
\newblock \showarticletitle{Efficient monitoring of parametric context-free patterns}.
\newblock \bibinfo{journal}{\emph{Autom. Softw. Eng.}} \bibinfo{volume}{17}, \bibinfo{number}{2} (\bibinfo{year}{2010}), \bibinfo{pages}{149--180}.
\newblock
\urldef\tempurl%
\url{https://doi.org/10.1007/S10515-010-0063-Y}
\showDOI{\tempurl}


\bibitem[Murali et~al\mbox{.}(2021)]%
        {Murali2021}
\bibfield{author}{\bibinfo{person}{Vijayaraghavan Murali}, \bibinfo{person}{Edward Yao}, \bibinfo{person}{Umang Mathur}, {and} \bibinfo{person}{Satish Chandra}.} \bibinfo{year}{2021}\natexlab{}.
\newblock \showarticletitle{Scalable Statistical Root Cause Analysis on App Telemetry}. In \bibinfo{booktitle}{\emph{Proceedings of the 43rd International Conference on Software Engineering: Software Engineering in Practice}} (Virtual Event, Spain) \emph{(\bibinfo{series}{ICSE-SEIP '21})}. \bibinfo{publisher}{IEEE Press}, \bibinfo{pages}{288–297}.
\newblock
\showISBNx{9780738146690}
\urldef\tempurl%
\url{https://doi.org/10.1109/ICSE-SEIP52600.2021.00038}
\showDOI{\tempurl}


\bibitem[Musuvathi and Qadeer(2006)]%
        {Musuvathi2007}
\bibfield{author}{\bibinfo{person}{Madan Musuvathi} {and} \bibinfo{person}{Shaz Qadeer}.} \bibinfo{year}{2006}\natexlab{}.
\newblock \showarticletitle{{CHESS:} Systematic Stress Testing of Concurrent Software}. In \bibinfo{booktitle}{\emph{Logic-Based Program Synthesis and Transformation, 16th International Symposium, {LOPSTR} 2006, Venice, Italy, July 12-14, 2006, Revised Selected Papers}} \emph{(\bibinfo{series}{Lecture Notes in Computer Science}, Vol.~\bibinfo{volume}{4407})}, \bibfield{editor}{\bibinfo{person}{Germ{\'{a}}n Puebla}} (Ed.). \bibinfo{publisher}{Springer}, \bibinfo{pages}{15--16}.
\newblock
\urldef\tempurl%
\url{https://doi.org/10.1007/978-3-540-71410-1\_2}
\showDOI{\tempurl}


\bibitem[Naik et~al\mbox{.}(2006)]%
        {Chord2006}
\bibfield{author}{\bibinfo{person}{Mayur Naik}, \bibinfo{person}{Alex Aiken}, {and} \bibinfo{person}{John Whaley}.} \bibinfo{year}{2006}\natexlab{}.
\newblock \showarticletitle{Effective Static Race Detection for Java}. In \bibinfo{booktitle}{\emph{Proceedings of the 27th ACM SIGPLAN Conference on Programming Language Design and Implementation}} (Ottawa, Ontario, Canada) \emph{(\bibinfo{series}{PLDI '06})}. \bibinfo{publisher}{Association for Computing Machinery}, \bibinfo{address}{New York, NY, USA}, \bibinfo{pages}{308–319}.
\newblock
\showISBNx{1595933204}
\urldef\tempurl%
\url{https://doi.org/10.1145/1133981.1134018}
\showDOI{\tempurl}


\bibitem[Oberhauser et~al\mbox{.}(2021)]%
        {oberhauser2021vsync}
\bibfield{author}{\bibinfo{person}{Jonas Oberhauser}, \bibinfo{person}{Rafael~Lourenco de Lima~Chehab}, \bibinfo{person}{Diogo Behrens}, \bibinfo{person}{Ming Fu}, \bibinfo{person}{Antonio Paolillo}, \bibinfo{person}{Lilith Oberhauser}, \bibinfo{person}{Koustubha Bhat}, \bibinfo{person}{Yuzhong Wen}, \bibinfo{person}{Haibo Chen}, \bibinfo{person}{Jaeho Kim}, {and} \bibinfo{person}{Viktor Vafeiadis}.} \bibinfo{year}{2021}\natexlab{}.
\newblock \showarticletitle{VSync: push-button verification and optimization for synchronization primitives on weak memory models}. In \bibinfo{booktitle}{\emph{{ASPLOS} '21: 26th {ACM} International Conference on Architectural Support for Programming Languages and Operating Systems, Virtual Event, USA, April 19-23, 2021}}, \bibfield{editor}{\bibinfo{person}{Tim Sherwood}, \bibinfo{person}{Emery~D. Berger}, {and} \bibinfo{person}{Christos Kozyrakis}} (Eds.). \bibinfo{publisher}{{ACM}}, \bibinfo{pages}{530--545}.
\newblock
\urldef\tempurl%
\url{https://doi.org/10.1145/3445814.3446748}
\showDOI{\tempurl}


\bibitem[Ochma{\'n}ski(1985)]%
        {Ochmanski85}
\bibfield{author}{\bibinfo{person}{Edward Ochma{\'n}ski}.} \bibinfo{year}{1985}\natexlab{}.
\newblock \showarticletitle{Regular behaviour of concurrent systems}.
\newblock \bibinfo{journal}{\emph{Bull. {EATCS}}}  \bibinfo{volume}{27} (\bibinfo{year}{1985}), \bibinfo{pages}{56--67}.
\newblock


\bibitem[Ozkan et~al\mbox{.}(2019)]%
        {ozkan2019trace}
\bibfield{author}{\bibinfo{person}{Burcu~Kulahcioglu Ozkan}, \bibinfo{person}{Rupak Majumdar}, {and} \bibinfo{person}{Simin Oraee}.} \bibinfo{year}{2019}\natexlab{}.
\newblock \showarticletitle{Trace aware random testing for distributed systems}.
\newblock \bibinfo{journal}{\emph{Proc. {ACM} Program. Lang.}} \bibinfo{volume}{3}, \bibinfo{number}{{OOPSLA}} (\bibinfo{year}{2019}), \bibinfo{pages}{180:1--180:29}.
\newblock
\urldef\tempurl%
\url{https://doi.org/10.1145/3360606}
\showDOI{\tempurl}


\bibitem[Pavlogiannis(2019)]%
        {Pavlogiannis2019}
\bibfield{author}{\bibinfo{person}{Andreas Pavlogiannis}.} \bibinfo{year}{2019}\natexlab{}.
\newblock \showarticletitle{Fast, Sound, and Effectively Complete Dynamic Race Prediction}.
\newblock \bibinfo{journal}{\emph{Proc. ACM Program. Lang.}} \bibinfo{volume}{4}, \bibinfo{number}{POPL}, Article \bibinfo{articleno}{17} (\bibinfo{date}{dec} \bibinfo{year}{2019}), \bibinfo{numpages}{29}~pages.
\newblock
\urldef\tempurl%
\url{https://doi.org/10.1145/3371085}
\showDOI{\tempurl}


\bibitem[Pnueli(1977)]%
        {Pnueli1977}
\bibfield{author}{\bibinfo{person}{Amir Pnueli}.} \bibinfo{year}{1977}\natexlab{}.
\newblock \showarticletitle{The Temporal Logic of Programs}. In \bibinfo{booktitle}{\emph{Proceedings of the 18th Annual Symposium on Foundations of Computer Science}} \emph{(\bibinfo{series}{SFCS '77})}. \bibinfo{publisher}{IEEE Computer Society}, \bibinfo{address}{USA}, \bibinfo{pages}{46–57}.
\newblock
\urldef\tempurl%
\url{https://doi.org/10.1109/SFCS.1977.32}
\showDOI{\tempurl}


\bibitem[Poznianski and Schuster(2003)]%
        {pozniansky2003efficient}
\bibfield{author}{\bibinfo{person}{Eli Poznianski} {and} \bibinfo{person}{Assaf Schuster}.} \bibinfo{year}{2003}\natexlab{}.
\newblock \showarticletitle{Efficient On-the-Fly Data Race Detection in Multithreaded {C++} Programs}. In \bibinfo{booktitle}{\emph{17th International Parallel and Distributed Processing Symposium {(IPDPS} 2003), 22-26 April 2003, Nice, France, CD-ROM/Abstracts Proceedings}}. \bibinfo{publisher}{{IEEE} Computer Society}, \bibinfo{pages}{287}.
\newblock
\urldef\tempurl%
\url{https://doi.org/10.1109/IPDPS.2003.1213513}
\showDOI{\tempurl}


\bibitem[Roemer and Bond(2019)]%
        {roemer2019online}
\bibfield{author}{\bibinfo{person}{Jake Roemer} {and} \bibinfo{person}{Michael~D. Bond}.} \bibinfo{year}{2019}\natexlab{}.
\newblock \showarticletitle{Online Set-Based Dynamic Analysis for Sound Predictive Race Detection}.
\newblock \bibinfo{journal}{\emph{CoRR}}  \bibinfo{volume}{abs/1907.08337} (\bibinfo{year}{2019}).
\newblock
\showeprint[arXiv]{1907.08337}
\urldef\tempurl%
\url{http://arxiv.org/abs/1907.08337}
\showURL{%
\tempurl}


\bibitem[Roemer et~al\mbox{.}(2018)]%
        {roemer2018high}
\bibfield{author}{\bibinfo{person}{Jake Roemer}, \bibinfo{person}{Kaan Gen{\c{c}}}, {and} \bibinfo{person}{Michael~D. Bond}.} \bibinfo{year}{2018}\natexlab{}.
\newblock \showarticletitle{High-coverage, unbounded sound predictive race detection}. In \bibinfo{booktitle}{\emph{Proceedings of the 39th {ACM} {SIGPLAN} Conference on Programming Language Design and Implementation, {PLDI} 2018, Philadelphia, PA, USA, June 18-22, 2018}}, \bibfield{editor}{\bibinfo{person}{Jeffrey~S. Foster} {and} \bibinfo{person}{Dan Grossman}} (Eds.). \bibinfo{publisher}{{ACM}}, \bibinfo{pages}{374--389}.
\newblock
\urldef\tempurl%
\url{https://doi.org/10.1145/3192366.3192385}
\showDOI{\tempurl}


\bibitem[Roemer et~al\mbox{.}(2020)]%
        {roemer2020smarttrack}
\bibfield{author}{\bibinfo{person}{Jake Roemer}, \bibinfo{person}{Kaan Gen{\c{c}}}, {and} \bibinfo{person}{Michael~D. Bond}.} \bibinfo{year}{2020}\natexlab{}.
\newblock \showarticletitle{SmartTrack: efficient predictive race detection}. In \bibinfo{booktitle}{\emph{Proceedings of the 41st {ACM} {SIGPLAN} International Conference on Programming Language Design and Implementation, {PLDI} 2020, London, UK, June 15-20, 2020}}, \bibfield{editor}{\bibinfo{person}{Alastair~F. Donaldson} {and} \bibinfo{person}{Emina Torlak}} (Eds.). \bibinfo{publisher}{{ACM}}, \bibinfo{pages}{747--762}.
\newblock
\urldef\tempurl%
\url{https://doi.org/10.1145/3385412.3385993}
\showDOI{\tempurl}


\bibitem[Rosu and Havelund(2005)]%
        {rocsu2005rewriting}
\bibfield{author}{\bibinfo{person}{Grigore Rosu} {and} \bibinfo{person}{Klaus Havelund}.} \bibinfo{year}{2005}\natexlab{}.
\newblock \showarticletitle{Rewriting-Based Techniques for Runtime Verification}.
\newblock \bibinfo{journal}{\emph{Autom. Softw. Eng.}} \bibinfo{volume}{12}, \bibinfo{number}{2} (\bibinfo{year}{2005}), \bibinfo{pages}{151--197}.
\newblock
\urldef\tempurl%
\url{https://doi.org/10.1007/S10515-005-6205-Y}
\showDOI{\tempurl}


\bibitem[Rosu and Viswanathan(2003)]%
        {RV2003}
\bibfield{author}{\bibinfo{person}{Grigore Rosu} {and} \bibinfo{person}{Mahesh Viswanathan}.} \bibinfo{year}{2003}\natexlab{}.
\newblock \showarticletitle{Testing Extended Regular Language Membership Incrementally by Rewriting}. In \bibinfo{booktitle}{\emph{Rewriting Techniques and Applications, 14th International Conference, {RTA} 2003, Valencia, Spain, June 9-11, 2003, Proceedings}} \emph{(\bibinfo{series}{Lecture Notes in Computer Science}, Vol.~\bibinfo{volume}{2706})}, \bibfield{editor}{\bibinfo{person}{Robert Nieuwenhuis}} (Ed.). \bibinfo{publisher}{Springer}, \bibinfo{pages}{499--514}.
\newblock
\urldef\tempurl%
\url{https://doi.org/10.1007/3-540-44881-0\_35}
\showDOI{\tempurl}


\bibitem[Sadowski et~al\mbox{.}(2018)]%
        {Sadowski2018}
\bibfield{author}{\bibinfo{person}{Caitlin Sadowski}, \bibinfo{person}{Edward Aftandilian}, \bibinfo{person}{Alex Eagle}, \bibinfo{person}{Liam Miller-Cushon}, {and} \bibinfo{person}{Ciera Jaspan}.} \bibinfo{year}{2018}\natexlab{}.
\newblock \showarticletitle{Lessons from Building Static Analysis Tools at Google}.
\newblock \bibinfo{journal}{\emph{Commun. ACM}} \bibinfo{volume}{61}, \bibinfo{number}{4} (\bibinfo{date}{mar} \bibinfo{year}{2018}), \bibinfo{pages}{58–66}.
\newblock
\showISSN{0001-0782}
\urldef\tempurl%
\url{https://doi.org/10.1145/3188720}
\showDOI{\tempurl}


\bibitem[Said et~al\mbox{.}(2011)]%
        {Said11}
\bibfield{author}{\bibinfo{person}{Mahmoud Said}, \bibinfo{person}{Chao Wang}, \bibinfo{person}{Zijiang Yang}, {and} \bibinfo{person}{Karem~A. Sakallah}.} \bibinfo{year}{2011}\natexlab{}.
\newblock \showarticletitle{Generating Data Race Witnesses by an SMT-Based Analysis}. In \bibinfo{booktitle}{\emph{{NASA} Formal Methods - Third International Symposium, {NFM} 2011, Pasadena, CA, USA, April 18-20, 2011. Proceedings}} \emph{(\bibinfo{series}{Lecture Notes in Computer Science}, Vol.~\bibinfo{volume}{6617})}, \bibfield{editor}{\bibinfo{person}{Mihaela~Gheorghiu Bobaru}, \bibinfo{person}{Klaus Havelund}, \bibinfo{person}{Gerard~J. Holzmann}, {and} \bibinfo{person}{Rajeev Joshi}} (Eds.). \bibinfo{publisher}{Springer}, \bibinfo{pages}{313--327}.
\newblock
\urldef\tempurl%
\url{https://doi.org/10.1007/978-3-642-20398-5\_23}
\showDOI{\tempurl}


\bibitem[Savage et~al\mbox{.}(1997)]%
        {savage1997eraser}
\bibfield{author}{\bibinfo{person}{Stefan Savage}, \bibinfo{person}{Michael Burrows}, \bibinfo{person}{Greg Nelson}, \bibinfo{person}{Patrick Sobalvarro}, {and} \bibinfo{person}{Thomas~E. Anderson}.} \bibinfo{year}{1997}\natexlab{}.
\newblock \showarticletitle{Eraser: {A} Dynamic Data Race Detector for Multithreaded Programs}.
\newblock \bibinfo{journal}{\emph{{ACM} Trans. Comput. Syst.}} \bibinfo{volume}{15}, \bibinfo{number}{4} (\bibinfo{year}{1997}), \bibinfo{pages}{391--411}.
\newblock
\urldef\tempurl%
\url{https://doi.org/10.1145/265924.265927}
\showDOI{\tempurl}


\bibitem[Schützenberger(1965)]%
        {Schutzenberger1965}
\bibfield{author}{\bibinfo{person}{M.P. Schützenberger}.} \bibinfo{year}{1965}\natexlab{}.
\newblock \showarticletitle{On finite monoids having only trivial subgroups}.
\newblock \bibinfo{journal}{\emph{Information and Control}} \bibinfo{volume}{8}, \bibinfo{number}{2} (\bibinfo{year}{1965}), \bibinfo{pages}{190--194}.
\newblock
\showISSN{0019-9958}
\urldef\tempurl%
\url{https://doi.org/10.1016/S0019-9958(65)90108-7}
\showDOI{\tempurl}


\bibitem[Sen et~al\mbox{.}(2005)]%
        {Sen2005}
\bibfield{author}{\bibinfo{person}{Koushik Sen}, \bibinfo{person}{Grigore Ro\c{s}u}, {and} \bibinfo{person}{Gul Agha}.} \bibinfo{year}{2005}\natexlab{}.
\newblock \showarticletitle{Detecting Errors in Multithreaded Programs by Generalized Predictive Analysis of Executions}. In \bibinfo{booktitle}{\emph{Proceedings of the 7th IFIP WG 6.1 International Conference on Formal Methods for Open Object-Based Distributed Systems}} (Athens, Greece) \emph{(\bibinfo{series}{FMOODS'05})}. \bibinfo{publisher}{Springer-Verlag}, \bibinfo{address}{Berlin, Heidelberg}, \bibinfo{pages}{211–226}.
\newblock
\showISBNx{3540261818}
\urldef\tempurl%
\url{https://doi.org/10.1007/11494881_14}
\showDOI{\tempurl}


\bibitem[Sen and Rosu(2003)]%
        {sen2003generating}
\bibfield{author}{\bibinfo{person}{Koushik Sen} {and} \bibinfo{person}{Grigore Rosu}.} \bibinfo{year}{2003}\natexlab{}.
\newblock \showarticletitle{Generating Optimal Monitors for Extended Regular Expressions}. In \bibinfo{booktitle}{\emph{Third Workshop on Run-time Verification, RV@CAV 2003, Boulder, Colorado, USA, July 14, 2003}} \emph{(\bibinfo{series}{Electronic Notes in Theoretical Computer Science}, Vol.~\bibinfo{volume}{89})}, \bibfield{editor}{\bibinfo{person}{Oleg Sokolsky} {and} \bibinfo{person}{Mahesh Viswanathan}} (Eds.). \bibinfo{publisher}{Elsevier}, \bibinfo{pages}{226--245}.
\newblock
\urldef\tempurl%
\url{https://doi.org/10.1016/S1571-0661(04)81051-X}
\showDOI{\tempurl}


\bibitem[Sen et~al\mbox{.}(2006)]%
        {sen2006online}
\bibfield{author}{\bibinfo{person}{Koushik Sen}, \bibinfo{person}{Grigore Rosu}, {and} \bibinfo{person}{Gul Agha}.} \bibinfo{year}{2006}\natexlab{}.
\newblock \showarticletitle{Online efficient predictive safety analysis of multithreaded programs}.
\newblock \bibinfo{journal}{\emph{Int. J. Softw. Tools Technol. Transf.}} \bibinfo{volume}{8}, \bibinfo{number}{3} (\bibinfo{year}{2006}), \bibinfo{pages}{248--260}.
\newblock
\urldef\tempurl%
\url{https://doi.org/10.1007/S10009-005-0192-Y}
\showDOI{\tempurl}


\bibitem[Serbanuta et~al\mbox{.}(2012)]%
        {MaximalCausalModel2013}
\bibfield{author}{\bibinfo{person}{Traian{-}Florin Serbanuta}, \bibinfo{person}{Feng Chen}, {and} \bibinfo{person}{Grigore Rosu}.} \bibinfo{year}{2012}\natexlab{}.
\newblock \showarticletitle{Maximal Causal Models for Sequentially Consistent Systems}. In \bibinfo{booktitle}{\emph{Runtime Verification, Third International Conference, {RV} 2012, Istanbul, Turkey, September 25-28, 2012, Revised Selected Papers}} \emph{(\bibinfo{series}{Lecture Notes in Computer Science}, Vol.~\bibinfo{volume}{7687})}, \bibfield{editor}{\bibinfo{person}{Shaz Qadeer} {and} \bibinfo{person}{Serdar Tasiran}} (Eds.). \bibinfo{publisher}{Springer}, \bibinfo{pages}{136--150}.
\newblock
\urldef\tempurl%
\url{https://doi.org/10.1007/978-3-642-35632-2\_16}
\showDOI{\tempurl}


\bibitem[Serebryany and Iskhodzhanov(2009)]%
        {threadsanitizer}
\bibfield{author}{\bibinfo{person}{Konstantin Serebryany} {and} \bibinfo{person}{Timur Iskhodzhanov}.} \bibinfo{year}{2009}\natexlab{}.
\newblock \showarticletitle{ThreadSanitizer: Data Race Detection in Practice}. In \bibinfo{booktitle}{\emph{Proceedings of the Workshop on Binary Instrumentation and Applications}} (New York, New York, USA) \emph{(\bibinfo{series}{WBIA '09})}. \bibinfo{publisher}{Association for Computing Machinery}, \bibinfo{address}{New York, NY, USA}, \bibinfo{pages}{62–71}.
\newblock
\showISBNx{9781605587936}
\urldef\tempurl%
\url{https://doi.org/10.1145/1791194.1791203}
\showDOI{\tempurl}


\bibitem[Sinha et~al\mbox{.}(2011)]%
        {sinha2012predicting}
\bibfield{author}{\bibinfo{person}{Arnab Sinha}, \bibinfo{person}{Sharad Malik}, \bibinfo{person}{Chao Wang}, {and} \bibinfo{person}{Aarti Gupta}.} \bibinfo{year}{2011}\natexlab{}.
\newblock \showarticletitle{Predicting Serializability Violations: SMT-Based Search vs. DPOR-Based Search}. In \bibinfo{booktitle}{\emph{Hardware and Software: Verification and Testing - 7th International Haifa Verification Conference, {HVC} 2011, Haifa, Israel, December 6-8, 2011, Revised Selected Papers}} \emph{(\bibinfo{series}{Lecture Notes in Computer Science}, Vol.~\bibinfo{volume}{7261})}, \bibfield{editor}{\bibinfo{person}{Kerstin Eder}, \bibinfo{person}{Jo{\~{a}}o Louren{\c{c}}o}, {and} \bibinfo{person}{Onn Shehory}} (Eds.). \bibinfo{publisher}{Springer}, \bibinfo{pages}{95--114}.
\newblock
\urldef\tempurl%
\url{https://doi.org/10.1007/978-3-642-34188-5\_11}
\showDOI{\tempurl}


\bibitem[Smaragdakis et~al\mbox{.}(2012)]%
        {cp2012}
\bibfield{author}{\bibinfo{person}{Yannis Smaragdakis}, \bibinfo{person}{Jacob Evans}, \bibinfo{person}{Caitlin Sadowski}, \bibinfo{person}{Jaeheon Yi}, {and} \bibinfo{person}{Cormac Flanagan}.} \bibinfo{year}{2012}\natexlab{}.
\newblock \showarticletitle{Sound predictive race detection in polynomial time}. In \bibinfo{booktitle}{\emph{Proceedings of the 39th {ACM} {SIGPLAN-SIGACT} Symposium on Principles of Programming Languages, {POPL} 2012, Philadelphia, Pennsylvania, USA, January 22-28, 2012}}, \bibfield{editor}{\bibinfo{person}{John Field} {and} \bibinfo{person}{Michael Hicks}} (Eds.). \bibinfo{publisher}{{ACM}}, \bibinfo{pages}{387--400}.
\newblock
\urldef\tempurl%
\url{https://doi.org/10.1145/2103656.2103702}
\showDOI{\tempurl}


\bibitem[Smith et~al\mbox{.}(2001)]%
        {Smith01}
\bibfield{author}{\bibinfo{person}{L.~A. Smith}, \bibinfo{person}{J.~M. Bull}, {and} \bibinfo{person}{J. Obdrz\'{a}lek}.} \bibinfo{year}{2001}\natexlab{}.
\newblock \showarticletitle{A Parallel Java Grande Benchmark Suite}. In \bibinfo{booktitle}{\emph{Proceedings of the 2001 ACM/IEEE Conference on Supercomputing}} (Denver, Colorado) \emph{(\bibinfo{series}{SC '01})}. \bibinfo{publisher}{ACM}, \bibinfo{address}{New York, NY, USA}, \bibinfo{pages}{8--8}.
\newblock
\showISBNx{1-58113-293-X}
\urldef\tempurl%
\url{https://doi.org/10.1145/582034.582042}
\showDOI{\tempurl}


\bibitem[Sorrentino et~al\mbox{.}(2010)]%
        {sorrentino2010penelope}
\bibfield{author}{\bibinfo{person}{Francesco Sorrentino}, \bibinfo{person}{Azadeh Farzan}, {and} \bibinfo{person}{P. Madhusudan}.} \bibinfo{year}{2010}\natexlab{}.
\newblock \showarticletitle{{PENELOPE:} weaving threads to expose atomicity violations}. In \bibinfo{booktitle}{\emph{Proceedings of the 18th {ACM} {SIGSOFT} International Symposium on Foundations of Software Engineering, 2010, Santa Fe, NM, USA, November 7-11, 2010}}, \bibfield{editor}{\bibinfo{person}{Gruia{-}Catalin Roman} {and} \bibinfo{person}{Andr{\'{e}} van~der Hoek}} (Eds.). \bibinfo{publisher}{{ACM}}, \bibinfo{pages}{37--46}.
\newblock
\urldef\tempurl%
\url{https://doi.org/10.1145/1882291.1882300}
\showDOI{\tempurl}


\bibitem[Tun{\c{c}} et~al\mbox{.}(2023)]%
        {tuncc2023sound}
\bibfield{author}{\bibinfo{person}{H{\"{u}}nkar~Can Tun{\c{c}}}, \bibinfo{person}{Umang Mathur}, \bibinfo{person}{Andreas Pavlogiannis}, {and} \bibinfo{person}{Mahesh Viswanathan}.} \bibinfo{year}{2023}\natexlab{}.
\newblock \showarticletitle{Sound Dynamic Deadlock Prediction in Linear Time}.
\newblock \bibinfo{journal}{\emph{Proc. {ACM} Program. Lang.}} \bibinfo{volume}{7}, \bibinfo{number}{{PLDI}} (\bibinfo{year}{2023}), \bibinfo{pages}{1733--1758}.
\newblock
\urldef\tempurl%
\url{https://doi.org/10.1145/3591291}
\showDOI{\tempurl}


\bibitem[Vaziri et~al\mbox{.}(2006)]%
        {vaziri2006associating}
\bibfield{author}{\bibinfo{person}{Mandana Vaziri}, \bibinfo{person}{Frank Tip}, {and} \bibinfo{person}{Julian Dolby}.} \bibinfo{year}{2006}\natexlab{}.
\newblock \showarticletitle{Associating synchronization constraints with data in an object-oriented language}. In \bibinfo{booktitle}{\emph{Proceedings of the 33rd {ACM} {SIGPLAN-SIGACT} Symposium on Principles of Programming Languages, {POPL} 2006, Charleston, South Carolina, USA, January 11-13, 2006}}, \bibfield{editor}{\bibinfo{person}{J.~Gregory Morrisett} {and} \bibinfo{person}{Simon L.~Peyton Jones}} (Eds.). \bibinfo{publisher}{{ACM}}, \bibinfo{pages}{334--345}.
\newblock
\urldef\tempurl%
\url{https://doi.org/10.1145/1111037.1111067}
\showDOI{\tempurl}


\bibitem[Voung et~al\mbox{.}(2007)]%
        {voung2007relay}
\bibfield{author}{\bibinfo{person}{Jan~Wen Voung}, \bibinfo{person}{Ranjit Jhala}, {and} \bibinfo{person}{Sorin Lerner}.} \bibinfo{year}{2007}\natexlab{}.
\newblock \showarticletitle{{RELAY:} static race detection on millions of lines of code}. In \bibinfo{booktitle}{\emph{Proceedings of the 6th joint meeting of the European Software Engineering Conference and the {ACM} {SIGSOFT} International Symposium on Foundations of Software Engineering, 2007, Dubrovnik, Croatia, September 3-7, 2007}}, \bibfield{editor}{\bibinfo{person}{Ivica Crnkovic} {and} \bibinfo{person}{Antonia Bertolino}} (Eds.). \bibinfo{publisher}{{ACM}}, \bibinfo{pages}{205--214}.
\newblock
\urldef\tempurl%
\url{https://doi.org/10.1145/1287624.1287654}
\showDOI{\tempurl}


\bibitem[Williams(2018)]%
        {williams2018}
\bibfield{author}{\bibinfo{person}{Virginia~Vassilevska Williams}.} \bibinfo{year}{2018}\natexlab{}.
\newblock \showarticletitle{{On some fine-grained questions in algorithms and complexity}}. In \bibinfo{booktitle}{\emph{Proceedings of the International Congress of Mathematicians: Rio de Janeiro 2018}}. World Scientific, \bibinfo{pages}{3447--3487}.
\newblock
\urldef\tempurl%
\url{https://doi.org/10.1142/9789813272880_0188}
\showDOI{\tempurl}


\bibitem[Yuan et~al\mbox{.}(2018)]%
        {yuan2018partial}
\bibfield{author}{\bibinfo{person}{Xinhao Yuan}, \bibinfo{person}{Junfeng Yang}, {and} \bibinfo{person}{Ronghui Gu}.} \bibinfo{year}{2018}\natexlab{}.
\newblock \showarticletitle{Partial Order Aware Concurrency Sampling}. In \bibinfo{booktitle}{\emph{Computer Aided Verification - 30th International Conference, {CAV} 2018, Held as Part of the Federated Logic Conference, FloC 2018, Oxford, UK, July 14-17, 2018, Proceedings, Part {II}}} \emph{(\bibinfo{series}{Lecture Notes in Computer Science}, Vol.~\bibinfo{volume}{10982})}, \bibfield{editor}{\bibinfo{person}{Hana Chockler} {and} \bibinfo{person}{Georg Weissenbacher}} (Eds.). \bibinfo{publisher}{Springer}, \bibinfo{pages}{317--335}.
\newblock
\urldef\tempurl%
\url{https://doi.org/10.1007/978-3-319-96142-2\_20}
\showDOI{\tempurl}


\end{thebibliography}


\iftoggle{appendix}{
\appendix

\section{Proofs from \secref{pattern-languages}}
\applabel{closure-property}

In this section, we prove the closure properties of 
pattern languages and generalized pattern languages
(\thmref{pat-clos} and \thmref{gener-clos}).


We first give proof of some closure properties of pattern languages $\patre$.

\begin{proof}[Proof of \thmref{pat-clos}]

We prove that the class of all pattern languages, $\patre$, is closed under concatenation,
but is not closed under union, intersection, complementation, and Kleene Star.

\myparagraph{Pattern languages are not closed under union}
We prove it by providing a counterexample.
Let us consider two patterns $\patts{a}, \patts{b}\in \patre$,
where $a \neq b \in \alphabet$.
To prove by contradiction, 
we assume there is a pattern $\patts{\seq{c}{k}}\in \patre$ such that 
$\patts{\seq{c}{k}} = \patts{a} \cup \patts{b}$.
Since all the strings contained in $\patts{\seq{c}{k}}$ 
have a minimum length of $k$,
and $a, b \in \patts{\seq{c}{k}}$,
it implies that $k = 1$.
If $c_1\not\in \set{a,b}$, then $c_1$ belongs to $\patts{c_1}$ 
but not to $\patts{a} \cup \patts{b}$.
If $c_1 = a$, then $b$ belongs to $\patts{b}$ 
but not to $\patts{c_1}$.
Similarly, $c_1 = b$ also leads to a contradiction.
From the above cases, we can conclude that 
no pattern equals the union of $\patts{a}$ and $\patts{b}$.

\myparagraph{Pattern languages are not closed under intersection}
A counterexample, in this case, is that 
$\patts{a} \cap \patts{b}$ cannot be written as a pattern language.
We assume that, by contradiction, 
there is a pattern $\patts{\seq{c}{k}}\in \patre$ such that 
$\patts{\seq{c}{k}} = \patts{a} \cap \patts{b}$.
Since $ab$ and $ba$ belong to $\patts{a} \cap \patts{b}$,
$k$ is at most 2.

If $k = 1$, $c_1 \in \patts{c_1} = \patts{a} \cap \patts{b}$.
But $c_1$ cannot be $a$ and $b$ at the same time.
Then $k = 2$. 
Observe that $ab, ba\in \patts{c_1, c_2}$.
However, $c_1, c_2$ do not exist because they should be $ab$ and $ba$ at the same time.
From the above contradictions, 
we conclude that $\patts{a} \cap \patts{b}$ is not a pattern language.
Thus pattern languages are not closed under intersection.

\myparagraph{Pattern languages are not closed under complementation}
Here is a counterexample.
Consider the pattern language $L = \patts{a}$.
$L^c = (\alphabet \setminus \set{a})^*$,
which contains all the strings without the occurrence of $a$.
By contradiction, we assume that there is a $\patts{\seq{b}{k}}\in \patre$ such that 
$\patts{\seq{b}{k}} = L^c$.
However, observe that 
$a\cdot b_1 \cdot \dots \cdot b_k \in \patts{\seq{b}{k}}$,
which gives us a contradiction.

\myparagraph{Pattern languages are closed under concatenation}
For any two pattern languages $L_1 = \patts{\seq{a}{d}}$
and $L_2 = \patts{\seq{b}{l}}$, 
we claim that the language $L = \patts{\seq{a}{d} \circ \seq{b}{l}}$ is indeed
the concatenation of $L_1$ and $L_2$, i.e.,
$L = L_1 \circ L_2$.

\myparagraph{Pattern languages are not closed under Kleene star}
We first observe that the only pattern language that contains $\emptystr$
is $\alphabet^*$, which is a pattern language when $d = 0$.
$\emptystr \in (\patts{a, a})^*$ implies that $(\patts{a, a})^* = \alphabet^*$.
However, $a\in \alphabet^*$ does not belong to $(\patts{a, a})^*$, 
which contains either an empty string or a string with at least two $a$s.
Thus, $(\patts{a, a})^*$ is not a generalized pattern language,
which gives a counterexample.

\end{proof}


In the following, we prove some closure properties of generalized pattern languages.

\begin{proof}[Proof of \thmref{gener-clos}]

We prove that the class of all generalized pattern languages is closed under 
union, intersection, concatenation, and Kleene Star,
but is not closed under complementation.

\myparagraph{Generalized pattern languages are closed under union}
Consider two generalized pattern languages $L_1 = \bigcup_{i\in I} L_{1,i}$
and $L_2 = \bigcup_{j\in J} L_{2,j}$, 
where each of $\set{L_{1,i}, L_{2, j}}_{i \in I, j \in J}$ is one of the following: 
pattern language, $\set{\emptystr}$, or $\emptyset$.
From the definition, we know that $L_1 \cup L_2 = (\bigcup_{i\in I} L_{1,i}) \bigcup (\bigcup_{j\in J} L_{2,j})$
is a generalized pattern language.

\myparagraph{Generalized pattern languages are closed under intersection}
We first consider the intersection of two pattern languages.
We have proved that pattern languages are not closed under intersection,
but we claim that the intersection of two pattern languages is a generalized pattern language.
For any two pattern languages $L_1 = \patts{\seq{a}{k}}$ and $L_2 = \patts{\seq{b}{l}}$,
$L_1 \cap L_2 = \bigcup_{w\in W} \patts{w}$,
where 
\[W = \seq{a}{d} \odot \seq{b}{l}\]
Notice that $|W|$ is bounded because the length of $w\in W$ is bounded by $k + l$.
Now we prove equality.
\begin{enumerate}
        \item ($L_1 \cap L_2 \subseteq \bigcup_{w\in W} \patts{w}$)
                For any $\sigma\in L_1 \cap L_2$, 
                there must be two subsequences $w_1$ and $w_2$ of $\sigma$
                such that $w_1 = \seq{a}{k}$ and $w_2 = \seq{b}{l}$.
                Consider the smallest subsequence $w'$ of $\sigma$ 
                such that $w_1$ and $w_2$ are subsequences of $w'$.
                Clearly, $w'\in W$,
                which implies that $\sigma \in \bigcup_{w\in W} \patts{w}$.
        \item ($\bigcup_{w\in W} \patts{w} \subseteq L_1 \cap L_2$)
                For any $\sigma \in \bigcup_{w\in W} \patts{w}$, 
                there exists a $w\in W$ such that $\sigma \in \patts{w}$.
                Since $\seq{a}{k}$ and $\seq{b}{l}$ are both subsequences of $w$,
                they are also subsequences of $\sigma$,
                which directly gives us $\sigma \in L_1 \cap L_2$.
\end{enumerate}

Now we turn our focus on the original closure property of generalized pattern languages.
Consider two generalized pattern languages $L_1 = \bigcup_{i\in I} L_{1,i}$
and $L_2 = \bigcup_{j\in J} L_{2,j}$, 
where each of $\set{L_{1,i}, L_{2, j}}_{i \in I, j \in J}$ is one of the following: 
pattern language, $\set{\emptystr}$, or $\emptyset$.
Their intersection is $L_1 \cap L_2 = \bigcup_{i \in I, j \in J} L_{1,i} \cap L_{2, j}$.
Now consider the following cases:
\begin{enumerate}
    \item $L_{1,i}$ and $L_{2,j}$ are both pattern languages.
            From above, $L_{1, i} \cap L_{2, j}$ is a union of pattern languages.
    \item $L_{1,i} = \set{\emptystr}$ and $L_{2,j} = \set{\emptystr}$.
            Then $L_{1,i} \cap L_{2, j} = \set{\emptystr}$. 
    \item One of $L_{1,i}, L_{2,j}$ is $\set{\emptystr}$ and the other is not.
            Then $L_{1,i} \cap L_{2, j} = \emptyset$.
    \item At least one of $L_{1,i}, L_{2,j}$ is $\emptyset$.
            Then $L_{1,i} \cap L_{2, j} = \emptyset$.
\end{enumerate}
We can conclude from the above cases that $L_{1,i} \cap L_{2, j}$ is one of the following:
a union of pattern languages, $\set{\emptystr}$, or $\emptyset$.
Hence, $L_1 \cap L_2$ is a generalized pattern language.

\myparagraph{Generalized pattern languages are not closed under complementation}
We present a similar counterexample here.
Consider the pattern language $L = \patts{a}$.
$L^c = (\alphabet \setminus \set{a})^*$,
which contains all the strings without the occurrence of $a$.
By contradiction, we assume that there is a 
generalized pattern language $L' = \bigcup_{i\in I} L_{i}$
where $L_i$ is one of the following: 
pattern language, $\set{\emptystr}$, or $\emptyset$.
If no $L_i$ is a pattern language, then $L' = \set{\emptystr}$,
which cannot be $\bigcup_{i\in I} L_{i}$. 
Then, without loss of generality,
$L_1$ is a pattern language and is $\patts{\seq{b}{l}}$.
$a\cdot b_1 \cdot \dots \cdot b_k$ belongs to $\patts{\seq{b}{k}}$, 
thus belongs to $L'$,
which gives a contradiction because, by assumption,
$L' = L^c = (\alphabet \setminus \set{a})$.

\myparagraph{Generalized pattern languages are closed under concatenation}
Consider two generalized pattern languages $L_1 = \bigcup_{i\in I} L_{1,i}$
and $L_2 = \bigcup_{j\in J} L_{2,j}$, 
where each of $\set{L_{1,i}, L_{2, j}}_{i \in I, j \in J}$ is one of the following: 
pattern language, $\set{\emptystr}$, or $\emptyset$.
Their concatenation is $L_1 \circ L_2 = \bigcup_{i \in I, j \in J} L_{1,i} \circ L_{2, j}$.
Now consider the following cases:
\begin{enumerate}
    \item $L_{1,i}$ and $L_{2,j}$ are both pattern languages.
            $L_{1, i} \circ L_{2, j}$ is also a pattern language because pattern languages are closed under concatenation.
    \item $L_{1,i} = \set{\emptystr}$ or $L_{2,j} = \set{\emptystr}$.
            We consider, without loss of generality, the first case that $L_{1, i} = \set{\emptystr}$. 
            Then $L_{1,i} \circ L_{2, j} = L_{2, j}$. 
    \item $L_{1,i} = \emptyset$ or $L_{2,j} = \emptyset$.
            For either case, we have $L_{1,i} \circ L_{2, j} = \emptyset$.
\end{enumerate}
We can conclude from the above cases that $L_{1,i} \circ L_{2, j}$ is still one of the following:
pattern language, $\set{\emptystr}$, or $\emptyset$.
Hence, $L_1 \circ L_2$ is a generalized pattern language.

\myparagraph{Generalized pattern languages are closed under Kleene star}
We claim $L^* = L\cup \set{\emptystr}$, where $L$ is a generalized pattern language.
One direction that $L\cup \set{\emptystr} \subseteq L^*$ is trivial.
Now consider $L^* \subseteq L\cup \set{\emptystr}$,
which will be proved by induction that 
for all $n > 0$, $L^n \subseteq L\cup \set{\emptystr}$.
\begin{enumerate}
    \item (Base case) $L^0 = \set{\emptystr}\subseteq L\cup \set{\emptystr}$.
    \item (Inductive case) Assume that $L^k \subseteq L\cup \set{\emptystr}$.
    $L^{k+1} = L^k \circ L$. Consider $\sigma \in L^{k+1} = L^k \circ L$.
    We can break down $\sigma$ into two parts $\sigma = \sigma_1 \circ \sigma_2$,
    such that $\sigma_1 \in L^k \subseteq L\cup \set{\emptystr}$ and $\sigma_2 \in L$.
    From the definition of generalized pattern language, 
    $\sigma_1$ must be either $\emptystr$ or a member of the pattern language $\patts{\seq{a}{d}} \subseteq L$.
    If $\sigma_1 = \emptystr$, then $\sigma = \sigma_2\in L \subseteq L\cup \set{\emptystr}$.
    In the other case that $\sigma_1 \in \patts{\seq{a}{d}}$, 
    observe that $\sigma = \sigma_1 \circ \sigma_2$ is also a member of $\patts{\seq{a}{d}}$,
    because $\sigma_2 \in \alphabet^*$ which is the last composition of $\patts{\seq{a}{d}}$.
    Thus, $\sigma \in L \subseteq L\cup \set{\emptystr}$.
\end{enumerate}
Notice that $L\cup \set{\emptystr}$ is also a generalized pattern language,
implying that generalized Pattern Languages are closed under the Kleene star.

\end{proof}

\section{Proofs from \secref{algo-pattern}}
\applabel{proof-algo}


\subsection{Proofs from \secref{candidate}}
\applabel{app-ad}
\begin{proof}[Proof of~\lemref{witness}]

    Let $\tau^\dagger= \sort{\tau}{\tuple{a_{j_1}, \dots, a_{j_m}}} = \tuple{e_{i^\dagger_1}, \dots, e_{i^\dagger_m}}$.
    What we need to prove is that 
    \[\exists \tr'\mazeq{\indep}\tr, \tau^\dagger\text{ is a subsequence of }\tr' \iff 
        \forall i^\dagger_r\neq i^\dagger_s, s < r \Rightarrow e_{i^\dagger_r}\nmazpo{\dep}{\tr}e_{i^\dagger_s}\]
    
    \noindent($\Rightarrow$) 
    We assume that, by contradiction, 
    there are $r, s$ such that $s < r$ and $e_{i^\dagger_r}\mazpo{\dep}{\tr}e_{i^\dagger_s}$.
    However, the latter assumption contradicts the fact that $e_{i^\dagger_r}$ occurs after $e_{i^\dagger_s}$ in $\tr'$
    and $\tr'$ follows the same partial order as $\tr$.
    
    \noindent($\Leftarrow$) 
    Consider the underlying directed acyclic graph $G_{\tr}$ of the partial order $\mazpo{\dep}{\tr}$ induced by $\tr$.
    The left-hand side is equivalent to say,
    adding edges $(e_{i^\dagger_j}, e_{i^\dagger_{j+1}})$, $1\le j \le m-1$,
    to $G_{\tr}$ does not change its acyclicity. 
    Then a linearization of the new graph is equivalent to $\tr$,
    and also, has a subsequence $\tau^\dagger$.
    
    Let us prove this by contradiction.
    Assume that the new graph $G_{\tr\cup \tau^\dagger}$ is not acyclic.
    Then We prove that there must be a \emph{bad cycle} that
    consists of edges from $\setpred{(e_{i^\dagger_j}, e_{i^\dagger_{j+1}})}{1\le j \le m-1}$,
    which is a contradiction since those edges form a chain.
    We do induction on the number of edges in the cycle that come from $G_{\tr}$.
    The base case when the number is zero is trivial. 
    We assume that if there is a cycle containing less than or equal to $k$ edges coming from $G_{\tr}$,
    then there is a bad cycle in the graph.
    Consider a cycle that contains $k+1$ coming from $G_{\tr}$.
    There must be two different nodes $e_{i^\dagger_r}$ and $e_{i^\dagger_{s}}$ linked by a chain of edges from $G_{\tr}$.
    Otherwise, the whole cycle consists of edges from $G_{\tr}$,
    which is acyclic.
    Then $e_{i^\dagger_r} \mazpo{\dep}{\tr} e_{i^\dagger_s}$,
    which implies $r < s$.
    Thus, there is a chain of edges $(e_{i^\dagger_r}, e_{i^\dagger_{r + 1}}), \dots, (e_{i^\dagger_{s-1}}, e_{i^\dagger_{s}})$
    from $e_{i^\dagger_r}$ to $e_{i^\dagger_{s}}$.
    Now we get a cycle containing less than or equal to $k$ edges coming from $G_{\tr}$,
    which implies the existence of a bad cycle.
    Consequently, if the new graph is not acyclic,
    there must be a bad cycle that cannot exist. 
\end{proof}


Now we prove the correctness of incremental updates of after sets.

\begin{proof}[Proof of \lemref{aft-incre}]

From the definition, $\aft{\rho\circ f}{e}$ is defined as 
\begin{align*}
    \aft{\rho\circ f}{e}=\begin{cases}
        \aft{\rho}{e}\cup \set{\lbl{f}} & \text{ if }e\mazpo{\dep}{\rho\circ f}f,\\
        \aft{\rho}{e} &\text{ otherwise}.
    \end{cases}
\end{align*}

We will prove that $e\mazpo{\dep}{\rho\circ f}f$ iff there exists $a\in \aft{\rho}{e}$ such that $(a,\lbl{f})\not\in\indep$.

\noindent($\Leftarrow$)
Assume that $a\in \aft{\rho}{e}$ such that $(a,\lbl{f})\not\in\indep$.
Since $a\in \aft{\rho}{e}$, from the definition, 
there is an event $e'\in \rho$ such that $e\mazpo{\dep}{\rho} e'$ and $\lbl{e'} = a$.
Moreover, we have $e'\mazpo{\dep}{\rho} f$ because $(a,\lbl{f})\not\in\indep$.
By transitivity, $e\mazpo{\dep}{\rho\circ f}f$.

\noindent($\Rightarrow$)
From the definition of the partial order $\mazpo{\dep}{\rho\circ f}$, 
$e\mazpo{\dep}{\rho\circ f}f$ implies that 
there is an event $e' \in \rho$ such that 
$e\mazpo{\dep}{\rho} e'$ and $(\lbl{e'},\lbl{f})\not\in\indep$.
So $\lbl{e'}\in \aft{\rho}{e}$ by definition of after set,
which provides an instance of $a$ on the right-hand side.

\end{proof}


\subsection{Proofs from \secref{alg-correct}}
\applabel{app-alg-correct}
Before we prove the existence of the maximum element in a complete set of partially admissible tuples,
we prove that the set is closed under the join operation.

\begin{proof}[Proof of \lemref{clojoin}]
Let $\tau_1 = \tuple{\seq{e^1}{m}}, \tau_2 = \tuple{\seq{e^2}{m}}$,
such that $\lbl{\tau_1} = \lbl{\tau_2} = \tuple{\seq{b}{m}}$.
Let $\tau_1\bigtriangledown\tau_2 = \tuple{\seq{e}{m}}$,
where for $1\le i\le m$, $e_i = e^2_i$ if $e^1_i\trord{\tr}e^2_i$, 
or $e^1_i$ otherwise.
It is obvious that $\lbl{\tau_1\bigtriangledown\tau_2} = \tuple{\seq{b}{m}}$. 
We will prove that $\tau_1\bigtriangledown\tau_2$ is a partially locally admissible tuple.

Let $\tuple{a_{j_1}, \dots, a_{j_m}}$ be a subsequence of $a_1, \dots, a_d$ and a permutation of $\tuple{\seq{b}{m}}$.
Notice that the reordering of $\mathsf{sort}$ only depends on the labels.
Therefore, let 
\[\tuple{e^1_{i^\dagger_1}, \dots, e^1_{i^\dagger_m}} = \sort{\tau_1}{\tuple{a_{j_1}, \dots, a_{j_m}}},\]
\[\tuple{e^2_{i^\dagger_1}, \dots, e^2_{i^\dagger_m}} = \sort{\tau_2}{\tuple{a_{j_1}, \dots, a_{j_m}}},\]
\[\tuple{e_{i^\dagger_1}, \dots, e_{i^\dagger_m}} = \sort{\tau_1\bigtriangledown\tau_2}{\tuple{a_{j_1}, \dots, a_{j_m}}}.\]
    
Since $\tau_1$ and $\tau_2$ are admissible, 
for all $i^\dagger_r\neq i^\dagger_s$, 
we have
$s < r \Rightarrow e^1_{i^\dagger_r}\nmazpo{\dep}{\tr}e^1_{i^\dagger_s}$
and $e^2_{i^\dagger_r}\nmazpo{\dep}{\tr}e^2_{i^\dagger_s}$.

For any $i^\dagger_r\neq i^\dagger_s$ such that $s < r$,
we prove that $e_{i^\dagger_r}\nmazpo{\dep}{\tr}e_{i^\dagger_s}$ by contradiction.
Assume that $e_{i^\dagger_r}\mazpo{\dep}{\tr}e_{i^\dagger_s}$.
It is impossible that both $e_{i^\dagger_r}$ and $e_{i^\dagger_s}$ come from the same source, $\tau_1$ or $\tau_2$.
Without loss of generality, let $e_{i^\dagger_r} = e^1_{i^\dagger_r}$ and $e_{i^\dagger_s} = e^2_{i^\dagger_s}$.
Then we have $e^2_{i^\dagger_r}\trord{\tr} e^1_{i^\dagger_r} \mazpo{\dep}{\tr} e^2_{i^\dagger_s}$,
which contradicts to the fact that $\tau_2$ is locally admissible.
This concludes that $\tau_1\bigtriangledown\tau_2$ is admissible.
\end{proof}

Now let us prove the existence of the maximum element.

\begin{proof}[Proof of \lemref{exMax}]
    In short, the set $\adset{\rho}{\tuple{\seq{b}{m}}}$ and the join operations give a join-semilattice, 
    resulting in the existence of the maximum element.
    Also, it is easy to check that the unique maximum element is exactly $\bigtriangledown_{\tau\in \adset{\rho}{\tuple{\seq{b}{m}}}}\tau$.
\end{proof}

Let us prove \lemref{comp-max} demonstrating the way we keep track of the maximum element of $\adset{\rho}{\tuple{\seq{b}{m}}}$ along the prefix $\rho$.

\begin{proof}[Proof of \lemref{comp-max}]
    If $\lbl{f} = b_m$ and $\max(\adset{\rho}{\tuple{\seq{b}{m-1}}}) \circ f$ is partially admissible,
    then it is obvious that 
    \[\max(\adset{\rho\circ f}{\tuple{\seq{b}{m}}}) = \max(\adset{\rho}{\tuple{\seq{b}{m-1}}}) \circ f,\]
    since for all $\tuple{\seq{e}{m}} \in \adset{\rho\circ f}{\tuple{\seq{b}{m}}}$, we have
    \[\tuple{\seq{e}{m-1}} \unlhd \max(\adset{\rho}{\tuple{\seq{b}{m-1}}})\text{ and }e_m \trord{\rho\circ f} f.\]

    If $\lbl{f} \neq b_m$, then the set of admissible tuples of label $\tuple{\seq{b}{m}}$ is unchanged from $\rho$ to $\rho\circ f$.
    The remaining is to prove that when $\lbl{f} = b_m$ and $\max(\adset{\rho}{\tuple{\seq{b}{m-1}}}) \circ f$ is not partially admissible,
    $\max(\adset{\rho\circ f}{\tuple{\seq{b}{m}}}) = \max(\adset{\rho}{\tuple{\seq{b}{m}}})$.

    By contraction, assume $\max(\adset{\rho\circ f}{\tuple{\seq{b}{m}}}) \neq \max(\adset{\rho}{\tuple{\seq{b}{m}}})$.
    Then $\max(\adset{\rho\circ f}{\tuple{\seq{b}{m}}})$ must be a tuple ending with $f$.
    
    Let $\max(\adset{\rho\circ f}{\tuple{\seq{b}{m}}}) = \tuple{\seq{e}{m-1}, f}$,
    where we also have $\tuple{\seq{e}{m-1}} \in \adset{\rho}{\tuple{\seq{b}{m-1}}}$.
    Let $\max(\adset{\rho}{\tuple{\seq{b}{m-1}}}) = \tuple{\seq{e'}{m-1}}$.
    Then for all $1\le i\le m-1$, $e_i \trord{\rho} e'_i$.

    Since $\max(\adset{\rho}{\tuple{\seq{b}{m-1}}}) \circ f$ is not partially admissible, 
    there must be a $e'_i$ which should occur after $f$ in pattern $\tuple{\seq{a}{m}}$ 
    but is partially ordered with $f$, $e'_i \mazpo{\dep}{\rho\circ f} f$.
    Then we have $e_i \trord{\rho} e'_i \mazpo{\dep}{\rho\circ f} f$,
    which contradicts with the fact that $\tuple{\seq{e}{m-1}, f}$ is admissible.
    Hence, we have $\max(\adset{\rho\circ f}{\tuple{\seq{b}{m}}}) = \max(\adset{\rho}{\tuple{\seq{b}{m}}})$. 

\end{proof}


\subsection{Proofs from \secref{comb}}
\applabel{comb}
Finally, we present the proof of the correctness of \algoref{detail}.

\begin{proof}[Proof of \thmref{algo-correct}]
    The correctness of the algorithm can be shown by a loop invariant:

    \begin{quote}
        For the prefix $\rho\circ f$ of $\tr$ and for any permutation of prefixes of $\seq{a}{d}$, $\tuple{\seq{b}{m}}$.
        \[M(\tuple{\seq{b}{m}}) = \tuple{\aft{\rho\circ f}{e_1}, \dots, \aft{\rho\circ f}{e_m}},\]
        where $\tuple{\seq{e}{m}} = \max(\adset{\rho\circ f}{\tuple{\seq{b}{m}}})$.
    \end{quote}

    The loop in the algorithm keeps track of the maximum element of each set in the form of after sets.
    The correctness of the loop invariant follows $\lemref{aft-incre}$ and $\lemref{comp-max}$.
    
    Finally, the predictive monitoring problem answers ``YES'' if and only if there is an admissible tuple of length $k$.
    This is equivalent to the existence of the maximum element of length $k$ after processing some prefix of $\tr$,
    which is also equivalent to the case that our algorithm returns ``YES''.

\end{proof}
}

\end{document}